\documentclass[11pt]{article}
\usepackage[T1]{fontenc}
\usepackage[utf8]{inputenc}
\usepackage{lmodern,microtype}
\usepackage[margin=1in]{geometry}
\usepackage{amsmath,amssymb,amsthm,mathtools}
\usepackage{booktabs,array,tabularx,enumitem}
\usepackage{float}
\floatstyle{ruled}
\newfloat{algorithm}{tbp}{loa}
\floatname{algorithm}{Algorithm}
\newcounter{algline}
\newlength{\algindent}
\newlength{\algorithmicindent}
\newenvironment{algorithmic}[1][]{%
 \small\setcounter{algline}{0}\setlength{\algindent}{0pt}%
 \begin{list}{}{\setlength{\leftmargin}{1.8em}\setlength{\labelwidth}{1.4em}%
 \setlength{\labelsep}{.4em}\setlength{\itemsep}{2pt}\setlength{\parsep}{0pt}}%
}{\end{list}}
\newcommand{\State}{\stepcounter{algline}\item[\scriptsize\thealgline:]\hspace*{\algindent}}
\newcommand{\Statex}{\item[]\hspace*{\algindent}}
\newcommand{\Require}[1]{\item[]\textbf{Input}\quad #1}
\newcommand{\Ensure}[1]{\item[]\textbf{Output}\quad #1}
\newcommand{\Return}{\textbf{return}\ }
\newcommand{\For}[1]{\State\textbf{for} #1 \textbf{do}\addtolength{\algindent}{1em}}
\newcommand{\EndFor}{\addtolength{\algindent}{-1em}\State\textbf{end for}}
\newcommand{\If}[1]{\State\textbf{if} #1 \textbf{then}\addtolength{\algindent}{1em}}
\newcommand{\Else}{\addtolength{\algindent}{-1em}\State\textbf{else}\addtolength{\algindent}{1em}}
\newcommand{\EndIf}{\addtolength{\algindent}{-1em}\State\textbf{end if}}
\newcommand{\Loop}{\State\textbf{repeat}\addtolength{\algindent}{1em}}
\newcommand{\EndLoop}{\addtolength{\algindent}{-1em}\State\textbf{end repeat}}
\usepackage{tikz}
\usetikzlibrary{arrows.meta,positioning,calc,patterns}
\usepackage[dvipsnames]{xcolor}
\usepackage[colorlinks=true,linkcolor=MidnightBlue,citecolor=MidnightBlue,urlcolor=MidnightBlue]{hyperref}
\usepackage{url}
\setlist{topsep=4pt,itemsep=2pt,parsep=0pt}
\newtheorem{theorem}{Theorem}[section]
\newtheorem{lemma}[theorem]{Lemma}

\theoremstyle{definition}

\theoremstyle{remark}

\newcommand{\diam}{\operatorname{diam}}
\newcommand{\poly}{\operatorname{poly}}
\newcommand{\supp}{\operatorname{supp}}

\newcommand{\E}{\mathbb E}

\newcommand{\abs}[1]{\left|#1\right|}

\newcommand{\Hull}{\mathcal H}

\newcommand*{\LeanRepository}{https://github.com/hajiaghayi/Diameter-Alg2} 
\title{On Euclidean Diameter in Dynamic Streams}
\author{Jeff Giliberti \quad MohammadTaghi Hajiaghayi\\[0.3em]
\normalsize University of Maryland, College Park\\
\small\texttt{\{jeffgili,hajiagha\}@umd.edu}}
\date{}
\begin{document}
\maketitle

\begin{abstract}
We give a streaming algorithm for maintaining the diameter of a high-dimensional Euclidean point set under arbitrary insertions and deletions that achieves update time polylogarithmic in the number of points and polynomial in the dimension. For every fixed $c>\sqrt3$, there is an $\epsilon_c > 0$ such that our one-pass algorithm uses $\tilde O(N^{1-\epsilon_c}\poly(d))$ space, has worst-case update time
$\poly(d,\log N)$, and maintains a factor-$c$ pair witness. 

On the hardness side, we prove that every constant-factor approximation requires polynomial memory even on short dynamic streams: a factor-$c$ approximation needs $N^{\Omega(1/(c^2\log(c+1)))}$ space on streams of at most $2N-1$ updates. 

Our upper bound combines a geometric hull, Gaussian hashing through
successively smaller caps, and local and global rebuilding schedules.
Every update changes only one leaf of the hash hierarchy, while the answer
is maintained through small variance certificates. The lower bound is
based on a communication complexity argument. Together, these results separate the polynomial memory regime from
the time needed to process an update.
\end{abstract}
\setcounter{tocdepth}{1}
\tableofcontents

\section{Introduction}\label{sec:introduction}

The diameter of a point set is determined by two points, yet maintaining
even a rough estimate can require information about many points that
are far from the current extremes. A deletion can remove a witness,
and a sequence of deletions can expose a point that the algorithm discarded long ago. In this work, we study Euclidean diameter in the fully dynamic streaming model, where an update inserts or deletes a point. Our goal is to maintain an estimate
and an actual witness pair, while minimizing the information that the algorithm must retain and the time it takes to process each update, with a focus on achieving update time polynomial in dimension and polylogarithmic in the number of points.

For insertion-only Euclidean streams, Agarwal and Sharathkumar~\cite{as}
obtained a $(\sqrt2+\varepsilon)$-approximation and showed that a
factor below $\sqrt2$ can require linear space in high dimension.
Halld\'orsson, Matsakis, and Vesel\'y~\cite{hmv} achieve the same upper
bound while storing only
$O(\varepsilon^{-2}\log(1/\varepsilon))$ points. 
To handle deletions, Indyk's
asymmetric embedding~\cite{indyk}, in the dynamic-streaming formulation
of Padaki, Singal, and Waingarten~\cite{psw}, gives space
$\widetilde O(n^{2/(c^2-1)})$ for approximation $c$. Its direct implementation updates every projection coordinate.
At factor two this yields an $n^{2/3}$ dependence in both memory
and update time, up to polynomial factors in dimension and
logarithmic parameters. In view of the lower bounds of Khanna, Padaki,
Singal, and Waingarten~\cite{kpsw} for general metrics and of Padaki,
Singal, and Waingarten~\cite{psw} for Euclidean space,  $n^{\widetilde\Omega(1/c^2)}$ space usage is required by any algorithm that is correct on streams of unrestricted
length and frequency (a fixed horizon otherwise would limit the deletion
histories the algorithm has to handle). Padaki et al. explicitely ask whether
polynomial space is still necessary with a polynomial
horizon~\cite[Section~1.2]{psw}.

Fully dynamic data structures offer another way to handle deletions.
Chan and Pathak~\cite{chanpathak} used enclosing balls in high dimension,
and Banihashem et al.~\cite{dynamic} maintain a factor-two diameter
with polynomial worst-case update time against adaptive updates.
These fully dynamic data structures retain the input. This leaves open the question whether a streaming algorithm can maintain an approximate farthest pair
using sublinear space and 
update time polynomial in the dimension and polylogarithmic in the stream parameters. 

\subsection{Our contribution}

Let $P_t$ denote the current support and $D_t=\diam(P_t)$, with
$D_t=0$ when $|P_t|\le1$. Write
\[
 \mathcal L=\bigl(d,\log(N+2),\log(\Delta+2),
                         \log((T+2)/\delta)\bigr).
\]
Here $N\ge2$ is a supplied upper bound on the maximum number of
simultaneously active distinct points, and $T$ is a supplied update
horizon. The approximation convention is
$D_t/c\le L_t\le D_t$. Space is measured in ordinary machine words;
the model and the treatment of arithmetic are specified below.

\begin{theorem}[Polynomial update time]\label{thm:main-upper}
For every fixed $c>\sqrt3$, there is a constant $\epsilon_c>0$
and a randomized one-pass algorithm with
\[
 \textnormal{space }N^{1-\epsilon_c}\poly_c(\mathcal L),
 \qquad
 \textnormal{worst-case update time }\poly_c(\mathcal L).
\]
For every oblivious strict-turnstile stream of at most $T$ updates,
with probability at least $1-\delta$ the algorithm maintains
$D_t/c\le L_t\le D_t$ simultaneously at all times.
The answer is cached. Every nonzero answer is the distance between
two current original points, on every random outcome.
The resource bounds also hold on every random outcome.
\end{theorem}

The theorem includes factor two. Section~\ref{sec:combined-quantitative-two}
gives a complete numerical choice with
$\epsilon_2=10^{-10}$. The small exponent is a conservative
existence bound; the contribution is the absence of a positive
power of $N$ in the update time. We do not optimize this exponent
or claim that the resulting algorithm is practical.
A query returns a pointer to the stored score and witness records
in constant time. Writing the exact score, point keys, or coordinates
takes the corresponding polynomial output time.

\begin{samepage}
\begin{theorem}[A short-stream space lower bound]\label{thm:main-lower}
For every fixed $c\ge1$, any randomized streaming algorithm that
returns a factor-$c$ estimate with error at most $1/10$ requires
\[
              N^{\Omega(1/(c^2\log(c+1)))}
\]
memory words in the worst case. The bound holds for streams with
peak cardinality at most $N$ and at most $2N-1$ updates,
with all insertions before all deletions and
frequencies in $\{0,1\}$ at every prefix. The points have dimension
$O_c(\log N)$ and integer coordinates of absolute value
$O_c(\sqrt{\log N})$.
Only a final query is required, and the algorithm may be nonlinear,
history dependent, and computationally unbounded.
\end{theorem}
\end{samepage}

The word size in the second theorem is polylogarithmic in $N$,
as it is for the standard RAM model on these instances.
The proof first bounds the information in the state; dividing by
the word length preserves the stated exponent. In particular,
the theorem excludes $\poly(d,\log N,\log\Delta,\log T)$ memory
for every fixed approximation factor.
It is a space lower bound, and gives no positive-power lower bound
on update time.

\begin{table}[t]
\centering
\small
\renewcommand{\arraystretch}{1.18}
\begin{tabularx}{\linewidth}{@{}>{\raggedright\arraybackslash}p{0.29\linewidth}
  >{\raggedright\arraybackslash}p{0.19\linewidth}
  >{\raggedright\arraybackslash}p{0.20\linewidth}X@{}}
\toprule
Factor-two guarantee & Memory & Update time & Output\\
\midrule
Gaussian-embedding baseline & $N^{2/3}\poly(\mathcal L)$
 & $N^{2/3}\poly(\mathcal L)$ & Cached scalar\\
Theorem~\ref{thm:main-upper} & $N^{1-\epsilon_2}\poly(\mathcal L)$
 & $\poly(\mathcal L)$ & Current witness pair\\
\bottomrule
\end{tabularx}
\caption{The two upper bounds address different time--space tradeoffs.
Both use the peak bound $N$ and an oblivious finite horizon in this
comparison. Appendix~\ref{sec:two-thirds-baseline} gives the
implementation of the established embedding baseline.
Theorem~\ref{thm:main-lower} rules out polylogarithmic memory, but
does not distinguish these update times.}
\label{tab:comparison}
\end{table}

\paragraph{What is new relative to the space lower bound.}
The lower-bound contribution is its model and proof, rather than an
improved dependence on $c$ over~\cite{psw}. We use a single insertion
prefix and a single deletion suffix, with no multiplicities or long
history. An arbitrary streaming state becomes a message in a finite
communication problem. Its conditional entropies give the required
submodular function directly, so there is no reduction to linear
sketches. The upper-bound contribution is complementary: a
constant-factor approximation can use polynomial memory without
spending polynomial time in $N$ on an update. The factor-two space
exponents in Table~\ref{tab:comparison} and the lower bound do not
match; an optimal factor-two time--space tradeoff remains open.

\section{Technical overview}\label{sec:techniques}

A small sample may miss an endpoint of the farthest pair. It can
still tell us where to look for such a point. We use the sample to
bound distances in a region of space, recover the points outside that
region, and maintain a current pair from the sample and the recovered
points. The lower bound uses a tight cluster and one distant point to
show how much the algorithm must remember about earlier insertions.

\subsection{A sample and its hull}

Take a small random sample $S$ of the current distinct points. A ball
that omits many points is unlikely to contain the whole sample. A
sampling bound for Euclidean balls makes this statement simultaneous
over all centers and radii, so it applies even to a ball chosen after
seeing $S$.

Fix an expansion factor $a>1$. Expand every ball containing $S$ about
its own center by $a$ and intersect the expanded balls. Call the
intersection $\Hull_a(S)$. If a point lies outside this hull, some ball
contains $S$ while the point is farther than $a$ times the ball's radius
from its center. The sampling bound says that this ball contains most
of the input. The other useful property is the diameter bound from
Lemma~\ref{lem:hull}.
\[
                 \diam(\Hull_a(S))\le\sqrt{2a^2-1}\,\diam(S).
\]
A hull point lies within
$\sqrt{(a^2-1)V}$ of the mean of a distribution on $S$ with variance
$V$. For two such distributions, the squared distance between their
means plus their variances equals the expected squared distance
between independent draws from them, at most $\diam(S)^2$.
Cauchy--Schwarz then gives the hull diameter bound.

The sample diameter accounts for pairs inside the hull. We recover
every exterior point into a candidate set, which accounts for pairs
outside the hull. For a mixed pair, viewing the hull from the exterior
endpoint bounds the distance to the interior endpoint using a farthest
point in $S$. To find the exterior points we use fixed filters drawn
independently of the sample, so we can analyze them for a witnessing
ball chosen after seeing $S$.

\subsection{Recovering the points outside the hull}

Consider one of the witnessing balls and one target outside it. The
filters should place that target in a bucket shared by few other
points. We use several independent Gaussian projections. Each
projection has a randomly shifted grid, and a hash maps its cells to
buckets. The tuple of bucket identifiers determines a single leaf
with a sparse-recovery sketch. This assignment depends only on the
point and the preselected maps, so an insertion or deletion changes
one leaf per scale and repetition.

\begin{figure}[t]
\centering
\begin{tikzpicture}[x=1.85cm,y=1.85cm,font=\small,>=Latex]
  \begin{scope}
    \clip (0,0) circle [radius=1];
    \fill[blue!18] (.6,-1.02) rectangle (1.02,1.02);
  \end{scope}
  \draw[thick] (0,0) circle [radius=1];
  \draw[blue!65!black,dashed,thick] (.6,0) circle [radius=.8];
  \draw[gray!65,->] (-1.13,0)--(1.94,0) node[right] {$u$};
  \draw[blue!65!black,dotted] (.6,-.8)--(.6,.8);
  \fill (0,0) circle (.023) node[below left] {$0$};
  \fill[blue!65!black] (.6,0) circle (.023) node[below left] {$0.6u$};
  \fill[BrickRed] (1.581,0) circle (.027) node[above=4pt] {$y$};
  \draw[gray!75,<-] (-.67,.74)--(-.92,1.02)--(-1.36,1.02)
      node[left,black] {$B(0,1)$};
  \draw[blue!65!black,<-] (1.11,.62)--(1.32,.90)--(1.57,.90)
      node[right] {$B(0.6u,0.8)$};
\end{tikzpicture}
\caption{A cross-section for the two-stage construction. The first grid's
difficult points lie in the shaded cap $t\ge0.6$ of the unit ball.
The dashed ball contains this entire cap and determines the smaller
radius of the second grid. Its center and the target direction $u$ are
used only in the analysis.}\label{fig:mask}
\end{figure}
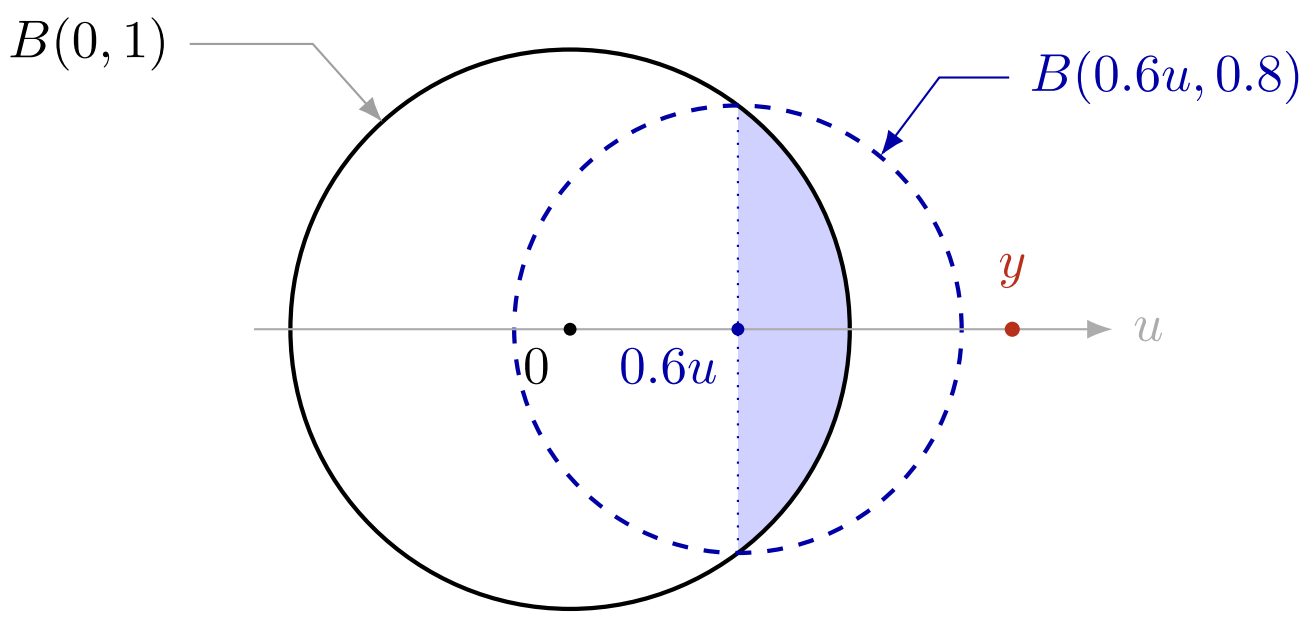

The analysis studies which points of the witnessing ball can follow the
target into the same projected cell. Normalize the ball to unit radius
and consider a target at distance at least $\lambda>\sqrt2$ from its
center. If a point in the ball has coordinate $t$ along the target
axis, its squared perpendicular component is at most $1-t^2$.
Conditional on the projection of the axis, that perpendicular
component supplies independent Gaussian noise. The squared axial
separation exceeds its variance scale by a positive margin, since
$(\lambda-t)^2-(1-t^2)\ge\lambda^2/2-1>0$. The collision analysis uses
this margin to enclose points with weaker same-cell bounds in a cap
facing the target. This cap fits in a smaller ball (Figure~\ref{fig:mask}).
The bound on the squared radius drops by a fixed amount at each stage,
so after a constant number of stages the difficult region is empty.
The analysis chooses a covering center on the target axis at each stage
while the algorithm uses the same maps for all targets.

To bound the load at a target's leaf, separate bulk points by the
first stage at which they leave the nested difficult regions. At
that stage, a point lies in the covering ball and has a small chance
of sharing the target's grid cell. With constant probability the hash
isolates the target's bucket from the other cells in the projected core.
Together these give an inverse bucket-count factor and an extra
saving exponential in the projection dimension. At later stages
the same-cell probability stays small, and distinct cells can share
a bucket only through a hash collision. Thus the inverse bucket-count
factors from the exit stage onward cancel the corresponding factors
when we sum capacity over all leaves. We increase the projection
dimensions geometrically so that the extra saving at the exit stage
pays for the bucket counts of earlier stages with a power of $N$ to spare.

The leaf capacity must also allow every point outside the original
ball to collide. With sampling probability of order $N^{s-1}$ for a
small constant $s>0$, the number of exceptions is at most $N^{1-s}$
times polynomial factors in the dimension and the logarithmic
parameters. Using $O(N^{s/2})$ leaves keeps the space for these
exceptions sublinear as well.
Section~\ref{sec:gaussian-main} gives the load and space calculations.

Our recovery argument uses separation above $\sqrt2$ in units of the
witness radius. Taking $a$ just above $\sqrt2$ gives a hull diameter
factor $\sqrt{2a^2-1}$ just above $\sqrt3$.

\subsection{Maintaining a current witness}

A sparse leaf can contain too many points to decode in one update.
We measure the age of a snapshot by updates to its own leaf. A leaf
with capacity $K$ takes a snapshot every $K$ local updates and spends
the next block decoding and replaying changes. Its serving snapshot
is at most $2K$ local updates old. If the leaf has at most $K$ points
now, that snapshot had at most $3K$ points, within the decoder's
budget. A leaf that receives no updates needs no further work to keep
its serving dictionary current.

The decoder certifies the identities and the completeness of its
snapshot from exact moments and total mass. A signed change log
converts each recovered multiplicity to its current value. When a
point appears in several leaves, its entries share a single current
count. Original deletions update this count before background work
runs, so even an unsuccessful decoding attempt can only affect which
current points are available as candidates. Section~\ref{sec:local-union}
gives the reconstruction and its schedule.

Write $C$ for the resulting set of current candidates. Scanning $C$
after every update would lose the time bound. A dynamic hierarchy on
$C$ maintains a pair that approximates $\diam(C)$ and exposes a small
family of covering balls. Each ball is described by a few current points, its
\emph{anchors}, with convex weights. Its center $q$ is their weighted
mean, and its radius is close to the square root of their variance
$V$. For a candidate $x$ covered by this ball and a hull point $y$,
the triangle inequality bounds $\|x-y\|$ by the ball's radius plus
$a$ times the farthest sample distance from $q$.

We can approximate that farthest distance by querying the sample for
a point $z$. To turn the reply into a pair of input points, take the
anchor $p$ farthest from $z$. The weighted average of the squared
anchor-to-$z$ distances gives
\[
                       \|p-z\|^2\ge V+\|q-z\|^2.
\]
The same pair thus accounts for both the radius and the distance from
the center to the sample. Cauchy--Schwarz gives a mixed-pair factor
arbitrarily close to $\sqrt{1+a^2}$ as the ball and lookup accuracies
improve. The hierarchy's factor for pairs in $C$ can be made
arbitrarily close to $\sqrt2$. Both fit within the hull factor when
$a>\sqrt2$.

Fast farthest-point lookup uses a second schedule, measured in stream
updates. At the start of a block we freeze the live sample. We compute
its diameter and build a farthest-point lookup table on a projected
grid during that block, then use them in the next block. The sample
has size on the order of $N^s$ up to polynomial factors. Taking $s$
small lets us spread its polynomial preprocessing over a block of
length on the order of $N^{1-s}$. A deletion of a sampled identity
invalidates the corresponding frozen generation immediately. On an
oblivious stream, the identities deleted during its two blocks are
fixed in advance. The probability that any is sampled is at most
their number times the sampling rate. With a small constant in the
block length, the sample survives with constant probability. Repeating
the whole construction independently gives the approximation
throughout the stream. The algorithm caches the longest of the sample
diameter pair, the candidate hierarchy's witness, and the mixed pairs
found by lookup (Section~\ref{sec:cached-main}).

\subsection{A direct lower bound on short set streams}

A deletion suffix can make a final diameter query reveal one bit of
the insertion prefix. Alice has one independent bit for each point on
a fine net of the unit sphere and inserts the points whose bits are
one. She sends the memory state to Bob. Bob is given a small cap and
its antipodal target. He knows which points were inserted outside the
cap and the target. He first puts a fresh anchor near the cap center,
then deletes the points he knows were inserted. The remaining set has
small diameter if the target is absent and large diameter if it is
present. A cap of radius proportional to $1/c$ separates the two cases
by more than a factor $c$.

Bob's knowledge of other bits matters because one message can encode
several bits together. Let $M$ be the message after fixing Alice's
randomness, and let $f(A)$ be the conditional entropy of $M$ when Bob
knows the bits outside $A$. This set function is monotone and
submodular, with total value at most the message length $s$. For a cap
$B$ and a target $z$, the gain $f(B\cup\{z\})-f(B)$ measures the
information the message gives about the target bit after Bob learns
the bits outside the cap and the target. This gain is zero if $z$ is
already in $B$. Bob decodes an antipodal target with bounded error,
so those queries have a constant average gain.

We relate these gains across chord lengths. Given several cap--target
pairs, form a table whose entry for cap $i$ and target $j$ is their
gain. A submodular inequality bounds the square of the sum along
the diagonal by $2s$ times the sum of the entire table. Its proof
charges the diagonal gains along a random chain of targets, with
cross entries accounting for the reduction in a gain when earlier
targets have already been added.

To use the inequality, place the paired cap centers and targets at
opposite ends of diameters of a randomly oriented affine subsphere
of diameter $t$. Every paired chord then has length $t$, while the
cross chords typically have length near $t/\sqrt2$
(Figure~\ref{fig:subsphere}). If $a(t)$ denotes the average gain at
chord length $t$, rotational averaging yields
$a(t)^2\le 2s\,\E[a(t\sqrt{Z_0})]$, where $Z_0$ is the squared ratio
of a random cross chord to $t$ and concentrates near $1/2$. The proof
allows repeated targets in the table, including those created by
rounding to the finite net.

Starting with the positive gain at the antipodes, the recurrence
forces positive gain at shorter chords if $s$ is small. Each step
squares the gain, so the proof adjusts the tail cutoff for $Z_0$ as
the gains become smaller. This reaches the band where the target
already lies in the cap and the gain is zero, unless the message has
$\exp(\Omega(d/c^2))$ bits. The finite sphere net has
$\exp(O(d\log(c+1)))$ points. Each site can be inserted once and
deleted once, so converting the bit bound to a space bound uses only
a linear-length set stream (Section~\ref{sec:sphere-lower-main}).

\subsection{Further related work}
\paragraph{Geometric sketches.}
Jiang, Krauthgamer, and Sapir~\cite{jks} use moderate dimension
reduction for dynamic-streaming $k$-center. On polynomial-length
streams they obtain an $O(\alpha)$ approximation in
$n^{1/\alpha^2}\poly(kd\log n)$ bits. The one-center radius and
diameter are within a factor of two, so their result also gives
an $O(\alpha)$ diameter estimate. Their two-level sampler can also return original input identities.
Our filters isolate exterior points at a prescribed separation and
require one leaf update per hierarchy. The hull then gives an explicit
diameter factor arbitrarily close to $\sqrt3$ from above. Geometric hashing has also been used for
dynamic-streaming facility location~\cite{facility}, and streaming
coresets address $k$-center with outliers~\cite{dbbm}.

\paragraph{Deletions and representatives.}
Banihashem et al.~\cite{dynamic} maintain a robust representative for a
fully stored dynamic set. Here a representative of the small sample
can disappear on a single deletion. We invalidate its generation and
bound the probability of this event on an oblivious stream. Blank et
al.~\cite{expiration} obtain a $(1+\sqrt3+\varepsilon)$ approximation
to Euclidean diameter in the expiration model, where an item's
deletion time is given on arrival. In our lower bound the choice of
which items expire encodes the decoder's question.

\paragraph{From a streaming state to a message.}
The general-metric and Euclidean lower bounds of
Khanna et al.~\cite{kpsw} and Padaki et al.~\cite{psw} connect
streaming to linear sketches and index coding. Li, Nguyen, and
Woodruff~\cite{lnw}, Kallaugher and Price~\cite{kp}, and Jiang, Liu,
and Yu~\cite{jly} study when streaming algorithms admit linear-sketch
simulations. The last work covers polynomial-length strict-turnstile
streams and allows intermediate multiplicities above one. We use the
state itself as a message on a stream that stays binary. Conditional
entropy is a familiar tool for nonlinear index coding~\cite{bkl}.
The geometric step here relates its marginals at two chord lengths
and survives rounding to a finite input universe.

\section{Model and preliminaries}\label{sec:model}

An update is $(x,\sigma)$ with $x\in[-\Delta,\Delta]^d\cap\mathbb Z^d$
and $\sigma\in\{-1,1\}$. The stream starts empty and all frequencies
remain nonnegative. The support $P$ consists of locations of positive
frequency. Each update supplies its point's coordinates. The algorithm
has no external copy of the support. An oblivious stream is fixed
independently of all random choices and answers. We assume $N\ge2$,
$T,\Delta,d\ge1$ and $0<\delta<1/10$. A bounded range of $N$ depending
only on the fixed approximation parameters is handled by storing the
set with its counts and computing its diameter directly.

\paragraph{Arithmetic and scheduling.}
In the technical sections we abbreviate the parameters by
\[
 \mathcal L=(d,\log(N+2),\log(\Delta+2),\log((T+2)/\delta)).
\]
We use the word RAM with word length
$w_{\rm word}=\Theta(\log(d+2)+\log(N+2)+\log(\Delta+2)+
\log((T+2)/\delta))$. We write $\ln$ for the natural logarithm,
$\log_2$ for the binary logarithm, and $\log$ for a fixed base in
asymptotic bounds. We charge multiword arithmetic, random bits,
balanced-dictionary operations and memory reclamation. A score is
stored as an exact squared distance and two point keys, with zero as
the default. Repetitions use independent seeds. Each budgeted
background job saves its position between updates. All processing of
one update finishes before its answer becomes available. Budgets and
sample sizes specified by asymptotic bounds use sufficiently large
fixed constants and are rounded up to integers unless stated
otherwise. Appendix~\ref{sec:finite-bit-gaussian-implementation} gives
the precision for the randomized geometric tests.

The upper construction also accepts canonical rational coordinates
with numerators and positive denominators of at most $b$ bits. Distinct
points are then separated by at least $2^{-2b}$ and lie in
$[-2^b,2^b]^d$. The first radius scale is $2^{-2b-1}$ and the number of
scales is $O_c(b+\log(d+2))$. Key lengths, exact rational arithmetic
and grid precision use $b$ in place of the integer coordinate bit
bound. We state the results for the integer grid for ease of comparison.

\paragraph{Tools used in the proofs.}
The following forms specify the sampling and minimax results that we
use. The other geometric and recovery routines are constructed below.

\begin{lemma}[Half-nets for balls, Haussler and Welzl~\cite{hw}]\label{lem:half-net}
Let $A$ be a finite set in $\mathbb R^d$ and $0<\xi<1$. A uniform
sample without replacement of size
$\min\{|A|,\lceil C(d+\log(1/\xi))\rceil\}$ has the following
property with probability at least $1-\xi$. Every Euclidean ball
containing the sample contains at least half of $A$.
\end{lemma}
Indeed, the complements of balls have VC dimension $v=d+1$. For
$m$ independent draws the double-sampling bound of Haussler and
Welzl~\cite[Theorem~3.7]{hw} bounds the probability of missing a
complement of relative size over $1/2$ by
$2\bigl(\sum_{j=0}^v\binom{2m}{j}\bigr)2^{-m/4}$ for $m\ge16$.
For $m\ge C v$ the sum is at most $(2em/v)^v\le 2^{m/8}$, with
$C$ an absolute constant. Taking $m\ge C(v+\log(1/\xi))$ gives
failure at most $\xi$. If $m\ge |A|$, we take all points.
Otherwise we can keep the distinct independent draws and complete
them uniformly to $m$ distinct points. By symmetry this is a uniform
sample without replacement, and completion preserves every hit.

\begin{lemma}[Limited-independence moment bound~\cite{sss}]\label{lem:moment-tool}
If $Y$ is a sum of $2k$-wise independent Bernoulli variables with
mean $\mu$, then for an absolute constant $C$ and integer $k\ge1$,
$\mathbb E(Y-\mu)^{2k}\le(Ck\mu+Ck^2)^k$.
\end{lemma}
The centered moment is a polynomial of degree at most $2k$ in the
indicators and thus agrees with its value under full independence.
For independent variables the usual two-sided Bernstein bound is at
most $2\exp(-t^2/(4\mu))+2\exp(-3t/4)$ at deviation $t>0$ when
$\mu>0$. Integration against $2k t^{2k-1}$ gives terms of order at
most $(Ck\mu)^k$ and $(Ck^2)^k$, respectively. The case $\mu=0$
is immediate. Section~\ref{sec:sample-net} gives the finite-bit
sampling family.

\begin{lemma}[Minimax, Sion~\cite{sion}]\label{lem:minimax-tool}
For compact convex sets $U,V$ in finite-dimensional real vector
spaces and a continuous function $g$ concave in $u\in U$ and convex
in $v\in V$, one has
$\max_{u\in U}\min_{v\in V}g(u,v)=
  \min_{v\in V}\max_{u\in U}g(u,v)$.
\end{lemma}
We apply this identity to the simplex of distributions on a sample and
a bounded ball of possible centers. It is used only in the analysis.

\section{Maintaining a diameter pair}\label{sec:cached-main}

The algorithm keeps two sets for different purposes. A frozen sample
$S$ witnesses pairs in its expanded hull. A dynamic set $C$ contains
the exterior points and may contain other current points. We maintain
a score for $C$ and compare its ball centers with $S$. The construction
below proves Theorem~\ref{thm:main-upper}. The exact approximation
factor and space exponents appear in Lemma~\ref{lem:cached-center-composition}.

\subsection{Algorithm}\label{sec:algorithm}

For the desired factor $c>\sqrt3$, we choose a rational
$\sqrt2<a<\sqrt{(c^2+1)/2}$ for the hull, a rational accuracy
$0<\eta\le1/4$ for the candidate balls, and a rational $\kappa>1$ for
farthest-point lookup. We choose $\eta$ sufficiently small and $\kappa$ sufficiently close to one, as justified in the proof of Theorem~\ref{thm:main-upper}.
If $B_\kappa\ge2$ is the preprocessing exponent of
Lemma~\ref{lem:static-oracle}, we fix rational exponents
\begin{equation}\label{eq:cached-parameter-conditions}
        0<s<1/(B_\kappa+1),\qquad 0<\rho\le s/2 .
\end{equation}
The sample hash selects each location with a fixed dyadic probability
$p\in[N^{s-1}/4,N^{s-1}]$. The input starts with an empty live sample,
an empty candidate hierarchy, and zero scores. We initialize the recovery maps, hashes and clocks before the first
update in $\poly(\mathcal L)$ time. Moment tables and lookup entries
are allocated as needed. We use a sufficiently large
constant times $\lceil\log((T+2)/\delta)\rceil$ complete copies.
The exact factor-two parameters are in Appendix~\ref{sec:combined-quantitative-two}.

\begin{algorithm}[htbp]
\caption{Maintaining the witness after a point update}\label{alg:update}
\begin{algorithmic}[1]
\Require{A legal update $(p,\sigma)$ and the state after the preceding update}
\Ensure{Updated state and a cached score with current endpoints whenever positive}
\For{each enabled independent copy}
  \State Adjust $p$'s global count if recorded, then update its live-sample reference
  \State Invalidate each frozen generation containing $p$ if $\sigma=-1$
  \State If the sample cap is exceeded, disable this copy and continue with the next
  \For{each scale and recovery repetition}
    \State Hash $p$ to its leaf and apply Algorithm~\ref{alg:leaf} to $(p,\sigma)$
  \EndFor
  \State Apply the same update to the global fallback bucket
  \State Advance outer reference retirement, then process all queued transitions in order in $C$'s hierarchy
  \State Advance the sample-oracle job and private reclamation by their separate budgets
  \If{an epoch ends}
    \State Install the completed generation with its validity flag and freeze the next sample
  \EndIf
  \State Start with the witnessed candidate score $L_C$ (zero if none)
  \If{a valid nonempty frozen sample $S$ is serving}
    \State Include its precomputed diameter and maximizing pair
    \For{each exposed anchor set $K$ with center $q$, including singleton leaves}
      \State Query $S$ at $q$ for $z$ and include $(x,z)$ for an $x\in K$ farthest from $z$
    \EndFor
  \EndIf
  \State Cache the largest included squared distance and its endpoints
\EndFor
\State Store the best cache over enabled copies, or zero if none has a positive score
\end{algorithmic}
\end{algorithm}

\noindent Algorithm~\ref{alg:update} uses the following interfaces.
\begin{itemize}
\item \emph{Sampling and farthest-point lookup}
(Lemmas~\ref{lem:ball-net} and~\ref{lem:static-oracle}).
We store the positive sampled locations and their counts. At each epoch
boundary $\tau$, including time zero, we freeze a persistent root
$S_\tau$. During the next $H_{\rm epoch}$ updates we compute its exact
diameter, a maximizing pair, and its lookup table. An empty sample
needs no oracle. The resulting generation serves from
$\tau+H_{\rm epoch}$ up to the next installation, where
\[
                 H_{\rm epoch}=\max\{1,\lfloor N^{1-s}/100\rfloor\}.
\]
A deletion of any frozen identity permanently invalidates that
generation, even if the location retains another occurrence. We test
both the building and the serving generations. A query to a valid
table returns a sample identity. It makes no change to the sample or
to any future update.
\item \emph{Fixed filters and local reconstruction}
(Theorem~\ref{thm:recovery} and Lemma~\ref{lem:local-compiler}).
We choose $\sqrt2<\lambda<\min\{a,\sqrt3\}$ and maintain geometric
scales of ratio $1<q_{\rm scale}\le a/\lambda$ from $1/2$ through the
first scale at or above $2\Delta\sqrt d/(a-1)$. Each update reaches
one leaf per scale and repetition. Every leaf reconstructs its
snapshot on its own update clock. The boundary exception budget
of Lemma~\ref{lem:ball-net} is
$b_0=C N^{1-s}((d+2)\log(N+2)+\log((T+2)/\delta)+1)$.
For all serving times we use
\begin{equation}\label{eq:frozen-exception-budget} b=b_0+2H_{\rm epoch}.\end{equation}
A fallback bucket of capacity $\lceil b+2\rceil$ sees every update.
Each live sampled point also has a candidate reference. All reference
changes, including retirements, enter the same ordered queue. A
transition retains its original point key until consumed.
\item \emph{Dynamic ball certificates} (Lemma~\ref{lem:explicit-certificate}).
We apply each queued candidate insertion or deletion to a separately
randomized hierarchy. It maintains a witnessed score $L_C$ and exposes
a polynomial number of certificates. Each has a few current anchors
$K$, rational convex weights, mean $q$, and variance $V$.
On the hierarchy's success event the balls $B(q,(1+\eta)\sqrt V)$
cover $C$, including singleton certificates with $q=x$ and $V=0$.
After the queue drains, we score the exposed certificates against the
current serving generation. This also covers newly exposed leaf
points and an epoch switch on the same original update.
\end{itemize}
A query reads the resulting maximum in constant time. Its handle
refers to the present cache. Copying both endpoints costs their output
length. A disabled inner hierarchy contributes neither centers nor a
score. If every inner hierarchy in a complete copy is unavailable,
we take $L_C=0$ and may still use that copy's valid sample diameter.
Inner failures have no effect on the order of candidate transitions.

\subsection{Analysis}

\begin{lemma}[Composition of the diameter witnesses]\label{lem:cached-center-composition}
For fixed rational $a>\sqrt2$, $0<\eta\le1/4$, $\kappa>1$ and rational exponents
satisfying \eqref{eq:cached-parameter-conditions}, there is an
$\epsilon_0>0$ such that Algorithm~\ref{alg:update} uses space
\begin{equation}\label{eq:cached-composition-space}
 \left(N^{sB_\kappa}+N^{1-s+\rho}+N^{1-\epsilon_0}\right)\poly(\mathcal L)
\end{equation}
and worst-case update time $\poly(\mathcal L)$ on every outcome.
With probability at least $1-\delta$, simultaneously throughout the
stream, its witnessed answer obeys
\begin{equation}\label{eq:cached-composition-factor}
 L\le\diam(P)\le c_*L,\qquad
 c_*=\max\left\{\sqrt{2a^2-1},\
 \sqrt{(1+\eta)^2+1},\ \sqrt{(1+\eta)^2+a^2\kappa^2}\right\}.
\end{equation}
\end{lemma}
\begin{proof}
We show first which points the certificates must cover, then which
pair witnesses each of the three factors. The last part of the proof
checks the conditioning and the two rebuilding deadlines.

\paragraph{Coverage at a serving time.}
By Lemma~\ref{lem:ball-net}, every ball containing $S_\tau$ excludes at
most $b_0$ points at its boundary. At most $2H_{\rm epoch}$ insertions
occur before the end of service, so \eqref{eq:frozen-exception-budget}
bounds its exceptions even if the ball is chosen later. Suppose
$|S_\tau|\ge2$. For each $y$ outside its hull, the radius bound
\eqref{eq:hull-witness-scale} gives a witness with radius $r$ in the
maintained range. Rounding up preserves the bulk of the ball and leaves
the target more than $\lambda$ rounded radii from its center. Conditional
on the original stream and sample, we can fix this witness before
drawing the grids. Independent grid repetitions and a union bound over
points, generations and times give a leaf of size at most its capacity
for every such $y$. Conditional on the grids, the local snapshot times
are fixed independently of their private decoding hashes.
Lemma~\ref{lem:local-compiler} therefore gives
\begin{equation}\label{eq:cached-current-outlier-containment}
              P\setminus\mathcal H_a(S_\tau)\subseteq C\subseteq P .
\end{equation}
The right inclusion holds on every outcome. Before the first oracle is
ready, $|P|\le H_{\rm epoch}$. For an empty or singleton boundary sample
the net event gives boundary support at most $b_0+1$, hence current
support at most $b+1$. In each case the fallback bucket gives $C=P$
on the same recovery event.

\paragraph{Witnessed scores.}\label{sec:center-score}
For a valid serving sample, let $s_*\in S_\tau$ be the oracle's reply
at a certificate center $q$. The contributions of that certificate are
\begin{equation}\label{eq:cached-center-score}
       F=\|q-s_*\|,\qquad
       L_{\rm node}=\max_{z\in K}\|z-s_*\|.
\end{equation}
We retain the pair attaining $L_{\rm node}$. Its squared distance
dominates the weighted average over the anchors, namely
\begin{equation}\label{eq:cached-center-variance}
 L_{\rm node}^2\ge\sum_{z\in K}\lambda_z\|z-s_*\|^2
                         =V+F^2.
\end{equation}
Thus for a surviving sample the cached answer is
\begin{equation}\label{eq:cached-answer-definition}
 L=\max\left\{L_C,\operatorname{diam}(S_\tau),
                     \max_{e\ {\rm exposed}} L_e\right\},
\end{equation}
where $L_e$ is the score of exposed certificate $e$. With an absent,
invalid or empty sample we use $L_C$ alone. Each available score has
current endpoints. On a successful explicit hierarchy,
$\diam(C)\le\sqrt{(1+\eta)^2+1}\,L_C$.

Fix a time when the frozen sample survives and its net, recovery,
explicit-hierarchy, and oracle events all hold.  For an empty or
singleton frozen sample, or during startup, the global fallback gives
$C=P$, and the explicit-point hierarchy suffices.  Otherwise consider
a diameter pair $x,y$.

If both points are in $\mathcal H_a(S_\tau)$, the hull lemma gives
\[
 \|x-y\|\le\sqrt{2a^2-1}\operatorname{diam}(S_\tau)
                                  \le\sqrt{2a^2-1}\,L.
\]
If both are outside the hull, both belong to $C$ by
\eqref{eq:cached-current-outlier-containment}, and
\[
              \|x-y\|\le\sqrt{(1+\eta)^2+1}\,L_C.
\]
For the remaining case suppose $x\in C\setminus \mathcal H_a(S_\tau)$ and
$y\in \mathcal H_a(S_\tau)$.  A valid certificate covers $x$. The
leaf case uses $q=x$ and $V=0$. Evaluating hull membership at the
certificate center $q$ gives $\|q-y\|\le aR_{S_\tau}(q)$. The farthest-point guarantee gives
\[
 \begin{split}
 \|x-y\|&\le(1+\eta)\sqrt V+aR_{S_\tau}(q)\\
        &\le(1+\eta)\sqrt V+a\kappa F\\
        &\le\sqrt{(1+\eta)^2+a^2\kappa^2}\sqrt{V+F^2}\\
        &\le\sqrt{(1+\eta)^2+a^2\kappa^2}\,L_{\rm node}.
 \end{split}
\]
The variance term suffers no oracle loss.  Taking the maximum of
these three cases proves \eqref{eq:cached-composition-factor}.

\paragraph{Conditioning and survival.}
At any fixed serving time the oblivious stream specifies at most
$2H_{\rm epoch}$ identities deleted since the boundary. Each is selected
with probability at most $p$. In the asymptotic range of $N$,
\begin{equation}\label{eq:frozen-sample-survival}
                  \Pr[\text{serving sample invalid}]\le2H_{\rm epoch}p\le1/50 .
\end{equation}
We use this unconditional bound together with the coverage probability.
An adversary that sees a witness could deliberately delete a sampled
endpoint, which is where this argument uses obliviousness.
We first fix the original oblivious stream and all sample, grid,
and sparse-recovery seeds.  These fix the candidate sequence and its operation schedule.
Each candidate operation is padded to the explicit primitive's known
worst-case running-time bound.  Outer reference retirement and inner
geometric retirement have separate queues and separate fixed budgets;
private inner allocations never change the outer cleanup order.
Thus the candidate sequence is independent of the hierarchy's private
randomness.  The explicit-point theorem applies to this fixed sequence
of at most $T\operatorname{poly}(\mathcal L)$ operations.

Next fix the hierarchy seeds.  They determine every exposed center,
its validity interval, and its query time.  None depends on a static
oracle answer.  Frozen-oracle construction uses its prescribed work
budget and installation deadline, and oracle operations may also be
padded.  Hence the entire finite list of queries to a given frozen
oracle is fixed independently of that oracle's seeds.  Whole-oracle
repetition gives the required guarantee for all these queries; the
logarithm of their number is polynomial in $\mathcal L$.  Its finite
precision is chosen for the bounded rational center domain.

We choose the net, recovery, hierarchy, and oracle failure budgets so that their total, at a fixed time, is at most $1/10$.  Combining
them with \eqref{eq:frozen-sample-survival} by a union bound gives a
constant success probability for one whole copy.  No independence
between these events is needed. We run
$O(\log((T+2)/\delta))$ independent whole copies and return the largest
available current-pair score.  The copies are independent for the
fixed stream, so all-copy failure at a fixed time is exponentially
small in the number of copies.  A union bound over the $T+1$ output
times proves the claimed simultaneous guarantee.  Taking the maximum
cannot create an overestimate.

\paragraph{Worst-case resource bounds.} Table~\ref{tab:costs} records the costs of the structures maintained in one copy.

\begin{table}[htbp]
\centering\small
\renewcommand{\arraystretch}{1.13}
\begin{tabularx}{\linewidth}{@{}l l >{\raggedright\arraybackslash}X@{}}
\toprule
Structure & Space up to $\poly(\mathcal L)$ & Work charged to one update\\
\midrule
Live sample & $N^s$ & One membership test and a count change\\
Frozen lookup & $N^{sB_\kappa}$ & One epoch's share of preprocessing, plus center lookups\\
Geometric leaves & $N^{1-s+\rho}+N^{1-\epsilon_0}$ & One touched leaf per scale and repetition\\
Fallback bucket & $N^{1-s}$ & One local reconstruction budget\\
Candidate hierarchy & $M$ & A polynomial number of transitions and exposed centers\\
\bottomrule
\end{tabularx}
\caption{Costs in one complete copy. $M$ bounds the number of current
candidates including intermediate transitions. Rebuilding is paced by
global updates for the frozen lookup and by updates to each leaf for its
decoder.}\label{tab:costs}
\end{table}

We take a deterministic bound $M$ on the size of the candidate stream,
including intermediate transitions, and bound its length by
$T_C\le(T+1)\poly(\mathcal L)$. Theorem~\ref{thm:recovery} and the
local compiler give
$M=O(N^{1-s+\rho}+N^{1-\epsilon_0})\poly(\mathcal L)$, including the
fallback of size $O(b)$ and the live sample. Here $s<1/(B_\kappa+1)$
ensures $s<1-s+\rho$. Lemma~\ref{lem:explicit-certificate} thus costs
$M\poly(\mathcal L)$ space, exposes
$O(R_{\rm hier}(\log(M+2)+H_{\rm leaf}))=\poly(\mathcal L)$ centers,
and processes each transition in worst-case polynomial time. Here
$R_{\rm hier}$ is the number of hierarchy copies and $H_{\rm leaf}$
bounds exact-leaf size up to a constant factor. Each center is a
bounded rational convex combination, so its encoding fits the query
contract of Lemma~\ref{lem:static-oracle}.
The sample cap is $N^s\operatorname{poly}(\mathcal L)$. A static
oracle, including exact sample diameter and repetition, therefore takes
$N^{sB_\kappa}\operatorname{poly}(\mathcal L)$ time and space to
construct. Bound its total work by $W=N^{sB_\kappa}P(\mathcal L)$
for a fixed polynomial $P$, and spread it over the next epoch. Since
$H_{\rm epoch}\ge N^{1-s}/200$, the per-update budget satisfies
\begin{equation}\label{eq:frozen-oracle-work-budget}
 \left\lceil W/H_{\rm epoch}\right\rceil
 \le 1+200 N^{s(B_\kappa+1)-1}P(\mathcal L)
 = O(1+P(\mathcal L))
\end{equation}
by \eqref{eq:cached-parameter-conditions}.  The bound includes snapshot
traversal, Gaussian-map generation, lookup enumeration, sample scans,
and all amplified tables. We retain a constant number of generations,
retire old roots in constant metadata work, and reclaim their records
under a separate fixed global budget.  Since
record allocation per update is polynomial, the burst-plus-rate
retirement argument adds at most another constant multiple of the
non-retired storage.

Leaf construction follows local clocks.  Each touched leaf gets a
fixed polynomial budget, independent of its capacity; its
$K\operatorname{poly}(\mathcal L)$ decoding and materialization are
spread over $K$ updates to that leaf.  There are polynomially many
touched leaves per original update.  The compiler produces only
polynomially many candidate transitions, each processed by the
worst-case polynomial explicit-point primitive. A full enumeration of
the exposed certificates, with one oracle query and a polynomial number
of anchor comparisons per certificate, also fits this bound. The lookup jobs
use the first term in \eqref{eq:cached-composition-space}, and the
filters and the candidate hierarchy use the other two. The caps on
samples, chain depth and work allocations apply before an overflow can
increase the storage bound. This proves the lemma.
\end{proof}

\begin{proof}[Proof of Theorem~\ref{thm:main-upper}]
For $\sqrt2<a<\sqrt{(c^2+1)/2}$ the hull factor is strictly below $c$.
When $\eta=0$ and $\kappa=1$, the squared mixed factor is
$1+a^2<2a^2-1$. Continuity gives fixed rational $\eta>0$ and $\kappa>1$
for which both other factors are smaller than the hull factor. With
$s,\rho$ as in \eqref{eq:cached-parameter-conditions}, we may take
\[
       \epsilon_c=\min\{1-sB_\kappa,\ s-\rho,\ \epsilon_0\}>0 .
\]
Lemma~\ref{lem:cached-center-composition} now gives the claimed space,
update time and simultaneous guarantee. The finite-bit implementation
is given in Appendix~\ref{sec:finite-bit-gaussian-implementation}.
\end{proof}

\section{The sample and its farthest-distance hull}\label{sec:hull-main}

The proof of Algorithm~\ref{alg:update} uses two properties of its
frozen sample. The hull bounds the length of pairs with two interior
endpoints. Sampling bounds the number of exceptions to every ball
that can witness an exterior endpoint. The hull and the witnessing
balls are analysis objects.

\subsection{Distances inside the hull}

For a nonempty finite $S\subset\mathbb R^d$ and $a>1$, write
\[
 R_S(q)=\max_{s\in S}\|s-q\|,\qquad
 \mathcal H_a(S)=\bigcap_{q\in\mathbb R^d}B(q,aR_S(q)).
\]
Since $S\subseteq\mathcal H_a(S)$, the question is how far this hull
can extend beyond the sample. Write $L_S=\operatorname{diam}(S)$.

For a fixed center the hull involves the farthest individual sample
point. Passing to distributions lets us choose a single mean and
variance for each hull point, and then compare two hull points by
drawing independently from their two distributions.
\begin{lemma}[Variance representation of the hull]\label{lem:hull}
For a distribution $\lambda$ on a nonempty finite set $S$, let
$\mu_\lambda$ be its mean and $V_\lambda=\mathbb E_\lambda
\|s-\mu_\lambda\|^2$ its variance. For every $a>1$,
\begin{equation}\label{eq:hull-union-balls}
 \mathcal H_a(S)=\bigcup_\lambda
    B\!\left(\mu_\lambda,\sqrt{(a^2-1)V_\lambda}\right).
\end{equation}
In particular,
\begin{equation}\label{eq:hull-diameter-allscale}
 \operatorname{diam}(\mathcal H_a(S))\le\sqrt{2a^2-1}\,L_S .
\end{equation}
\end{lemma}
\begin{proof}
For a fixed point $y$, membership in the hull is equivalent to
\[
 \sup_q\min_{\lambda}
 \left(\|y-q\|^2-a^2\sum_s\lambda_s\|s-q\|^2\right)\leq0.
\]
The expression is affine in $\lambda$ and concave in $q$, with
quadratic part $-(a^2-1)\|q\|^2$.  Since $\lambda$ belongs to a compact
simplex and the linear coefficients in $q$ are uniformly bounded, all
relevant maximizers lie in one compact ball. Lemma~\ref{lem:minimax-tool} therefore interchanges the supremum and minimum.  Completing the square
then gives the value
\[
 \min_\lambda a^2\left(
 \frac{\|y-\mu_\lambda\|^2}{a^2-1}-V_\lambda\right).
\]
The minimum is attained, proving \eqref{eq:hull-union-balls}.

Consider two hull points represented by distributions $\lambda,\gamma$,
with means $\mu,\xi$ and variances $V,W$.  Put
$h=\|\mu-\xi\|$.  Independent draws from these two distributions give
\[
 h^2+V+W=\mathbb E\|s-t\|^2\leq L_S^2.
\]
The two hull points are consequently at distance at most
\[
 h+\sqrt{a^2-1}(\sqrt V+\sqrt W)
 \leq h+\sqrt{2(a^2-1)}\sqrt{L_S^2-h^2}
 \leq\sqrt{2a^2-1}\,L_S.
\]
This proves the lemma, including arbitrary affine dimension or
coincident points in $S$.
\end{proof}

\paragraph{The hull factor is attained.}
Let $h=(2a^2-1)^{-1/2}$ and $V=(1-h^2)/2=(a^2-1)h^2$.
We take two regular simplices, each with $m\ge\max\{2,h^{-2}\}$ vertices
and circumradius $\sqrt V$. We place their centers at distance $h$ on a
line perpendicular to both simplices, with the two simplex direction
spaces mutually orthogonal. Let $S$ consist of all their vertices.
Every cross-simplex distance has square $h^2+2V=1$. Within either
simplex the squared edge length is $2Vm/(m-1)\le1$, so $L_S=1$.
The uniform distributions on the two simplices have variance $V$.
By \eqref{eq:hull-union-balls}, the hull contains a ball of radius
$\sqrt{(a^2-1)V}$ about each center. Their two outermost points on
the line of centers are at distance
$h+2\sqrt{(a^2-1)V}=1/h=\sqrt{2a^2-1}$.

\paragraph{Bounded witness radii.}
For an input point $y\notin\mathcal H_a(S)$, some $q$ satisfies
\[
 \|y-q\|>aR_S(q).
\]
Let $D_{\rm univ}=2\sqrt d\Delta$.  For any $s\in S$,
\[
 \|y-q\|\leq\|y-s\|+\|s-q\|
            \leq D_{\rm univ}+R_S(q).
\]
Therefore every such witnessing ball has
\begin{equation}\label{eq:hull-witness-scale}
 R_S(q)<D_{\rm univ}/(a-1).
\end{equation}
If $S$ contains two distinct integer points, also
$R_S(q)\geq L_S/2\geq1/2$.  Thus only logarithmically many geometric
radius scales are needed, even though the hull definition quantifies
over all centers in $\mathbb R^d$.  A witnessing center need not be
computed by the algorithm.  For probabilistic arguments it can be
chosen as a deterministic function of $S$ and $y$; strict inequality
also permits choosing one from a fixed enumeration of rational centers.
\subsection{Why recovering the outside of the hull is sufficient}

Suppose $S\subseteq P$, and a set $C\subseteq P$ contains every point
of $P\setminus\mathcal H_a(S)$.  Define the cross score
\[
 L_{\rm cross}=\max_{p\in C}R_S(p),
\]
with value zero if $C$ is empty.  If a witnessed lower estimate
$L_C$ satisfies $\operatorname{diam}(C)\leq fL_C$, then
\begin{equation}\label{eq:hull-three-cases}
 \operatorname{diam}(P)
 \leq\max\{\sqrt{2a^2-1},f\}
       \max\{L_S,L_{\rm cross},L_C\}.
\end{equation}
Indeed, take any two points of $P$.  If both belong to the hull, use
\eqref{eq:hull-diameter-allscale}.  If both are outside the hull, both
belong to $C$, and use the estimate on $C$.  If $x\in C$ is outside
the hull and $y$ is inside it, the particular center $q=x$ in the hull
definition gives
\[
 \|x-y\|\leq aR_S(x)\leq aL_{\rm cross}.
\]

Since $a>1$, the mixed-pair factor $a$ is smaller than $\sqrt{2a^2-1}$.
For a nonempty sample, an explicit computation of $\operatorname{diam}(S\cup C)$
therefore gives factor $\sqrt{2a^2-1}$.
Algorithm~\ref{alg:update} attains the same factor with suitable
certificate and oracle accuracies, while inspecting only a polynomial
number of certificate centers per update.

\subsection{Sampling all containing balls}\label{sec:sample-net}

\paragraph{Algorithm.}
We set $m=N^s$ and select each point identity with common probability
$p\in[m/(4N),m/N]$ using a fixed hash. A dyadic threshold in this
interval is sufficient. We store the selected locations with their
exact frequencies and use a deterministic size cap. Let
\[
 Q=(d+2)\log(N+2)+\log((T+2)/\delta)+1,\qquad b=C N^{1-s}Q .
\]
We use $2k$-wise independence with $k=\lceil C_1 Q\rceil$.
To evaluate the hash we embed the keys in a binary field with enough
bits for one key and the dyadic threshold. A random polynomial of
degree at most $2k-1$ is evaluated at the key and its output is
compared with the threshold. The seed consists of $2k$ field elements.
A field of adequate bit length can be obtained by successive quadratic
extensions.\footnote{Starting with $a=1$ in $\mathbb F_2$, maintain
an element of absolute trace one. Then $X^2+X+a$ is irreducible and a
root $z$ has relative trace one. The element $az$ has absolute trace
one in the extension. The bit degrees double, and arithmetic in this
tower takes polynomially many bit operations.}

\paragraph{Analysis.}
\begin{lemma}[Simultaneous ball net]\label{lem:ball-net}
For any fixed $0<s<1$, the sample maintained above has
$N^s\poly(\mathcal L)$ space and polynomial update time.
With probability $1-\delta$, at each prescribed snapshot time
every ball containing the sample excludes at most
$C N^{1-s}Q$ current points.
For any epoch length $H_{\rm epoch}\ge1$, if the frozen snapshot is
used for at most $2H_{\rm epoch}$ further updates,
the corresponding exception bound is $C N^{1-s}Q+2H_{\rm epoch}$.
\end{lemma}
\begin{proof}
The lifts $(x,\|x\|^2)$ express ball membership as a halfspace test
in $\mathbb R^{d+1}$. On a support of size at most $N$ there are at
most $(N+1)^{d+2}$ traces of balls. For the complement of any one
trace of size greater than $b$, its sampled count has mean
$\mu\ge C Q/4$. Lemma~\ref{lem:moment-tool} and Markov's inequality
give $\Pr[Y=0]\le 2^{-k}$ for $C$ sufficiently large relative to $C_1$.
Taking $C_1$ sufficiently large first, a union bound over the traces
and the $T+1$ times leaves failure probability at most $\delta/2$.
Thus even a ball chosen after seeing the sample satisfies
\begin{equation}\label{eq:ball-net-main}
 S\subseteq B(q,r)\quad\Longrightarrow\quad
               |P\setminus B(q,r)|\le b
 \qquad\text{for every ball }B(q,r).
\end{equation}

For the total sample size the mean is at most $m$. Applying the same
moment bound at order $2k'$, for $k'=\Theta(\log((T+2)/\delta))\le k$,
bounds the probability of exceeding $m+C_2(m+k')$ by $2^{-k'}$ for
large enough $C_2$. We disable an overflowing copy. This charges at
most $\delta/2$ over all updates while enforcing the space bound on
every outcome. Sampling a key and updating its dictionary take
polynomial time.

For a frozen sample the number of exceptions to any containing ball
can increase by at most one per intervening insertion. If the sample
is empty we may use a ball disjoint from the boundary support. For
a singleton we may use the radius-zero ball at its unique point.
These cases give boundary support at most $b$ and $b+1$, respectively.
\end{proof}

\section{Fixed filters for exterior points}\label{sec:gaussian-main}
\label{sec:hierarchical-recovery}

The update routine needs two properties from a family of filters.
It must find the filters containing the updated point quickly, and
a distant target must belong to a filter sparse enough to decode.
Only the second property involves a witnessing ball. The first holds
for every choice of the random maps.

\subsection{Algorithm}

At one base radius $w$ a hierarchy consists of $h$ stages. At stage $i$
we independently choose a $k_i\times d$ Gaussian matrix $A_i$, a
uniformly shifted grid in its image of side $\delta_g\widetilde r_iw$,
and a pairwise-independent hash from cell identifiers to $B_i$ buckets.
For point $x$, its leaf is the tuple of the bucket identifiers of the
cells containing $A_i x$ at all stages. An update changes one certified
moment sketch at that leaf (Lemma~\ref{lem:certified-recovery}).
The stages share their maps across all prefixes of the tuple.

Here are the parameters for the general construction. We assume
$\sqrt2<\lambda<\sqrt3$, set the mask margin
$0<\eta_{\rm mask}\le\min\{m_\lambda/4,1/4\}$, and take
\[
 m_\lambda=\lambda^2/2-1,\qquad r_i^2=1-(i-1)m_\lambda/2 .
\]
The last stage $h$ is the first for which
$(1+\eta_{\rm mask})r_h^2<m_\lambda$. This gives $h=O_\lambda(1)$
and $r_h>0$. Each $\widetilde r_i$ is a fixed rational in
$[r_i,(1+\zeta)r_i]$. We take $\xi,\delta_g,\zeta>0$ small,
$\beta>\sqrt{1+\xi}$ close to one, and set
\[
 C_g=\left\lceil\sqrt{2\pi e}\,(\beta+\delta_g)/\delta_g\right\rceil,
 \qquad A_0=100h .
\]
We choose $B_i$ to be the least power of two at least $A_0 C_g^{k_i}$.
The proof supplies a constant $\tau>0$ that depends only on these
choices. For a budget exponent $0<\rho<s$ we set
\[
 G>2+4\ln(C_g)/\tau,\qquad
 W=\sum_{i=1}^hG^{i-1},\qquad
 0<\theta<\frac{\rho}{W\ln C_g},
 \qquad k_i=\left\lceil\theta G^{i-1}\ln(N+2)\right\rceil.
\]
The leaf capacity is given by \eqref{eq:hier-leaf-capacity} below.
Equivalently the implementation can use the explicit coefficient stated
there. Appendix~\ref{sec:combined-quantitative-two} uses a sharper two-stage calculation and the corresponding capacity \eqref{eq:factor-two-capacity} for the explicit factor-two exponent. All choices precede the stream.

We maintain independent repetitions at each prescribed base scale.
The fixed-filter interface returns the touched leaf and its capacity.
For stand-alone recovery, decoding all leaves returns the union of
their certified dictionaries. Algorithm~\ref{alg:update} instead
feeds the same updates to the local clocks of
Lemma~\ref{lem:local-compiler}.

\subsection{Analysis}

\begin{theorem}[Sparse recovery outside a ball]\label{thm:recovery}
Fix $\sqrt2<\lambda<\sqrt3$, $0<s<1$, and $0<\rho<s$. There is
$\epsilon_0>0$ such that at one prescribed positive rational radius
$w$ the fixed filters have total leaf capacity at most
\[
 (N^{1-s+\rho}+N^{1-\epsilon_0})\poly_{\lambda,s,\rho}(\mathcal L).
\]
They use space of the same form and take $\poly_{\lambda,s,\rho}(\mathcal L)$
time per update, provided the scale encoding and $|\log w|$ are
polynomially bounded in $\mathcal L$. Suppose all but
$b=O(N^{1-s}\poly(\mathcal L))$ current points lie in $B(c,w)$ and a
current point $y$ satisfies $\|y-c\|\ge\lambda w$. For this fixed
input, target and witness, with constant probability the leaf of $y$
has support at most its assigned capacity. Its certified snapshot can
then recover $y$ with constant success probability. The update uses
only the point identity and the preselected filters.
For a prescribed list of radii the costs add. Independent repetitions
give simultaneous coverage for any prescribed finite family of targets
and times, at a logarithmic cost in the family size and inverse failure
probability.
\end{theorem}

\begin{proof}
We fix the witness ball and the target independently of the grids.
The argument has three steps. We identify the part of the ball that
may collide too often, bound collisions at the first stage where a
point leaves that part, and sum its contribution over the leaves.

\subsubsection*{The cap at each scale}
Fix a witness ball $B(c,w)$. In the geometric analysis use normalized
coordinates $z=(p-c)/w$ for an input point $p$, and write $P^\circ$
for the normalized input set. The target $y$ has normalized coordinates
$Ru$, where $u$ is a unit vector. Write a bulk point as $z=t u+v$, where $v\perp u$,
$q=\|v\|^2$, and $t^2+q\leq1$.  The separation hypothesis gives
$R\geq\lambda$. Define
\[
 \phi(t,q)=q\exp\left(\frac{(\lambda-t)^2}{q}-1\right),
\]
with value $+\infty$ at $q=0$.  This convention is appropriate because
$\lambda-t\ge\lambda-1>0$. A collinear point has normalized projected
separation $(R-t)\|A_i u\|$ from the target, which exceeds the cell
diameter $\delta_g\widetilde r_i\sqrt{k_i}$ on the axial norm event below
once the constant $\delta_g$ is sufficiently small.
With $m=m_\lambda$ and $\eta=\eta_{\rm mask}$ for the following
calculation, define $R_0$ to be the original unit ball and set
$1\leq i\leq h$,
\[
 R_i=\{t u+v:\ t^2+\|v\|^2\leq1,
                    \phi(t,\|v\|^2)\leq(1+\eta)r_i^2\}.
\]
These sets are nested.  The last set $R_h$ is empty.

For every $i<h$, the set $R_i$ is contained in a ball of radius
$r_{i+1}$ whose center lies on the original target axis.  To prove
this, first observe
\[
 (\lambda-t)^2-q
 \geq \lambda^2-2\lambda t+2t^2-1
 \geq m.
\]
Consequently $\phi(t,q)\geq q+m$ and, for fixed $t$, the function
$q\mapsto\phi(t,q)$ decreases on $0<q\leq1-t^2$.
The possible $t$ values in $R_i$ therefore satisfy
\[
 F(t):=(1-t^2)
       \exp\left(\frac{(\lambda-t)^2}{1-t^2}-1\right)
       \leq(1+\eta)r_i^2.
\]
If this set is nonempty, let $t_i$ be its smallest element.
It is positive: for $t\leq0$, $F(t)\geq(\lambda-t)^2>2$, whereas
the right side is at most $1+\eta<2$.
Continuity, together with $F(t)\to\infty$ at the endpoints, gives
$F(t_i)=(1+\eta)r_i^2$.  Every point of $R_i$ satisfies $t\geq t_i$,
so
\[
 \|t u+v-t_i u\|^2
 \leq1-2t_i t+t_i^2\leq1-t_i^2.
\]
At the lower boundary, applying $\phi(t,q)\ge q+m$ at $q=1-t_i^2$
gives $F(t_i)=(1+\eta)r_i^2\ge 1-t_i^2+m$. Thus
\[
 1-t_i^2\le (1+\eta)r_i^2-m
 \le r_i^2-m/2=r_{i+1}^2,
\]
where $r_i\le1$ and $\eta\le m/4$.
This proves the claimed covering ball.  If a mask is already empty,
all subsequent masks are empty and their covering balls may be chosen
arbitrarily.  Finally $\phi\geq q+m>m$ proves $R_h=\varnothing$.

For stage $i$, let $c_i$ denote an analysis-only center of a radius
$r_i$ ball covering $R_{i-1}$.  Thus $c_1=0$, and later centers can be
the corresponding $t_{i-1}u$.  All these centers lie on the same axis;
neither they nor the target direction are computed by the algorithm.

\subsubsection*{Projection along the target direction}
Fix a unit vector $u$ and a Gaussian
matrix $A\in\mathbb R^{k\times d}$. Conditional on $Au$, a vector
$v\perp u$ has $Av\sim N(0,\|v\|^2 I_k)$ independently of $Au$.
The usual chi-square moment generating function gives exponential tails
for $\|Au\|^2/k$ about one.
\paragraph{The probability of sharing a shifted cell.}
\label{sec:grid-kernel}

In a grid of side $\ell$ with independent uniform coordinate shifts,
two projected points with difference vector $\omega\in\mathbb R^k$
share a cell with probability
\[
                   \prod_{j=1}^k(1-|\omega_j|/\ell)_+.
\]
This triangular kernel has total integral $\ell^k$ and is supported on
$[-\ell,\ell]^k$. If the random difference is Gaussian with mean $\mu$
and covariance $\sigma^2 I_k$ with $\sigma>0$, integrating against
its density gives
\[
 \Pr(\text{same cell})\le
 \left(\frac{\ell}{\sqrt{2\pi}\sigma}\right)^k
 \exp\left(-\frac{(\|\mu\|-\ell\sqrt k)_+^2}{2\sigma^2}\right).
\]
In the hierarchy, $z=tu+v$ and the target is $Ru$, so conditional on
$Au$ the difference has mean $(t-R)Au$ and covariance $qI_k$, where
$q=\|v\|^2$. On $\|Au\|^2\ge(1-\xi)k$, the lower mean separation
retains the \emph{signed} axial coordinate $t$. Replacing it by an
absolute value would lose the cap structure needed below.

\paragraph{Projecting a containing ball.}
\label{sec:grid-tail}

Translate the containing ball to the origin and scale its radius to one.
Write its point as $tu+v$, where $v\perp u$ and $t^2+\|v\|^2\le1$.
Fix $Au=g$ with $\|g\|^2\le(1+\xi)k$. Conditional on this vector,
$A(tu+v)$ is Gaussian with mean $tg$ and covariance $\|v\|^2 I_k$.

We divide each projected coordinate by $\sqrt{1+\xi}$. The coordinates now have common variance
$v_0=\|v\|^2/(1+\xi)$ and means $m_j=tg_j/\sqrt{1+\xi}$. Their
\emph{average} squared mean satisfies
$\overline m^2=k^{-1}\sum_jm_j^2\le t^2\le1-v_0$.
For $0<\theta<1/2$, the conditional log moment-generating function of
the squared norm, divided by $k$, is
\[
 -\tfrac12\ln(1-2\theta v_0)
       +\frac{\theta\overline m^2}{1-2\theta v_0}
 \le -\tfrac12\ln(1-2\theta v_0)
       +\frac{\theta(1-v_0)}{1-2\theta v_0}.
\]
Here independence lets us add the coordinate log moment-generating
functions, each affine in $m_j^2$. The derivative of the right side
in $v_0$ is $2\theta^2(1-v_0)/(1-2\theta v_0)^2\ge0$, so its maximum
is $-\tfrac12\ln(1-2\theta)$, the central $\chi^2_k$ bound per coordinate.
Consequently for $\beta>\sqrt{1+\xi}$,
\[
 \Pr\bigl(\|A(tu+v)\|>\beta\sqrt k\mid Au\bigr)
 \le e^{-Jk},\qquad
 J=\frac12\left(\frac{\beta^2}{1+\xi}-1-
                       \ln\frac{\beta^2}{1+\xi}\right)>0.
\]
The same calculation applies after translation by any center on the
original target axis and rescaling by its containing radius.

A Euclidean ball of radius $\beta r\sqrt k$ meets at most
\[
       \left(\sqrt{2\pi e}\frac{\beta+\delta_g}{\delta_g}\right)^k
\]
cells of side $\delta_g r$. Indeed, their disjoint interiors lie in the
ball enlarged by one cell diameter, and
the volume $v_k$ of the $k$-dimensional Euclidean unit ball satisfies
$v_k\le(2\pi e/k)^{k/2}$. These two bounds separate same-cell collisions
from other projected tail points in the proof that follows.

\paragraph{Isolating the projected core.}
We condition on
\[
 (1-\xi)k_i\leq\|A_i u\|^2\leq(1+\xi)k_i
 \quad\text{for every }i.
\]
For all sufficiently large $N$, this event has probability at least
$0.99$; all $k_i$ will be positive fixed multiples of $\log N$.
Bounded $N$ can be handled by full storage, with constants depending
on the chosen approximation.
In the original grid the projected core is the ball of radius
$\beta\widetilde r_iw\sqrt{k_i}$ about $A_i(c+w c_i)$. By the packing bound above it meets at
most $C_g^{k_i}$ cells, called the core cells. Write $\operatorname{cell}_i(z)$ for the
cell of $A_i(c+wz)$ when $z$ is in normalized coordinates, and use
$\operatorname{cell}_i(Ru)$ for the target cell. Conditional on the
projection and shift, the independent outer hash avoids sending any
other core cell to the target bucket with probability at least
$1-1/A_0$.  Let $I_i$ denote this isolation event and
$I=\bigcap_i I_i$.  Conditional on the Gaussian norm events,
$\Pr[I]\geq0.99$ by a union bound.
\subsubsection*{The collision probability outside a mask}

There are fixed choices of $\xi,\beta-1,\delta_g,\zeta>0$ and a
fixed $\gamma_{\rm col}>0$ such that, at every stage $i$, every point of the
original unit ball satisfying
\[
 \phi(t,q)>(1+\eta)r_i^2
\]
has conditional same-cell probability at most
\begin{equation}\label{eq:hier-safe-collision}
 \Pr[\operatorname{cell}_i(z)=\operatorname{cell}_i(Ru)
               \mid A_i u]
 \leq C_g^{-k_i}e^{-\gamma_{\rm col} k_i}.
\end{equation}
This statement does not require $z$ to lie in the stage-$i$ covering
ball; belonging to the original unit ball and lying outside the mask
is enough.

The shifted-grid estimate bounds the collision
probability by the $k_i$th power of
\[
 \frac{\delta_g\widetilde r_i}{\sqrt{2\pi q}}
 \exp\left[-\frac{
   (\sqrt{1-\xi}(\lambda-t)-\delta_g\widetilde r_i)^2}{2q}\right].
\]
We choose the constants so that the expression in parentheses in the
exponent stays positive.  The actual target ratio $R\geq\lambda$
can only reduce this bound.  Multiplying the base by $C_g$, and taking
the limit $\xi,\beta-1,\delta_g,\zeta\to0$, gives exactly
\[
 \frac{r_i}{\sqrt{\phi(t,q)}}
 \leq(1+\eta)^{-1/2}<1.
\]
The bound extends continuously by zero at $q=0$, uniformly because
$\lambda-t\geq\lambda-1>0$.  Compactness of the original unit ball
and finiteness of the stage set therefore provide common fixed small
parameters and a positive $\gamma_{\rm col}$ for
\eqref{eq:hier-safe-collision}.  One may equivalently separate a small
$q$ interval, where exponential decay is uniform, and use ordinary
uniform continuity on its compact complement.

For a normalized point $z$ in the stage-$i$ covering ball,
apply the conditional projection bound to $(z-c_i)/\widetilde r_i$.
In normalized coordinates the projected-core tail is
\begin{equation}\label{eq:hier-core-tail}
 \Pr[\|A_i(z-c_i)\|>
        \beta\widetilde r_i\sqrt{k_i}\mid A_i u]
 \leq e^{-Jk_i},
\end{equation}
where
\[
 J=\frac12\left[
 \frac{\beta^2}{1+\xi}-1-
 \ln\left(\frac{\beta^2}{1+\xi}\right)\right]>0.
\]
\subsubsection*{Load at a target leaf}

Partition the bulk points by their first departure from the nested
masks:
\[
 G_i=(P^\circ\cap R_{i-1})\setminus R_i,\qquad1\leq i\leq h.
\]
These classes cover all points in the original witness ball because
$R_h=\varnothing$.  They are determined by the fixed input, witness,
and target, independently of all stage randomness.

Fix $z\in G_i$ and condition on $A_i u$ in its prescribed norm range.
Let $E_i$ be the event that $z$ and the target have the same outer bucket,
$S_i$ the event that they have the same cell, and $T_i$ the projected-core
tail in \eqref{eq:hier-core-tail}. Core isolation gives the inclusion
\[
 E_i\cap I_i\ \subseteq\ S_i\ \cup\ (T_i\cap S_i^{\,c}\cap E_i).
\]
Indeed, a different cell meeting the projected core is excluded by $I_i$.
Conditional on the projection and shift, two \emph{different} cell keys
have outer-hash collision probability $1/B_i$. Integrating this fact
over $T_i\cap S_i^{\,c}$ and applying \eqref{eq:hier-safe-collision}
and \eqref{eq:hier-core-tail} gives
\[
 \begin{aligned}
 \Pr(E_i\cap I_i\mid A_i u)
 &\le \Pr(S_i\mid A_i u)+B_i^{-1}\Pr(T_i\mid A_i u)\\
 &\le \frac{2A_0e^{-\gamma_{\rm col} k_i}+e^{-Jk_i}}{B_i}
 \le \frac{(2A_0+1)e^{-\tau k_i}}{B_i},
 \end{aligned}
\]
for $0<\tau\le\min\{\gamma_{\rm col},J\}$, using $B_i<2A_0C_g^{k_i}$.
This bounds the joint event $E_i\cap I_i$.

At every later stage $j>i$, the same point is still outside the mask:
$\phi(t,q)>(1+\eta)r_i^2\geq(1+\eta)r_j^2$.
It therefore shares the geometric cell with probability at most
$C_g^{-k_j}e^{-\gamma_{\rm col} k_j}$.  If the cells differ, pairwise hashing
passes it with probability $1/B_j$.  Thus, without requiring any
isolation event at stage $j$,
\[
 \Pr[\text{$z$ passes stage $j$}\mid A_ju]
 \leq\frac{1+2A_0e^{-\gamma_{\rm col} k_j}}{B_j}
 \leq\frac2{B_j}
\]
for sufficiently large $N$.

Stages are independent even after conditioning on their separate norm
events. Discarding the tests before $i$ can only increase the bound. The probability that $z$
reaches the target leaf \emph{and} all isolation events hold is at
most
\[
 C_h\frac{e^{-\tau k_i}}{B_iB_{i+1}\cdots B_h},
\]
where $C_h$ depends only on the fixed depth.  This avoids an incorrect
analysis in which exceptional load at an early level is carried
unchanged through all later levels: once a point is outside a mask,
later independent hashes dilute it by $O(1/B_j)$.

By linearity of expectation, conditional on all norm events,
\[
 \mathbb E\left[\mathbf1_I\cdot
       \#\{\text{bulk points in the target leaf}\}\right]
 \leq C_h N\sum_{i=1}^h
          \frac{e^{-\tau k_i}}{B_i\cdots B_h}.
\]
Markov's inequality bounds the probability that $I$ holds but this
load is more than one hundred times the displayed expectation by
$0.01$. We charge the norm and isolation failures separately.  The
original $b$ exceptional points may all share the target leaf.
Consequently a leaf capacity
\begin{equation}\label{eq:hier-leaf-capacity}
 K=\left\lceil C'_h\left[
 b+1+N\sum_{i=1}^h
       \frac{e^{-\tau k_i}}{B_i\cdots B_h}\right]\right\rceil
\end{equation}
recovers the actual target with constant probability. One may take
$C'_h=100(2A_0+1)2^{h-1}$: the first-stage coefficient is at most
$2A_0+1$, the later coefficients contribute at most $2^{h-1}$,
and the factor $100$ is the Markov allowance. No independence
between points is assumed anywhere.

\subsubsection*{Space, update time, and repetition}
The growing dimensions in the algorithm imply
\[
 B_{\rm tot}:=\prod_{i=1}^hB_i=O_{\lambda,\rho}(N^\rho),
\]
and, for every stage $i$,
\[
 e^{-\tau k_i}\prod_{j<i}B_j
 \leq O_{\lambda,\rho}(N^{-\epsilon_0}),
 \qquad\epsilon_0=\tau\theta/2>0.
\]
The latter inequality follows from
$\ln C_g\sum_{j<i}G^{j-1}\leq(\tau/4)G^{i-1}$;
rounding and the $A_0$ factors contribute only fixed constants.
Multiplying \eqref{eq:hier-leaf-capacity} by the number of leaves gives
\[
 B_{\rm tot}K
 \leq O_{\lambda,s,\rho}\left(
      N^\rho(b+1)+N^{1-\epsilon_0}\right).
\]
For $b=N^{1-s}\operatorname{poly}$ and $\rho<s$, both powers of $N$
are strictly smaller than one.

Each original update computes $h$ Gaussian projections and grid/hash
IDs, and changes one sparse-recovery sketch at the resulting leaf.
Since $h$ is fixed and each $k_i=O_\lambda(\log N)$, this costs
polynomial time in $d$ and the logarithmic parameters, independent of
the number of leaves and the leaf recovery capacity.
Only the original witness radius must be guessed.  At each of the
$O(\log(d\Delta+2))$ original geometric scales $w$, all stage grid
widths are the fixed constants $\delta_g\widetilde r_iw$.
There is no data-dependent reclassification and no product of scale
choices. At each base scale we use $O(\log(NT/\delta))$ independent
hierarchies. A query that decodes all leaves takes the union of their
certified dictionaries.  A union bound over
targets and times gives the simultaneous recovery guarantee.

For a finite description we choose rational
$\lambda,\eta,C_g,G,\theta$ satisfying the displayed strict
inequalities. If the specified separation $\lambda$ is irrational, we use any rational separation strictly between $\sqrt2$ and $\lambda$. We choose a rational $q_0\in(0,1)$ strictly larger than
both $e^{-\gamma_{\rm col}}$ and $e^{-J}$, and set $\tau=-\ln q_0$; the
capacity terms $e^{-\tau k_i}=q_0^{k_i}$ are then rational and may be
rounded upward exactly. We take $G$ rational above its lower bound and
$\theta$ rational with strict aggregate bucket exponent
$\theta W\ln C_g<\rho$.  Approximating $\ln(N+2)$ from above
within a fixed additive constant changes each $k_i$ by only a fixed
constant.  The bucket ranges are powers of two and the grid radius
multipliers $\widetilde r_i$ are rational upper approximations to the
positive $r_i$.  These choices leave the positive exponent savings
intact.

Appendix~\ref{sec:finite-bit-gaussian-implementation} proves the membership coupling. With rational scale multipliers and sufficiently fine Gaussian and
shift rounding, every streamed point keeps its ideal grid cell except
with the allocated coupling-failure probability.  Only polynomially many
bits in the allowed parameters are needed.
\end{proof}

\subsection{Application to the hull}
We maintain a live Bernoulli sample of expected size at most $N^s$ independently of the hierarchy.  Its simultaneous ball-net property gives
$b=N^{1-s}\operatorname{poly}(d,\log N,\log(T/\delta))$ exceptions
for every ball containing the sample.  Empty and singleton samples use
the global fallback bucket in Algorithm~\ref{alg:update}.

Fix $c>\sqrt3$ and choose
\[
 \sqrt2<\lambda<\min\{a,\sqrt3\},\qquad
 \sqrt2<a<\sqrt{(c^2+1)/2},\qquad 1<q_{\rm scale}\le a/\lambda .
\]
For a witness $B(c_0,r)$ with $\|y-c_0\|>ar$, round $r$ up to the
first maintained scale $w\ge r$. Since $w\le q_{\rm scale}r$,
the larger ball still contains the sample and
$\|y-c_0\|/w>a/q_{\rm scale}\ge\lambda$.
Every point outside the expanded farthest-distance hull
$\mathcal H_a(S)$ is then recovered by some hierarchy.  If $Z$ is the
union of recovered points, the exact query
$L=\operatorname{diam}(S\cup Z)$ satisfies
\[
 L\leq\operatorname{diam}(P)
       \leq\sqrt{2a^2-1}\,L\le cL.
\]
Thus every fixed factor strictly larger than $\sqrt3$, in particular
factor two, has a fixed-power sublinear-space construction with
polynomial update time and an expensive query.  Algorithm~\ref{alg:update} maintains this approximation by querying only the exposed centers.
Including the live sample, one may take
\[
 \alpha=\max\{s,1-s+\rho,1-\epsilon_0\}<1
\]
for this slow-query consequence, with the displayed polynomial factors.
For the integer grid, \eqref{eq:hull-witness-scale} justifies the scale
endpoints in Algorithm~\ref{alg:update}.

\section{From sparse leaves to a current candidate set}\label{sec:local-union}

A filter that is sparse now may have received its last update long
before the current global epoch. To recover all its points in time to
score them, we measure the age of its snapshot in updates to that
filter. The resulting routine translates original updates into an
ordered sequence of insertions and deletions of explicit candidate
points. A second ingredient certifies each snapshot before any stored
entry is used as a witness.

\subsection{Algorithm}

Every bucket $i$ is a fixed filter of point identities, with capacity
$K_i\ge1$. Each of its selected points has its original nonnegative
frequency. Each original update reaches at most $r=\poly(\mathcal L)$
buckets. Conditional on the fixed stream and these filters, the
sequence in each bucket is independent of its private recovery hashes.

\subsubsection{Scheduling and maintaining reference counts}
\begin{figure}[htbp]
\centering
\begin{tikzpicture}[x=1cm,y=.8cm,font=\small,>=Latex]
 \draw[->] (0,0)--(10.4,0);
 \foreach \x/\lab in {0/{\tau},3/{\tau+K},6/{\tau+2K},9/{\tau+3K}} {
   \draw (\x,.08)--(\x,-.08);
   \node[below=5pt] at (\x,0) {$\lab$};
   \draw[gray!35,dotted] (\x,.18)--(\x,2.8);
 }
 \node[anchor=east,align=right] at (-.25,2.2) {Snapshot\\at $\tau$};
 \node[anchor=east,align=right] at (-.25,1.1) {Snapshot\\at $\tau+K$};
 \filldraw[fill=blue!10,draw=blue!65!black] (0,1.85) rectangle (3,2.55);
 \node at (1.5,2.2) {Build and replay};
 \filldraw[fill=green!10,draw=green!45!black] (3,1.85) rectangle (6,2.55);
 \node at (4.5,2.2) {Serve};
 \draw[dashed] (6,1.85)--(9.4,1.85)--(9.4,2.55)--(6,2.55);
 \node at (7.7,2.2) {Retire lazily};
 \draw[dashed,->] (9.4,2.2)--(10.15,2.2);
 \filldraw[fill=blue!10,draw=blue!65!black] (3,.75) rectangle (6,1.45);
 \node at (4.5,1.1) {Build and replay};
 \filldraw[fill=green!10,draw=green!45!black] (6,.75) rectangle (9,1.45);
 \node at (7.5,1.1) {Serve};
\end{tikzpicture}
\caption{Reconstruction follows a leaf's local update count.
During a generation's serving interval, its snapshot is at most
$2K$ local updates old. A currently sparse leaf therefore has a
recoverable serving snapshot. Retirement is handled by a separate
global work queue; the dashed arrow imposes no local deadline.
An idle leaf therefore cannot accumulate unbounded unreclaimed records.}
\label{fig:local-clock}
\end{figure}
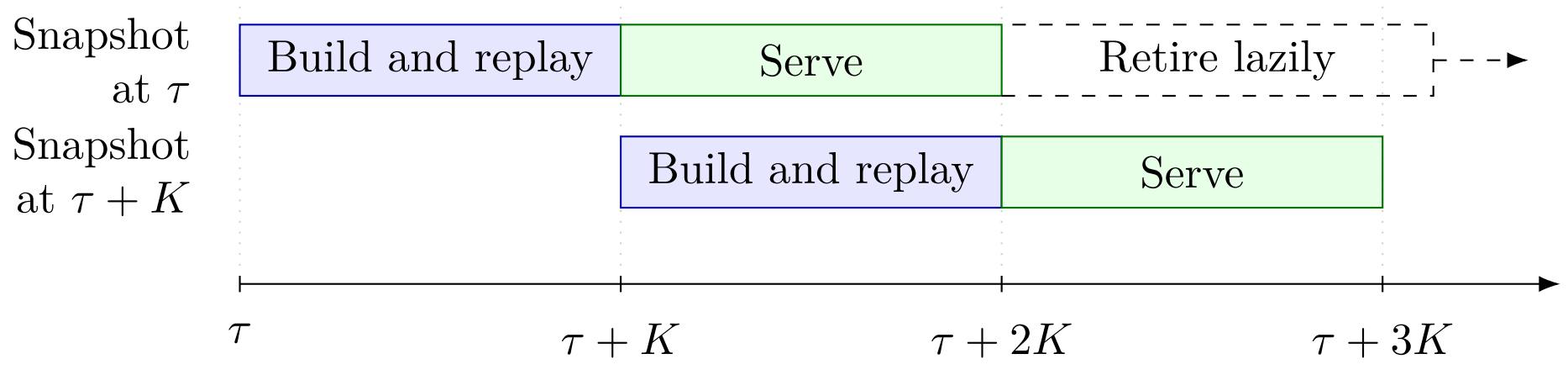

Figure~\ref{fig:local-clock} shows the build and serving intervals for two successive generations.
Bucket $i$ counts only its own updates.
Its blocks have length $K_i$ in this local clock.  At every local block
boundary, the bucket captures the root of its persistent moment sketch
and its total mass. It decodes that snapshot during the next block,
installs the resulting dictionary at the following boundary and retires
it one block later.  If decoding fails or the certified support exceeds $3K_i$, the
generation is unavailable.  Timing does not depend on recovery
success or on the identities of the recovered points.

Initially the empty dictionary serves the bucket exactly. It can remain
in service through the first two local blocks, recording at most $2K_i$
distinct keys. The job for the snapshot at time $K_i$ starts then and installs its output at $2K_i$. Subsequent jobs follow the same schedule.
Small constant $K_i$ can perform all bounded work immediately, without
altering the asymptotic bounds.
\begin{algorithm}[htbp]
\caption{One update of a leaf with capacity $K$}
\label{alg:leaf}
\begin{algorithmic}[1]
\Require{Point update $(p,\sigma)$, local time $u$, and current generations}
\Ensure{Updated sketch and generations, with candidate transitions added to the queue}
\State Update the persistent moment sketch and its exact total
\State Increment $u$ and, if a job is building, log $(p,\sigma)$ for it
\State If the serving generation is available, maintain its complete dictionary,
       adding a newly inserted key if needed
\State Refresh materialized references for $p$ using its shared global current count
\State Advance the building job, if present, by its fixed work budget
\Statex \hspace{\algorithmicindent}Decode and certify the snapshot, then scan snapshot and change-map keys
\Statex \hspace{\algorithmicindent}For each scanned key $x$, materialize its reference using
       $F_\tau(x)+\Delta_\tau(x)$
\If{$u$ is a multiple of $K$}
  \State If a job was due, install its result and enqueue the previous root for retirement
  \State Capture a snapshot of the current sketch and start the next job
\EndIf
\State Append the resulting membership changes, in order, to the copy's candidate queue
\end{algorithmic}
\end{algorithm}
A failed or oversized recovery produces
an unavailable generation, and the schedule continues unchanged.
There is at most one serving and one building generation to update.
Retired generations receive no local log maintenance: their references
read the global multiplicity and are removed by the retirement queue.
An entry of a complete serving dictionary that is absent before an
insertion has previous multiplicity zero. Its new reference can
therefore be initialized exactly. Existing entries share the global
current count; they never increment that count again.

We store a signed change map for each building job. It serves a
different purpose from the shared multiplicity of a point: the map
lets us initialize a point correctly the first time any dictionary
discovers it. In particular it handles a point deleted and reinserted
during discovery.
For a job whose snapshot is at local time $\tau$, retain a dictionary
of signed changes
\[
 \Delta_\tau(p)=\sum_{\tau<v\leq u}
             \text{update to point }p\text{ at local time }v.
\]
There are at most $K_i$ updates during the job's build interval, so this
map and its chronological log occupy $O(K_i\operatorname{poly})$
space.  Once certified decoding has supplied the complete snapshot
dictionary $F_\tau$, the current multiplicity of any relevant key is
exactly
\[
 f_u(p)=F_\tau(p)+\Delta_\tau(p),
\]
where a missing key of the complete snapshot dictionary has value zero.
Each local update receives a sufficiently large fixed polynomial work budget.  The first part of the block completes decoding.  The
remaining work scans the keys of $F_\tau$ and of the change map, and
materializes their current multiplicities using the displayed identity.
New change-map keys that arrive during this scan are enqueued.  There
are only $O(K_i)$ total keys to process during a block; processing
them faster than new keys arrive completes the scan before the fixed
deadline.  After the scan has caught up, a newly introduced key can be
materialized immediately in polynomial time, because the complete
snapshot dictionary is already available.  Every already materialized
key receives subsequent original frequency updates normally.

We keep a global dictionary keyed by the original point identity.  A record
stores its exact current original multiplicity, a reference count, and
one identity in the explicit-point geometric structure.  References
come from materialized entries of active, pending, and not-yet-cleaned
retired bucket generations.  The geometric candidate set contains this
point exactly when its current multiplicity and reference count are
both positive.  Thus repeated discoveries in many filters do not
produce many geometric identities for the same location.
When we first materialize a point, we initialize its global multiplicity
from the certified snapshot-plus-change formula.  If a global record
already exists, its current multiplicity is already known and agrees
with that formula.  Thereafter, every original update to that identity
updates the global multiplicity exactly once, before any background
materialization for that update.  A positive-to-zero transition removes
the point from the geometric structure once, irrespective of the number
of dictionary references.  A zero-to-positive transition adds it once
if references remain.  Original strictness guarantees that these
counts remain nonnegative.

Adding a reference creates a geometric candidate only if it is the
first reference to a currently positive point.  Dropping the final
reference removes a positive point from the candidate structure and
permits its global record to be freed.  If that point is later
rediscovered, the new certified snapshot and change map initialize its
then-current multiplicity afresh.  There is no need to maintain counts
for original points that have no stored references.
A new live-sample identity can initialize the same record from its
exact sample frequency. Every membership change appends a transition
to the common queue. The outer reference-retirement queue has a fixed
budget separate from any private work done by the candidate hierarchy.

\subsubsection{The snapshot decoder}\label{sec:exact-recovery}
The decoder receives a read-only snapshot. Its success condition needs
a check for a key's identity and a separate check that every key was
found. A collision can prevent discovery, so both checks must succeed.
We encode each original point as a distinct nonnegative integer key $u$
of $b_{\rm key}=O(d\log(\Delta+2))$ bits. The same construction applies
to any injective key encoding of polynomial bit length. We use $h_{\rm rec}$
independent pairwise-independent hashes into $B$ buckets, with $B$ the
power of two in $[24K,48K)$. A random binary matrix and independent
offset on fixed-length key strings supply each hash. In each bucket
we store three exact integer moments of its frequencies,
\[
 C_0=\sum_u f_u,\qquad C_1=\sum_u uf_u,\qquad C_2=\sum_u u^2f_u .
\]
An update of sign $\sigma$ increments $F$ by $\sigma$ once and changes
one moment triple in each table. A snapshot captures $F$ and all the
tables at the same instant. For a filtered stream, $F$ counts precisely
the mass in that filter. In a single table it also equals the sum of
$C_0$ over its buckets.
We take $h_{\rm rec}=\lceil C(\log(K+2)+\log(M_{\rm snap}/\delta+2))\rceil$
tables for a prescribed family of $M_{\rm snap}$ snapshots. Within
one fixed filter hierarchy the same recovery hashes may be used in
every leaf. The snapshots are determined by the stream and the outer
filters, and the argument union-bounds over them. Separate complete
copies use independent seeds.
Algorithm~\ref{alg:recover} scans them incrementally. Its only possible
outputs are a complete dictionary and failure.
\begin{algorithm}[htbp]
\caption{Certified recovery of a nonnegative bucket snapshot}
\label{alg:recover}
\begin{algorithmic}[1]
\Require{Moment tables and exact mass $F$ from the same filter snapshot, capacity $K$}
\Ensure{Complete snapshot dictionary with exact multiplicities, or failure}
\State Initialize an empty dictionary $D$
\For{each stored counter triple in each hash table}
  \If{$C_0>0$ and $C_0C_2=C_1^2$}
    \State $u\gets C_1/C_0$
    \State Store the entry $D[u]\gets C_0$ once, deduplicating by key
  \EndIf
\EndFor
\If{$|D|\le 3K$ and $\sum_{u\in D}D[u]=F$}
  \State \Return $D$
\Else
  \State \Return unavailable
\EndIf
\end{algorithmic}
\end{algorithm}

\subsection{Analysis}

\begin{lemma}[Certified sparse recovery]\label{lem:certified-recovery}
For a nonnegative frequency vector of peak support at most $N$ and total mass at most $T$,
a capacity-$K$ sketch uses $K\poly(\mathcal L)$ space, has
$\poly(\mathcal L)$ update time, and can be decoded in
$K\poly(\mathcal L)$ time. Every accepted output is the complete
snapshot dictionary with exact multiplicities. The sketch also maintains
$F=\sum_u f_u$, the total mass of its own frequency vector.
For a prescribed family of $M_{\rm snap}$ snapshots independent of
the private hashes, every snapshot of support at most $3K$ is accepted
simultaneously with probability at least $1-\delta$. The resource
bounds above assume $\log(M_{\rm snap}+2)+\log(K+2)=\poly(\mathcal L)$, as in the leaf and fallback buckets.
\end{lemma}
\begin{proof}
The test for a singleton follows from the identity
\begin{equation}\label{eq:purity-exact}
 C_0C_2-C_1^2=\sum_{u<v}f_uf_v(u-v)^2 .
\end{equation}
Every term on the right is nonnegative. Hence $C_0>0$ and
$C_0C_2=C_1^2$ hold exactly for a bucket with one occupied key.
Its key is $C_1/C_0$ and its snapshot multiplicity is $C_0$.
The decoder keeps one entry per recovered key across the pure buckets.
It accepts if the dictionary has at most $3K$ entries and their multiplicities
sum to $F$. Every recovered count is exact, so any omitted positive
frequency would make the sum strictly smaller than $F$. This also
certifies the empty dictionary when $F=0$.
For support at most $3K$, a fixed key fails to be isolated in one hash
table with probability below $1/8$. Therefore complete recovery fails
with probability at most $3K8^{-h_{\rm rec}}$. Taking
$h_{\rm rec}$ to be a sufficiently large multiple of
$\log(K+2)+\log(M_{\rm snap}/\delta+2)$ makes all needed sparse
recoveries succeed simultaneously. The snapshot family may be fixed
by conditioning on the stream and the outer filters. The union bound
uses only independence from the private recovery hashes.

Since the total mass is at most $T$, even $C_2$ has at most
$2b_{\rm key}+\lceil\log_2(T+1)\rceil+1$ bits. Thus each counter
occupies $O(1+(b_{\rm key}+\log(T+2))/w_{\rm word})$ words.
The parity operations, products, and exact divisions on these
multiword keys and counters take $\poly(\mathcal L)$ word operations. Lazy initialization avoids clearing a large table in one
update. A fixed scan budget, followed by deterministic sorting or
balanced-tree deduplication, gives the stated decoding bound. The
sketch is linear in the frequency vector and is valid under deletions
because every bucket remains a nonnegative-frequency stream.
\end{proof}

A rejected snapshot creates no new reference. To see why accepting a
snapshot suffices even while later updates arrive, we use the fixed
local deadlines and the exact frequency shared by all references to
the same point.
\begin{lemma}[Local-clock compiler]\label{lem:local-compiler}
There is a continuously maintained candidate set $C$ with these properties:
\begin{enumerate}
\item Every candidate is a current point of the original stream, on
      every random outcome.
\item On the recovery success event, $C$ contains every point in every
      bucket whose current support has size at most $K_i$.
\item At most $O(\sum_i K_i)\operatorname{poly}$ records are stored,
      including pending and retiring records.
\item Every original update causes only a polynomial number of
      insertions or deletions in $C$, and only polynomial other work,
      in the worst case.
\end{enumerate}
\end{lemma}
\begin{proof}
At any later time, the active generation is based on a snapshot at most
$2K_i$ local updates old.  Therefore, if the current support is at most
$K_i$, that snapshot had at most $3K_i$ keys and is accepted on the
success event.  An accepted generation has at most $5K_i$ current or
zero-frequency tombstone entries over its entire serving interval:
it starts with at most $3K_i$ keys, and at most $2K_i$ later local
updates can introduce new identities.  Thus the local readiness
invariant and space bound hold independently of how slowly or quickly
that bucket's local clock advances in global time.

\paragraph{Current identities at installation.}
There are at most $3K_i$ decoded keys and at most $K_i$ distinct keys
changed during a build. We reserve a fixed first fraction of the block
for decoding, then scan these keys at a service rate exceeding the
arrival rate. A newly encountered key is initialized with
$F_\tau(p)+\Delta_\tau(p)$. Once the scan catches up, a newly inserted
key is processed immediately. Increasing the per-update constant
reserves enough work for all $4K_i$ keys before the deadline, including
dictionary operations. An installed dictionary therefore already
contains a reference for each key of positive current frequency. Swapping
the serving root costs constant metadata work.
Generation retirement is lazy.  Its entries remain references until a
background server decrements them, and may keep still-present points
as extra candidates.  Such extra candidates can only help a diameter
lower estimate.  If an original point is deleted before its retired
references have been processed, its global multiplicity becomes zero
and it is removed geometrically immediately; old references cannot
produce a stale distance.  Reinsertions are equally safe because the
global frequency record continues to receive the original updates.

Zero-count tombstones and unprocessed retired entries are included in
the storage bound.  Each local dictionary entry contributes only one
reference.  Point records with no references can be freed in constant
record work, with standard generation/version checks if needed.

\paragraph{Work, memory, and independence.}
Each original update advances at most $r$ local clocks.  In each of
those buckets it updates the sparse sketch, bounded local bookkeeping,
and a fixed polynomial amount of decoding/materialization work.
Hence the number of new references and global point creations per
original update is bounded by a polynomial $A_0$ in the allowed
parameters.  The original frequency update causes at most one further
geometric insertion/deletion for its own identity.  All these operations
are performed in a fixed order.

Let $S_0=O(\sum_i K_i)\operatorname{poly}$ bound the non-retired
bucket, snapshot, and global-reference storage.  Retiring an entire
generation enqueues its root in constant work. We process at least
$2A_0$ constant-record cleanup actions per original update, increasing
the constant to cover reference decrements and geometric updates.
Over any interval of $u$ updates, at most $S_0+A_0u$ records can enter
retirement: records live at its start plus newly allocated records.
The retirement backlog is therefore at most $S_0$.  In particular,
many buckets may become idle immediately after retirement without
retaining unbounded garbage; the cleanup clock is global, not local.

Shared persistent snapshots use reference counting with deferred edge
decrements.  Every allocated pointer edge is processed only a constant
number of times, so these actions obey the same charge.  The geometric
structure has near-linear memory in the candidate records, including
its own already-bounded retirement overhead.  Total memory remains
$O(\sum_i K_i)\operatorname{poly}$ and total update work is
worst-case polynomial on every random outcome.

To preserve the oblivious-input premise of the explicit-point
primitive, pad each of its update calls to its known worst-case bound
and count it as one scheduled candidate operation.  Conditional on
the outer projection, hashing, and sparse-recovery randomness, the
candidate sequence and the times of its insertions/deletions then
depend only on the original fixed stream, not on the geometric
primitive's private random choices.  Its failure budget can be set
using the polynomial upper bound on the number of such internal
updates.  The logarithm of that bound is still polynomial in the
allowed logarithmic parameters.

The outer reference-retirement queue is separate from the geometric
primitive's private retirement queues.  Each receives its own fixed
work budget.  An outer cleanup action that inserts or removes a
candidate makes one padded primitive call; private allocations inside
that call do not consume or reorder the outer cleanup budget.  This
prevents private geometric randomness from changing the timing of
outer reference drops, and makes the preceding independence claim an
explicit implementation invariant.
\end{proof}

\section{Diameter certificates for explicit points}\label{sec:explicit}

The input to this routine is the ordered sequence of candidate
transitions. Its two outputs have different roles. A witnessed score
handles pairs of candidates. A short list of balls, each backed by
current anchors, lets Algorithm~\ref{alg:update} compare candidates
with the frozen sample.

\subsection{Algorithm}

\subsubsection{A chain of balls and an exact leaf}
A foreground copy is a single chain.  A node has an explicit input set
$P_i$, a fixed ball $B_i=B(c_i,(1+\eta)\sqrt{V_i})$, and a child
whose input is exactly $P_i\setminus B_i$.  Small leaves are stored and
processed directly.  The center and variance have a convex certificate
\[
 c_i=\sum_{a\in K_i}\lambda_a a,
 \qquad V_i=\sum_{a\in K_i}\lambda_a\|a\|^2-\|c_i\|^2,
 \quad \lambda_a\geq0,
 \quad\sum_a\lambda_a=1,
\]
where $K_i$ consists of $O(\eta^{-2})$ points from a uniform sample
$S_i$ and $B_i$ contains the whole sample.  The next subsection gives an
explicit rational construction. Let
$R_{\rm hier}=\lceil C_R\ln((T_C+2)/\delta)\rceil$ be the number of
independent copies used for amplification. At each nonterminal node take
\[
 s_{\rm node}=\left\lceil C_s\bigl(d+\ln((T_C+2) R_{\rm hier}
                       \log(M+2)/\delta)\bigr)\right\rceil
\]
sample points, or the whole input if smaller.
At every draw, conditional on the history defining its input snapshot,
the half-net guarantee of Lemma~\ref{lem:half-net} applies. Each of
the $T_C$ candidate transitions visits at most $O(\log(M+2))$ jobs per
copy, with at most $O(\log(M+2))$ sampled levels in each job. Thus
$O((T_C+1)R_{\rm hier}\log^2(M+2))$ bounds the number of samples
over all copies even if jobs are canceled. With sufficiently large
$C_s$, a union bound makes the probability of any failed half-net test
at most $\delta/8$ in the ideal uniform-rank experiment. We check the
sample ball against the same input snapshot that produced it. An affected hierarchy copy
becomes unavailable on failed half coverage, chain-depth overflow,
or an unfinished job at a fixed deadline. The depth cap is a
sufficiently large constant times $\log(M+2)$. These checks bound the
resources on every outcome. Each copy starts with an empty exact leaf.
For each anchor $a\in K_i$, store a distance heap to every current point
of $P_i$.  The node score is
\[
 H_i=\max_{a\in K_i,\,p\in P_i}\|a-p\|.
\]
If any point of the \emph{whole sample} $S_i$ is deleted after sampling,
permanently mark this node's score unavailable for that generation.  This
is deliberately stronger than necessary; it suffices to retain the
simple uniform-sampling analysis.  A valid node has $K_i\subseteq P_i$,
so $H_i$ is a current pair distance.  Balls remain fixed for routing even
when their score is unavailable.  All heaps continue to track the current
node input.  The copy reports the maximum available node score, including
its exact leaf diameter.
Each such node sample is monitored from the moment it is drawn. A deletion
during a build is recorded when routed or replayed, so an invalidated
node carries that flag into service. A disabled copy contributes no
score or certificates.

\subsubsection{Constructing a rational certificate}
Given a uniform sample, we need both a ball containing it and a few
original points witnessing its variance. Algorithm~\ref{alg:certificate}
starts with a pair and transfers a small amount of weight to a farthest
point whenever the ball still misses a sample point.
\begin{algorithm}[htbp]
\caption{A small rational variance certificate for a finite sample}
\label{alg:certificate}
\begin{algorithmic}[1]
\Require{Nonempty finite sample $S$ and rational $0<\eta\le1/4$}
\Ensure{Anchor set $K$, convex weights, mean $c$, and variance $V$ with $S\subseteq B(c,(1+\eta)\sqrt V)$}
\State Choose $a_0\in S$ and $b\in\arg\max_{x\in S}\|x-a_0\|$
\If{$b=a_0$}
  \State \Return $(K=\{a_0\},\lambda_{a_0}=1,c=a_0,V=0)$
\EndIf
\State $\lambda\gets(e_{a_0}+e_b)/2$;
       $\theta\gets(1+\eta)^2-1$; $\gamma\gets\theta/32$
\Loop
  \State $c\gets\sum_x\lambda_xx$;
         $V\gets\sum_x\lambda_x\|x\|^2-\|c\|^2$
  \State $a\gets\arg\max_{x\in S}\|x-c\|$; $R^2\gets\|a-c\|^2$
  \If{$R^2\le(1+\eta)^2V$}
    \State \Return $(K=\supp(\lambda),\lambda,c,V)$
  \EndIf
  \State $\lambda\gets(1-\gamma)\lambda+\gamma e_a$
\EndLoop
\end{algorithmic}
\end{algorithm}

\subsubsection{Rebuilding a suffix at a fixed deadline}
Once a ball is installed it stays fixed for routing. Every foreground
node owns at most one job, which rebuilds its suffix from top to bottom.
Its deadline must be fixed before its fresh samples are drawn. This
keeps rebuild decisions independent of whether those samples survive.
We use the integer scheduling parameters
\[
 B_{\rm sched}=\lceil C_B s_{\rm node}\log(M+2)\rceil,\qquad
 H_{\rm leaf}=128B_{\rm sched},
\]
where $C_B$ is sufficiently large and $B_{\rm sched}\ge128$.
Thus an exact leaf of size $O(H_{\rm leaf})$ fits a polynomial work
budget even if all of its pairwise distances are recomputed.
Immediately after a foreground node is installed, let $m_{\rm start}$ be the current
size of its input. Its next rebuild starts at installation with a deadline after
\[
 \ell=\lfloor m_{\rm start}/B_{\rm sched}\rfloor
\]
updates of that node's input.  The job rebuilds its whole suffix.  If an
ancestor replaces this node first, cancel this job.  Rebuild timing never
uses whether the node's sample survived, whether its score is available,
or the identities in that sample.  An exact leaf starts a suffix job when it first exceeds $H_{\rm leaf}$,
or immediately upon installation if it already exceeds that threshold.
It stays exact until that job commits. Its initial size is at most
$2H_{\rm leaf}$ and it receives at most $2H_{\rm leaf}/B_{\rm sched}$
more input updates, so direct processing remains polynomial.
Each foreground node may own one job.  Crucially, nodes inside an
\emph{uninstalled pending chain do not launch rebuild jobs}.  This rule
avoids the recurrence $U(m)\simeq2U(m/2)$ that would arise from recursively
replaying both an old and a pending hierarchy.

When deadlines coincide, we first route the update through the active
foreground chain and its pending prefixes and perform their budgeted
work. We then commit due jobs in ancestor-first order, canceling jobs
in each replaced suffix.  A newly installed node already includes the current
update through its replay, starts its own clock at zero, and receives
only future updates.  Thus a suffix swap never duplicates one input
update or requires a bulk replay into newly started jobs.

The initiating node's input updates are the job's host clock.  Its
fixed deadline is $D$.  The job consists of a completed prefix of new
nodes and one exact current residual dictionary $Q$; the latter is the
set not served by that prefix.  Before any prefix exists, it can alias the
host dictionary.  A host update changes $|Q|$ by at most one. The job updates the heaps
in its completed prefix and routes the change into $Q$ when it lies
outside every completed ball.  There is never a second recursively running pending
hierarchy inside this job.

Let $m=|Q|$ and let $r$ count the host updates remaining before $D$.
For $m>H_{\rm leaf}$ the job waits while $r>m/B_{\rm sched}$.
At the first crossing it draws a fresh uniform sample from the
\emph{current} $Q$, computes its anchors, and starts the heaps and
child residual dictionary for that level.
Sample computation is polynomial in $s_{\rm node},d$; scan/copy/heap construction and
catch-up take $mW$ work for a fixed polynomial $W$ in the displayed
parameters. We allocate a sufficiently large constant multiple of
$B_{\rm sched}W$ work per host update so that one level finishes within
$\lceil m/(16 B_{\rm sched})\rceil$ host updates. The job traverses a
persistent snapshot while logging new changes, replays those changes
in order faster than they arrive, and then switches its completed
prefix to direct updates. It waits for the next crossing on the
residual. If $m\le H_{\rm leaf}$ and $r>2H_{\rm leaf}/B_{\rm sched}$,
it maintains only the residual dictionary. Once both
$m\le H_{\rm leaf}$ and $r\le2H_{\rm leaf}/B_{\rm sched}$ hold, it
builds an exact leaf in polynomial work and keeps it current through
the deadline. A residual that grows past $H_{\rm leaf}$ while
waiting resumes the large-level rule.

\paragraph{Uniform draws with bounded work.}
For a strict worst-case running-time bound, the uniform ranks can be
generated with a fixed number of random bits instead of an unbounded
rejection loop.  For a current dictionary of size $m$, take a uniform
$b_0$-bit integer $U$ and use rank
$1+\lfloor mU/2^{b_0}\rfloor$.  Its total-variation distance from the
uniform rank distribution is at most $m/2^{b_0}$. We take
$b_0=\lceil\log_2(100(M+1)(D+1)/\delta)\rceil$, where $D$ bounds
the number of sample draws in all copies and is polynomial in the
update horizon and the work parameters. Sampling
without replacement can be implemented by drawing from the remaining
ranks, with temporary order-statistic exclusions.  A union coupling
then makes all samples equal to ideal uniform samples except with the
separately allocated small failure probability.  The work per draw and
its random-bit consumption are deterministically polynomial.

\subsection{Analysis}

\begin{lemma}[Explicit-point certificate hierarchy]\label{lem:explicit-certificate}
Let an oblivious sequence of at most $T_C$ insertions and deletions
maintain a set of at most $M$ explicit points in $\mathbb Q^d$, with
coordinate bit length at most $b$. For rational $0<\eta\le1/4$, a
randomized data structure maintains a score $L$ with a current witness
whenever $L>0$. With probability at least $1-\delta$, simultaneously
at every update,
\[
 L\le\operatorname{diam}(P)\le\sqrt{(1+\eta)^2+1}\,L.
\]
On this event at each time at least one copy covers $P$ by valid node
balls and the singleton points of its exact leaf. Each copy exposes only $O(\log(M+2))$ node balls and
polynomially many exact-leaf points. A valid exposed ball has
$O(\eta^{-2})$ current anchors with nonnegative rational weights
summing to one. Their mean $q$ and variance $V$ give the radius
$(1+\eta)\sqrt V$. The encodings of the weights, $q$, and $V$ have
length polynomial in the parameters of the update bound. A leaf point
exposes itself with unit weight and zero variance.
Update time is worst-case polynomial in $d,b,\eta^{-1}$, the encoding
length of $\eta$, $\log(M+2)$, and $\log((T_C+2)/\delta)$. Space is
$M$ times a polynomial in these parameters, and the answer is cached.
The guarantee for $\eta>1/4$ follows by running the structure with $1/4$.
\end{lemma}
\begin{proof}
We establish the quality of a valid chain first. We then show that
its fixed deadlines leave enough time to build and that at any fixed
output time an independently sampled chain is valid with constant
probability.

\paragraph{Progress in the variance construction.}
If the initial farthest point is the starting point, the sample is a
singleton. Otherwise, let $D_0=\|a_0-b\|$ for the initial pair in
Algorithm~\ref{alg:certificate}. The initial variance is $V_0=D_0^2/4$.
With $\theta=(1+\eta)^2-1$ and $\gamma=\theta/32$, a failed stopping
test moves the center from $c$ to $(1-\gamma)c+\gamma a$, where
$R=\|a-c\|$ is the current farthest distance.
The variance change is exactly
\[
 V'-V=\gamma(R^2-V)-\gamma^2R^2.
\]
The diameter of $S$ is at most $2D_0$.  Since $c\in\operatorname{conv}(S)$,
$R^2\leq4D_0^2=16V_0\leq16V$.  While the stopping test fails,
$R^2-V>\theta V$, so
\[
 V'-V\geq\gamma\theta V-16\gamma^2V
       =\theta^2V/64.
\]
On the other hand, every convex distribution on $S$ has
\[
 V=\frac12\sum_{a,b\in S}\lambda_a\lambda_b\|a-b\|^2
 \leq\frac12\operatorname{diam}(S)^2
 \leq2D_0^2=8V_0.
\]
Consequently the loop stops after $O(\theta^{-2})=O(\eta^{-2})$
iterations.  Each iteration adds at most one support point and scans only
$S$, so the running time is polynomial in $s_{\rm node},d,\eta^{-1}$ and input bit
length.  Because the mixing coefficient $\gamma$ is a fixed rational,
the common denominator of all weights grows by only a fixed factor per
iteration.  Its bit length, and those of $c,V$, are therefore polynomial
in $b,d,\eta^{-1}$ and the encoding length of $\eta$.  All comparisons
use squared rational distances.  In particular the serving ball has
rational center and rational squared radius.

\paragraph{A witnessed score and a cover.}
On a valid chain, every point is in the first sample ball that covers
it, or remains in the exact leaf. At a covering node the weighted
squared distance from its anchors to any point $y$ averages to
$V_i+\|c_i-y\|^2$. Thus the same anchors certify both the node score and
the cross score used in Section~\ref{sec:cached-main}.
If every foreground node score is valid, the copy has the desired
approximation.  For a diameter pair $x,y$, take the first node where one
endpoint, say $x$, lies in its serving ball.  Both endpoints still belong
to its node input.  If $c=c_i$, $r=\sqrt{V_i}$, and $t=\|c-y\|$, the convex
certificate gives
$H_i^2\geq\sum_a\lambda_a\|a-y\|^2=V_i+t^2$.  Thus
\[
 \|x-y\|\leq(1+\eta)r+t
 \leq\sqrt{(1+\eta)^2+1}\,H_i.
\]
The exact leaf handles the case where neither endpoint was served earlier.

\paragraph{Completion and geometric shrinkage.}
A large-level job scans and catches up in $mW$ work for a fixed
polynomial $W$, including heap and dictionary operations. With the
chosen budget its elapsed time in host updates is at most
\[
 \left\lceil\frac{m}{16B_{\rm sched}}\right\rceil
 \le\frac{m}{16B_{\rm sched}}+1
 \le\frac{m}{8B_{\rm sched}}\qquad(m\ge64B_{\rm sched})
\]
host updates.  Replay is processed faster than new changes arrive, so the
finished heaps and residual are current.
There is adequate time for each phase.  Since $m$ changes by at most one
per host update, at first crossing
\[
 r\geq m/B_{\rm sched}-2.
\]
For $m\geq64B_{\rm sched}$, after the phase at least $0.8m/B_{\rm sched}$ updates remain.  On
the coverage event the new residual has at most
$m/2+m/(8B_{\rm sched})<0.51m$ points (for our large $B_{\rm sched}$).  Thus there is time to
wait again until that residual's own threshold $r\leq|Q|/B_{\rm sched}$ is reached.
If the residual is small enough to install an exact leaf, its size
through the deadline is at most
$H_{\rm leaf}+2H_{\rm leaf}/B_{\rm sched}<2H_{\rm leaf}$.
If a waiting small residual grows past $H_{\rm leaf}$, the preceding
step had $r>2H_{\rm leaf}/B_{\rm sched}$. Since $m$ changes by at most
one, there is still more than $m/B_{\rm sched}$ time at that crossing.
The same first-crossing estimate therefore applies to the next large
level. These alternatives handle the final update before installation.
Every new nonterminal node sampled from $m$ points is therefore sampled
at most $m/B_{\rm sched}+2$ host updates before the \emph{fixed} install deadline.
The input of that node changes by at most this many points before
installation.  After installation, its next job commits within
$m_{\rm install}/B_{\rm sched}$ updates of its own input, where
$m_{\rm install}\leq m+m/B_{\rm sched}+2$.  Consequently, from its sample time until
its replacement, at most
\[
 m/B_{\rm sched}+2+m_{\rm install}/B_{\rm sched}\leq3m/B_{\rm sched}
\]
relevant deletions can occur, enlarging the small-set threshold if
necessary.

The input and child sizes satisfy, throughout the node's active lifetime,
\[
 |P_i|\geq m-3m/B_{\rm sched},
 \qquad |P_{i+1}|\leq m/2+3m/B_{\rm sched}.
\]
For $B_{\rm sched}\geq128$ this gives $|P_{i+1}|<0.6|P_i|$.  Thus all foreground
chains and all pending prefixes have depth $O(\log(M+2))$.  A pending
prefix's construction sizes decrease geometrically, apart from its
controlled small changes, so its total explicit storage is linear in its
host input size up to polynomial bookkeeping factors.

\paragraph{Survival under fixed deadlines.}
Fix an output time $t$ in the oblivious original sequence.  For a node
that is active at $t$, condition on the samples in all its ancestors and
on the scheduling history preceding that node's fresh sample.  Its input
snapshot is then fixed.  Its sample is uniform on that snapshot.  The
sample time and the original job's fixed installation deadline were
determined before this sample; the later rebuild deadline depends on its
input size and clock, which depend on ancestors but not on this node's
own sample.  Child construction times, child sizes, and the node's own
sample survival cannot change these deadlines.  An ancestor replacement
can only shorten its lifetime.  The node's own sample affects its serving
ball and its descendants, but does not affect its own input sequence.

The union of the pre-installation host-update window and the
post-installation own-input window therefore identifies, independently
of this sample, at most $3m/B_{\rm sched}$ snapshot points that could be deleted while
the node is active.  A uniform sample of $s_{\rm node}$ points hits this set with
probability at most $3s_{\rm node}/B_{\rm sched}$.  Conditioning instead on the node having
survived would be unnecessary and should be avoided.  Fixed deadlines
are essential: committing a whole pending chain whenever it happened to
finish could select sample-dependent lifetimes and would need a different
argument.

To account for sample-dependent coverage, we include the logical slot
at its fixed installation deadline even if the coverage test fails.
The event of passing coverage and losing a sample point is contained
in the unconditional sample-hit event. We charge failed coverage
separately.  Conditioning
on ancestor samples is legitimate, since these are independent of this
node's fresh sample.  This separates the global coverage-failure event
from the survival argument without biasing a sample.

At a fixed output time, expose the active chain from the root down.
At each depth, the ancestor histories and the input clock identify the
logical slot in service. Its sample-hit bound is $3s_{\rm node}/B_{\rm sched}$.
Summing over the $O(\log(M+2))$ depths gives probability at most $1/8$
of a hit in a slot with passing coverage, for large enough $C_B$.
We take the total probability of a failed coverage test or finite-rank
coupling in a copy to be at most $1/8$. Thus a given copy has an
entire valid cover at this time with probability at least $3/4$.
The $R_{\rm hier}$ copies use independent seeds on the fixed input.
The probability that every copy fails at this time is at most
$4^{-R_{\rm hier}}$. A union bound over $T_C+1$ outputs proves the
simultaneous guarantee for the maximum available score. Every available
score is a current pair distance on every outcome.

\paragraph{Worst-case work and storage.}
One foreground update follows only the single foreground chain of a copy.
At each visited node it advances that node's sole pending job by a fixed
polynomial budget and updates that job's completed prefix, whose depth is
$O(\log M)$.  Thus only $O(\log^2 M)$ prefix/node incidences are processed
per point update; no pending job recursively invokes another pending
job.  Sample selection uses an order-statistic dictionary.  Snapshot
capture, heap scans, replay queues, and retirement use persistence or
copy-on-write with background work.  Replacing a suffix is a pointer
switch; its old arenas can be reclaimed lazily under the same polynomial
budget (or recycled with generation stamps).  Initial and final small
sets are processed directly.

Here is an explicit global retirement charge.  Perform only a bounded
polynomial number $a$ of word allocations/initializations in a foreground
update; large snapshots and arrays are allocated incrementally, or use
lazy generation stamps. We keep all retired records in a reclamation queue
and process at least $2a$ units of reclamation work per update, adjusting
constants for pointer/reference cleanup.  If non-retired storage is
bounded by $S=M\operatorname{poly}(d,b,s_{\rm node},\log M)$, the volume that can
enter the retirement queue over any interval of $u$ updates is at most
$S+au$: it comprises records alive at the start plus records allocated
during the interval.  A server with rate $2a$ therefore has backlog at
most $S$ at every time (take the supremum over intervals ending at that
time).  Thus retired storage adds only another $S$, while reclamation
cost is worst-case polynomial.  Shared immutable snapshots can be
reference-counted with deferred decrement work; each allocated pointer
edge contributes only constant total cleanup work.  Cancellation of many
child jobs at an ancestor commit does not violate this burst-plus-rate
bound.

The geometrically decreasing foreground input sizes have sum $O(M)$.
Each foreground node owns one pending chain with storage linear in its
input size up to polynomial factors.  Hence a copy uses
$M\operatorname{poly}(d,b,s_{\rm node},\log M)$ space, including current and retiring
arenas.  Replication adds only $O(\log(T_C/\delta))$.  All foreground work
is explicitly capped by a polynomial budget, giving a worst-case bound
regardless of the random choices.  Heap distances and ball tests can be
compared using rational squared distances.  The final output can retain
an exact squared witness distance, so arbitrary-real precision is not
being hidden in a machine word.
\end{proof}

\section{Farthest-point lookup in a frozen sample}\label{sec:one-sided-static-oracle}

The main algorithm can afford polynomial preprocessing in the sample
size, but it must answer each center query in time polynomial in the
dimension and logarithmic parameters. We precompute a farthest sample
point for each of a bounded number of projected grid sites. Accuracy
at a given query requires an upper distortion bound for the whole
sample and a lower bound for just one farthest point.

\subsection{Algorithm}

For $m=1$ we return the unique sample point. Otherwise we compute the
diameter $D$ and its witnesses by examining all sample pairs, and fix
$s_0\in S$. For $R=\|q-s_0\|\ge D/(\kappa-1)$ the answer $s_0$ suffices,
since every sample point is at distance at most $R+D\le\kappa R$ from
$q$. Other queries use the following table.

We choose fixed rationals $b_0>1$, $e_0,\nu>0$ with
$\kappa(1-\nu-2e_0)>b_0+\nu$. Such constants exist for every fixed
$\kappa>1$, using a smaller rational factor internally if needed.
For $I(b)=\tfrac12(b^2-1-\ln b^2)$ we use a rational
$A_\kappa>I(b_0)^{-1}$ and take $k=\lceil A_\kappa\ln(240m)\rceil$,
using an upper logarithm approximation within one. We choose a
rational grid step $e_0D/(2\sqrt k)\le h\le e_0D/\sqrt k$ and a
constant $R_0>(b_0+\nu)/(\kappa-1)+e_0$. A table uses a normalized
Gaussian projection $A$ implemented by a rational matrix $\widetilde A$,
with entry error at most $\nu/\sqrt{kd}$ except with probability
$1/480$. Appendix~\ref{sec:finite-bit-gaussian-implementation} gives
the finite-bit sampling procedure.

We enumerate lattice sites $\widetilde A s_0+h j$ within a rational
radius in $[R_0D,(R_0+1)D]$ about $\widetilde A s_0$. At each site we
store the original identity farthest from it in projected distance.
For a query we round $\widetilde A q$ to its nearest site, with a
fixed tie rule, and return the stored point or $s_0$ if the site is
absent. We build $O(\log((Q+1)/\delta))$ independent tables and return
the farthest of their replies in original distance.

\subsection{Analysis}

\begin{lemma}[Farthest-point lookup]\label{lem:static-oracle}
For every fixed $\kappa>1$, a nonempty set of $m$ bounded integer points
admits a randomized structure with preprocessing time and space
$m^{B_\kappa}\poly_\kappa(\mathcal L,b_q,\log(Q+2))$ for a fixed
$B_\kappa\ge2$. Query time is
$\poly_\kappa(\mathcal L,b_q,\log(m+2),\log(Q+2))$.
For any prescribed list of $Q$ rational queries in $[-\Delta,\Delta]^d$ of coordinate
encoding length at most $b_q$, independent of its seeds, it
returns, for every query $q$, an identity $z$ with $\|z-q\|\ge\max_{x\in S}\|x-q\|/\kappa$
with probability $1-\delta$, using a logarithmic repetition factor.
Every reply is an original sample identity on every random outcome.
For $m=1$ the unique sample point answers every query exactly.
The same statement holds for canonical rational input coordinates of
bounded bit length, with their bit length included in the polynomial
factors. In the diameter application, $b_q$ and $\log(Q+2)$ are
polynomial in $\mathcal L$.
\end{lemma}
\begin{proof}
We bound distortion for a fixed query and then account for the number of grid sites.

\paragraph{Projection of a true farthest point.}
Let $Y\sim\chi_k^2$ and $X=Y-k$, for any integer $k\ge1$.  Its moments are
\[
 \mathbb EX=0,\qquad \mathbb EX^2=2k,\qquad
 \mathbb EX^4=12k^2+48k\le60k^2.
\]
Interpolation between the first, second, and fourth absolute moments,
followed by Cauchy--Schwarz, gives
\[
 \mathbb E|X|\ge
       \frac{(\mathbb EX^2)^{3/2}}{(\mathbb EX^4)^{1/2}},
 \qquad
 \mathbb EX_+=\tfrac12\mathbb E|X|,
 \qquad
 (\mathbb EX_+)^2
       \le \Pr[X\ge0]\,\mathbb EX^2.
\]
Consequently
\begin{equation}\label{eq:one-sided-chi-lower}
       \Pr[Y\ge k]\ge
        \frac{(\mathbb EX^2)^2}{4\mathbb EX^4}
                       \ge\frac1{60}.
\end{equation}
For a normalized Gaussian map $A:\mathbb R^d\to\mathbb R^k$ and any
fixed nonzero vector $v$, this says
$\Pr[\|Av\|\ge\|v\|]\ge1/60$.

For an upper multiplier $b_0>1$, the usual exact Chernoff bound is
\[
 \Pr[\|Av\|>b_0\|v\|]
       \le e^{-kI(b_0)},\qquad
 I(b_0)=\tfrac12(b_0^2-1-\ln b_0^2).
\]
The elementary inequality
$x-\ln(1+x)\ge x^2/[2(1+x)]$ gives the useful explicit lower bound
\[
                   I(b_0)\ge
       \frac{(b_0^2-1)^2}{4b_0^2}.
\]

\paragraph{Lookup accuracy at a fixed query.}
The entry error gives $\|\widetilde A-A\|_{\rm F}\le\nu$. It changes
the norm of any projected difference $v$ by at most $\nu\|v\|$.
For a fixed query, let $p_*\in S$ be a canonically selected true
farthest point and put $R=\|p_*-q\|$.  There are only $m$ vectors
$p-q$, including $s_0-q$. Our choice of $k$ bounds their combined
ideal upper-tail failure probability by $1/240$.  By
\eqref{eq:one-sided-chi-lower} and a union bound, the joint event
\[
 \|A(p_*-q)\|\ge R,\qquad
 \|A(p-q)\|\le b_0\|p-q\|\quad(p\in S),
\]
together with successful finite-bit coupling, has probability at least
$1/60-1/240-1/480>1/200$.
No independence between the upper and lower events is used.
Let $u$ be the rounded query site. Its distance from $\widetilde A q$
is at most $h\sqrt k/2\le e_0D/2$, so $u$ lies in the table on the good
event by the upper estimate for $s_0-q$ and the definition of $R_0$.
If the stored identity is $x$, then
$\|\widetilde A x-u\|\ge\|\widetilde A p_*-u\|$. Applying the rounding
bound at both ends gives
\[
 (b_0+\nu)\|x-q\|
 \ge (1-\nu)R-e_0D
 \ge (1-\nu-2e_0)R,
\]
where $D\le2R$.  Thus it suffices that
\begin{equation}\label{eq:one-sided-oracle-accuracy}
       \kappa(1-\nu-2e_0)\ge b_0+\nu.
\end{equation}
Since $\kappa(1-\nu-2e_0)>b_0+\nu$, this proves the lookup
guarantee for the fixed query. It holds with probability at least $1/200$ for each
independent table. Taking the farthest actual reply preserves any
success. A union bound over the $Q$ prescribed queries proves the
simultaneous guarantee. On every outcome the reply is an input point.

\paragraph{Enumeration and query costs.}
A table key satisfies $\|j\|\le C_\kappa\sqrt{k}$ for a fixed
$C_\kappa$. The enumeration grows feasible integer prefixes recursively
and discards a prefix once its squared norm exceeds the squared radius. At depth $i$, disjoint unit cubes about the feasible
prefixes lie in an $i$-dimensional ball of radius
$C_\kappa\sqrt{k}+\sqrt{i}/2$. The volume bound gives at most
$(C'_\kappa\sqrt{k/i})^i$ prefixes. Since
$i\ln(k/i)\le k/e$, the total enumeration takes
$\exp(O_\kappa(k))\poly(k)=m^{O_\kappa(1)}$ operations.
For each full key the algorithm scans the sample and stores the maximizing
original identity. Computing the sample diameter, generating the
map, and building the lookup dictionary have the same polynomial
bound after increasing $B_\kappa$.

A query projects its center, rounds its grid key, and looks up
the stored identity in a balanced dictionary. It then evaluates
the original distance. All of this takes polynomial time in
$\mathcal L$ and the query encoding bound. Independent table
repetition multiplies preprocessing and query time by only a
logarithmic factor. The explicit choice $B_{1.09}=2000$ is
verified in Appendix~\ref{sec:combined-quantitative-two}.
\end{proof}

\section{A lower bound on linear-length set streams}\label{sec:sphere-lower-main}

A single final diameter query can test one bit of a long insertion
prefix. We first specify the bit bound that suffices for
Theorem~\ref{thm:main-lower}. We then construct the finite instances
and derive the inequality on information that forces this bound.

Write $S$ for the bit length of an encoding of the state after the
insertion prefix, padding shorter descriptions to the same length.
Thus $W$ machine words give $S=O(W w_{\rm word})$. For fixed $c\ge1$
put $g=\lceil c\rceil+1$, $r=1/g$, $\delta_0=1/(100g)$,
$r_0=r-2\delta_0$, and $\eta=1-h_2(1/10)$, where $h_2$ is the
binary entropy. All entropies are in bits.

\begin{theorem}[Information in the insertion prefix]\label{thm:lb-dimension}
For every integer $d\ge4$ there is a finite rational set
$\mathcal P\subset\mathbb R^d$ of size $U$ with
\begin{equation}\label{eq:sphere-net-cardinality}
 (25g)^{d-1}\leq U\leq(802g)^d,
\end{equation}
such that the following holds. Suppose a randomized one-pass algorithm
returns $D/c\le L\le D$ with error at most $1/10$ on every stream in
this family. Then its state after the insertion prefix satisfies
\begin{equation}\label{eq:spherical-space-main}
 S+\lceil\log_2(U+1)\rceil
 >\frac\eta4
 \exp\left(\frac{(d-1)r_0^2}{1000}\right).
\end{equation}
The streams have dimension $d+1$, maximum support $U+1$ or less, and
length at most $2U+1$. Each point is inserted at most once and deleted at
most once, with all insertions first. Scaling every coordinate by the
same factor yields integers of magnitude $O(g^2\sqrt d)$. The
logarithmic term in \eqref{eq:spherical-space-main} allows a free
update clock and may be dropped when the clock is counted in the state.
\end{theorem}
\begin{proof}
Section~\ref{sec:finite-net-lb} constructs the net, and
Section~\ref{sec:cap-decoder} turns a streaming state into a message of
$s=S+\lceil\log_2(U+1)\rceil$ bits. The decoder forces a marginal of
at least $\eta$ at chord length two and zero throughout $[0,r_0]$.
The diagonal inequality and rotational averaging in
Section~\ref{sec:chord-lb} constrain the values between these endpoints
by \eqref{eq:sphere-recurrence}. Section~\ref{sec:iterate-lb} shows that
this recurrence is inconsistent with the two endpoint conditions unless
$s>(\eta/4)\exp((d-1)r_0^2/1000)$, as claimed.
\end{proof}

\subsection{The finite input and the decoder}

\subsubsection{A rational antipodal net}\label{sec:finite-net-lb}
We take a real $\delta_0/2$-net $\mathcal V$ of the unit sphere
$\mathbb S^{d-1}$ with
\[
 |\mathcal V|\leq(1+4/\delta_0)^d.
\]
For example a maximal $\delta_0/2$-separated set has this size bound:
the disjoint radius-$\delta_0/4$ balls about its points fit in the
radius-$(1+\delta_0/4)$ ball.  Maximality gives the covering property.
We adjoin the negatives of all selected points. From each antipodal pair
we round one representative to a nearest grid point $p$ and use $-p$ for
its antipode. All sites use the same grid with denominator
\[
 J=\left\lceil\frac{2\sqrt d}{\delta_0}\right\rceil.
\]
The rounding error is at most $\sqrt d/(2J)\leq\delta_0/4$.
After duplicates are removed, the symmetric rational set $\mathcal P$
has cardinality at most $2(1+4/\delta_0)^d\leq(802g)^d$, and every
unit vector is within $\delta_0$ of some point of $\mathcal P$.
Every point of $\mathcal P$ has norm between $1-\delta_0$ and
$1+\delta_0$.

We fix a measurable nearest-net map
\[
 Q:\mathbb S^{d-1}\longrightarrow\mathcal P
\]
with $\|Q(x)-x\|\leq\delta_0$ and $Q(-x)=-Q(x)$.  Such an odd
tie rule exists: first choose the first nearest antipodal pair in a
fixed ordering, then choose the nearer sign; on a sign tie use any
fixed odd hemisphere convention.  Ties are immaterial under the
rotation distributions used below, but this makes the map exact.

For completeness, any $\delta_0$-net of the sphere must also be
exponentially large in $d$.  The balls of radius $2\delta_0$ around
its sites cover the shell with radii in $[1,1+\delta_0]$.  Comparing
volumes gives
\[
 U(2\delta_0)^d\geq(1+\delta_0)^d-1\geq\delta_0.
\]
For $d\geq2$, this implies
$U\geq(1/(4\delta_0))^{d-1}=(25g)^{d-1}$, proving
\eqref{eq:sphere-net-cardinality}.

All coordinates of $\mathcal P$ have denominator $J$.  The additional
anchor height used below is $r=1/g$.  Scaling all coordinates by $Jg$
makes every point integral, with absolute coordinate bound
$O(gJ)=O(g^2\sqrt d)$.  This scaling preserves all diameter ratios.
Rounding only the chosen net sites keeps the cardinality bound
independent of the size of the ambient rational grid.

\subsubsection{A cap, a target, and a fresh anchor}\label{sec:cap-decoder}
Alice has independent fair bits $X_p$, one for every $p\in\mathcal P$.
She inserts $(p,0)$ when $X_p=1$, in canonical order, and sends the
streaming state to Bob, together with her insertion count if needed to
specify a free clock.  Independently choose $x$ uniformly on
$\mathbb S^{d-1}$.  Bob's target is
\[
 y=Q(-x)=-Q(x).
\]
He knows the bits outside
\[
 \mathcal B_x\cup\{y\},\qquad
 \mathcal B_x=\mathcal P\cap B(Q(x),r).
\]
The target lies outside $\mathcal B_x$, since $2\|Q(x)\|\ge
2(1-\delta_0)>r$. Bob first inserts the fresh anchor $(Q(x),r)$, whose
last coordinate distinguishes it from every Alice point. He then deletes
every present point in that known complement.  Thus the complete stream
has all its insertions before all its deletions, and no point is ever
reinserted.

There are at most $2U+1$ updates, every prefix is a set, and at most
$U+1$ points are active.  Once Alice's bits and Bob's query are chosen,
the entire update sequence is fixed independently of the algorithm's
random coins; thus this lower bound already holds for oblivious
streams. If $X_y=0$, only the anchor and the inserted cap points
survive. With no cap point the anchor is a singleton and the diameter
is zero. Otherwise any two cap points are within $2r$ of one another,
and any cap point is within $\sqrt2\,r$ of the anchor. Thus $D\le2r$.
If $X_y=1$, both $(y,0)=(-Q(x),0)$ and the anchor survive, so
\[
 D\ge\sqrt{4\|Q(x)\|^2+r^2}\ge 2\|Q(x)\|\ge 2(1-\delta_0).
\]
The ratio between the lower bound in the second case and the upper
bound in the first is
\[
 \frac{2(1-\delta_0)}{2r}=\frac{1-\delta_0}{r}=g-1/100>c .
\]
On the approximation event $D/c\le L\le D$, Bob thus decodes $X_y=1$
exactly when $L>2r$ (in the unscaled coordinates).
Although $x$ was used to describe the query distribution, Bob only
needs the finite data $Q(x)$ and $y$; no real-valued side channel is
provided to the decoder.

Fix public randomness and Alice's private coins by averaging so that
the average error remains at most $1/10$.  Bob may use fresh independent
randomness, or replace it by an optimal deterministic decoder for this
distribution.  The resulting deterministic message $M=M(X)$ has at
most
\[
 s=S+\lceil\log_2(U+1)\rceil
\]
bits.  Define
\[
 f(A)=H(M\mid X_{\mathcal P\setminus A}),\qquad A\subseteq\mathcal P.
\]
The function $f$ is normalized, monotone, and submodular, with
$f(\mathcal P)\le s$ and singleton marginals in $[0,1]$.
Normalization uses that $M$ is a deterministic function of the bits;
monotonicity follows by removing conditioning. For submodularity, write
\[
 f(A)=H(M,X_{\mathcal P\setminus A})-|\mathcal P\setminus A|.
\]
Entropy is submodular, complementation interchanges union and
intersection, and the subtracted cardinality is modular. A singleton
marginal is the conditional mutual information with one independent
fair bit, and so lies between zero and one.

To make the randomness bookkeeping explicit, Alice's private coins may
be represented by an independent random transition-coin table indexed
by update time, current state, and update symbol.  Along any one prefix,
each time index is used only once, so this has exactly the law of fresh
private coins.  Fixing the public seed and this prefix table makes
$M$ a deterministic function of the still-independent input bits.
Bob uses fresh transition coins for his continuation and the transmitted
prefix length to determine a free clock, if one is present.  Averaging
over the prefix table and public seed guarantees a fixed choice with
the stipulated average error.  No private random-tape head or other
unaccounted history is needed.

For continuous sphere points $x,z$, put
\[
 a(x,z)=f(\mathcal B_x\cup\{Q(z)\})-f(\mathcal B_x).
\]
If $\|x-z\|\leq r_0=r-2\delta_0$, then
$\|Q(x)-Q(z)\|\leq r$, so $a(x,z)=0$.
Let $a(t)$ be the average of $a(x,z)$ over rotation-invariant ordered
pairs on the sphere at chord distance $t\in[0,2]$.
Thus $a(t)\in[0,1]$ and $a(t)=0$ for $t\leq r_0$.
For $z=-x$, the marginal equals
$I(M;X_y\mid X_{\mathcal P\setminus(\mathcal B_x\cup\{y\})})$
at fixed $x$. Before seeing $M$, the target bit is fair even given
this side information. After seeing $M$, Bob decodes it with average
error at most $1/10$. For a fixed $x$ let $e_x$ be the conditional
error of the deterministic decoder. Given the message and the side
information, the target bit is determined by its error indicator, so
its residual entropy is at most $h_2(e_x)$. Averaging and using
$\mathbb E e_x\le1/10$ and concavity of $h_2$ gives the binary Fano
bound~\cite{ct}
\begin{equation}\label{eq:sphere-fano}
 a(2)\geq\eta=1-h_2(1/10).
\end{equation}

\subsection{From antipodes to shorter chords}\label{sec:chord-lb}

\subsubsection{A diagonal inequality for submodular marginals}
An entry on the diagonal represents information the decoder needs
about its own target. An off-diagonal entry measures how much that
gain can decrease if an earlier target has already been revealed.
Selecting a random subcollection lets us charge the latter losses
only when both indices are selected.
\begin{lemma}[Weighted diagonal inequality]\label{lem:weighted-diagonal}
For any normalized monotone submodular $f$ with total value at most
$s$, any sets $B_1,\ldots,B_m$, and any elements $z_1,\ldots,z_m$
\emph{with repetitions allowed}, set
\[
 A_{ij}=f(B_i\cup\{z_j\})-f(B_i),\qquad
 D=\sum_i A_{ii},\quad Z=\sum_{i,j}A_{ij}.
\]
Then $D^2\leq2sZ$.
\end{lemma}
\begin{proof}
Select a subset of indices and order it. Let $U_i$ contain the
elements selected before index $i$. Diminishing returns gives the
first inequality below. For the second, contract $B_i$ by writing
$g_i(U)=f(B_i\cup U)-f(B_i)$. Monotonicity gives
$g_i(U_i\cup\{z_i\})\ge g_i(\{z_i\})=A_{ii}$, while telescoping the
gains along the preceding indices gives
$g_i(U_i)\le\sum_{j\text{ preceding }i}g_i(\{z_j\})$.
Thus
\[
\begin{aligned}
 f(U_i\cup\{z_i\})-f(U_i)
 &\ge f(B_i\cup U_i\cup\{z_i\})-f(B_i\cup U_i)\\
 &\ge A_{ii}-\sum_{j\text{ preceding }i}A_{ij}.
\end{aligned}
\]
This remains valid if $z_i$ already appeared: the chain increment is
zero, and the preceding identical column subtracts $A_{ii}$ from the
right side. We select indices independently with probability $q$ and
order them uniformly. Each diagonal gain is selected with probability
$q$. Each ordered pair of distinct indices is selected in that order
with probability $q^2/2$. The chain telescopes to at most $s$, giving
$s\ge qD-q^2(Z-D)/2\ge qD-q^2 Z/2$. If $D>0$, use $q=D/Z\le1$
to obtain $D^2\le2sZ$. The case $D=0$ is immediate.
\end{proof}

\subsubsection{Averaging on an affine subsphere}
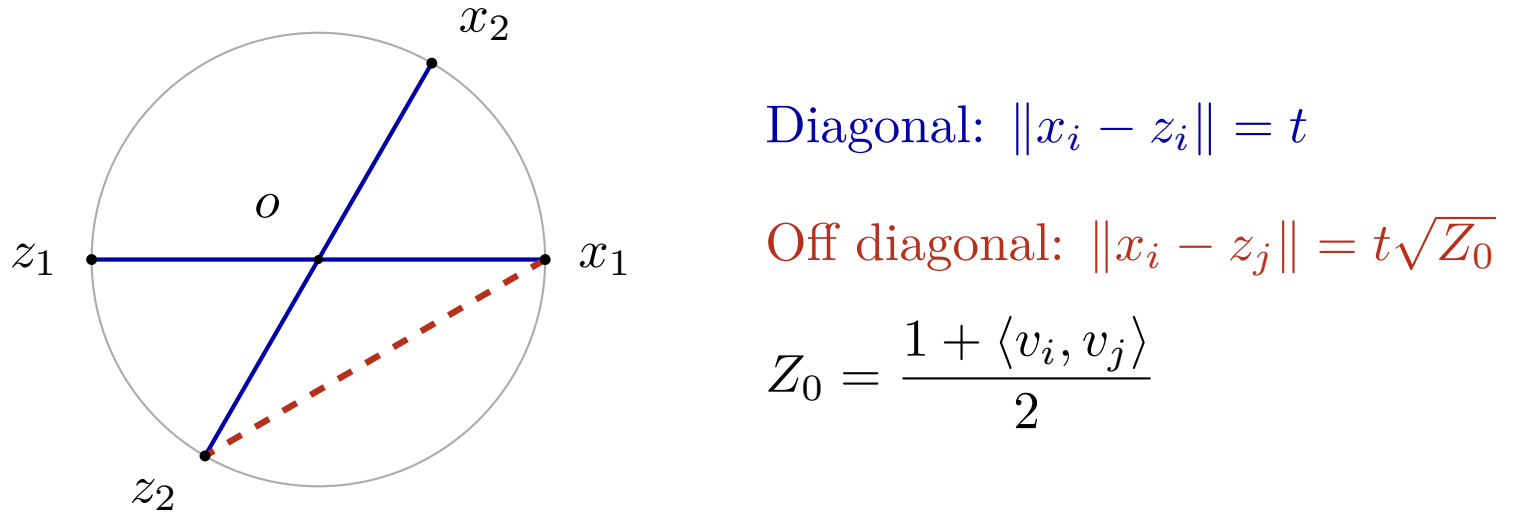
\begin{figure}[htbp]
\centering
\begin{tikzpicture}[x=1.2cm,y=1.2cm,font=\small]
 \coordinate (xone) at (1.25,0);
 \coordinate (zone) at (-1.25,0);
 \coordinate (xtwo) at (.625,1.0825);
 \coordinate (ztwo) at (-.625,-1.0825);
 \draw[gray!65] (0,0) circle (1.25);
 \draw[blue!65!black,thick] (xone)--(zone);
 \draw[blue!65!black,thick] (xtwo)--(ztwo);
 \draw[BrickRed,very thick,dashed] (xone)--(ztwo);
 \fill (0,0) circle (.025);
 \node[anchor=south east] at (-.10,.12) {$o$};
 \foreach \p in {xone,zone,xtwo,ztwo} {\fill (\p) circle (.03);}
 \node[anchor=west] at (1.32,0) {$x_1$};
 \node[anchor=east] at (-1.32,0) {$z_1$};
 \node[anchor=south west] at (.66,1.10) {$x_2$};
 \node[anchor=north east] at (-.66,-1.10) {$z_2$};
 \node[anchor=west,text=blue!65!black] at (2.35,.72)
       {Diagonal: $\|x_i-z_i\|=t$};
 \node[anchor=west,text=BrickRed] at (2.35,.08)
       {Off diagonal: $\|x_i-z_j\|=t\sqrt{Z_0}$};
 \node[anchor=west] at (2.35,-.62)
       {$\displaystyle Z_0=\frac{1+\langle v_i,v_j\rangle}{2}$};
\end{tikzpicture}
\caption{A two-dimensional cross-section of the affine subsphere,
with center $o=\sqrt{1-R^2}\,u$ and radius $R=t/2$.
Each pair $(x_i,z_i)$ is antipodal within this subsphere, so its
distance is exactly $t$. For independent directions, an off-diagonal
distance is $t\sqrt{Z_0}$, where $Z_0$ is concentrated near $1/2$.
This is the geometric step that propagates an entropy marginal
from distance $t$ toward distance $t/\sqrt2$.}
\label{fig:subsphere}
\end{figure}

We realize the diagonal at any prescribed chord length using the
diameters of one affine subsphere. Its center is common to all pairs,
so choosing two different indices changes their separation in a way
that depends only on two independent directions (Figure~\ref{fig:subsphere}).

Fix $t\in[0,2]$, write $R=t/2$, and choose a uniform unit vector $u$.
In the orthogonal hyperplane $u^\perp$, take independent uniform
vectors $v_1,\ldots,v_m$ on its unit sphere $\mathbb S^{d-2}$.
Define
\[
 x_i=\sqrt{1-R^2}\,u+Rv_i,
 \qquad z_i=\sqrt{1-R^2}\,u-Rv_i.
\]
These points lie on the original unit sphere.  Each diagonal pair
$(x_i,z_i)$ has chord distance $t$, and its law is rotationally
invariant. We apply Lemma~\ref{lem:weighted-diagonal} to the finite table
\[
 A_{ij}=a(x_i,z_j).
\]
Its underlying elements $Q(z_j)$ may coincide; the preceding lemma
explicitly permits this.

For $i\ne j$,
\[
 \|x_i-z_j\|=R\|v_i+v_j\|
   =t\sqrt{Z_0},\qquad
 Z_0=\frac{1+\langle v_i,v_j\rangle}{2}.
\]
The inner-product density on $\mathbb S^{d-2}$ shows that
\[
 Z_0\sim\operatorname{Beta}(\alpha,\alpha),
 \qquad\alpha=(d-2)/2.
\]
By rotation invariance, conditional on the chord length $v$ the
expected entry is $a(v)$. Averaging over $v=t\sqrt{Z_0}$ gives
$\mathbb E a(t\sqrt{Z_0})$ for each off-diagonal entry.
If $D_m,Z_m$ denote the diagonal and total table sums, then
\[
 \mathbb E D_m=ma(t),\qquad
 \mathbb E Z_m=ma(t)+m(m-1)\mathbb E a(t\sqrt{Z_0}).
\]
Using $(\mathbb E D_m)^2\leq\mathbb E D_m^2\leq2s\mathbb E Z_m$
and letting the auxiliary table size $m$ tend to infinity gives
\begin{equation}\label{eq:sphere-recurrence}
 a(t)^2\leq2s\,\mathbb E a(t\sqrt{Z_0}).
\end{equation}
The limit only concerns a mathematical table with repeated entries;
it introduces no long stream or enlarged input universe.

\subsection{Iterating to the zero band}\label{sec:iterate-lb}

\subsubsection{A tail bound for a random chord}
The recurrence averages over the length of the next chord. We bound
the probability that it is too long using the even moments of one
spherical coordinate.
Write $n=d-1$ and $W=2Z_0-1$.  The variable $W$ has the law of
the first coordinate of a uniform point on $\mathbb S^{n-1}$.
Symmetry gives zero odd moments.  The beta integral gives, for each
integer $j\geq1$,
\[
 \mathbb E W^{2j}
 =\frac{\Gamma(j+1/2)\Gamma(n/2)}
        {\sqrt\pi\,\Gamma(j+n/2)}
 =\frac{(2j-1)!!}{n(n+2)\cdots(n+2j-2)}
 \leq\frac{(2j-1)!!}{n^j}.
\]
Expanding the hyperbolic cosine and using monotone convergence therefore
shows that for every real $\lambda$,
\[
 \mathbb E e^{\lambda W}
 =\mathbb E\cosh(\lambda W)
 \leq\sum_{j=0}^{\infty}
       \frac{\lambda^{2j}(2j-1)!!}{(2j)!n^j}
 =\exp\left(\frac{\lambda^2}{2n}\right).
\]
Here the $j=0$ term is interpreted as $1$.
Chernoff's inequality, with $\lambda=2n\delta$, now gives the useful
dimension-dependent upper tail
\begin{equation}\label{eq:sphere-beta-tail}
 \Pr[Z_0>1/2+\delta]
 \leq\exp(-2n\delta^2),\qquad \delta\geq0.
\end{equation}

\subsubsection{Choosing the truncation at each step}
Each application of the recurrence squares the marginal, so the error
from a fixed tail cutoff could exceed the quantity it is meant to
bound. We increase the cutoff as the marginal shrinks and keep track
of the accumulated loss in distance.
Define the monotone envelope
\[
 A(t)=\sup_{0\leq v\leq t}a(v),\qquad 0\leq t\leq2.
\]
It is monotone and hence measurable, lies in $[0,1]$, vanishes on
$[0,r_0]$, and satisfies $A(2)\geq\eta$.
Taking the supremum in \eqref{eq:sphere-recurrence}, and using
monotonicity of $A$, gives
\begin{equation}\label{eq:sphere-envelope-recurrence}
 A(t)^2\leq2s\,\mathbb E A(t\sqrt{Z_0}).
\end{equation}
The number of message bits $s$ is an integer and must be positive by
\eqref{eq:sphere-fano}; thus $s\geq1$.
Put
\[
 L=\ln(4s/\eta),\qquad
 h=\left\lceil\log_2\frac{16}{r_0^2}\right\rceil,
 \qquad \ell_j=4s\exp(-2^jL),
 \qquad \delta_j=\sqrt{\frac{2^jL}{n}}.
\]
Notice that $\ell_0=\eta$ and
$\ell_{j+1}=\ell_j^2/(4s)$.
Suppose, for a contradiction, that
\begin{equation}\label{eq:sphere-small-space-assumption}
 L\leq\frac{nr_0^2}{1000}.
\end{equation}
Since $2^h<32/r_0^2$,
\begin{align*}
 \sum_{j=0}^{h-1}\delta_j
 &=\sqrt{L/n}\,
       \frac{2^{h/2}-1}{\sqrt2-1}\\
 &<(\sqrt2+1)\sqrt{2^hL/n}\\
 &<(\sqrt2+1)\sqrt{32/1000}<\tfrac12<\ln2.
\end{align*}
In particular each $\delta_j<1/2$ for $j<h$.

Starting from $t_0=2$, define
\[
 t_{j+1}=t_j\sqrt{1/2+\delta_j},
 \qquad 0\leq j<h.
\]
All these values stay in $[0,2]$.  Splitting the expectation in
\eqref{eq:sphere-envelope-recurrence} according to
$Z_0\leq1/2+\delta_j$, and using $A\leq1$, yields
\[
 A(t_{j+1})\geq\frac{A(t_j)^2}{2s}
                  -\Pr[Z_0>1/2+\delta_j].
\]
By \eqref{eq:sphere-beta-tail}, the subtracted term is at most
\[
 \exp(-2^{j+1}L)
 \leq4s\exp(-2^{j+1}L)
 =\frac{\ell_j^2}{4s}.
\]
If $A(t_j)\ge\ell_j$, the main term exceeds the tail by at least
\[
 A(t_{j+1})\ge\frac{\ell_j^2}{2s}-\frac{\ell_j^2}{4s}
              =\frac{\ell_j^2}{4s}=\ell_{j+1}>0 .
\]
Induction from $A(2)\geq\eta=\ell_0$ gives $A(t_h)>0$.
On the other hand, $\ln(1+2\delta)\leq2\delta$ gives
\[
 t_h
 =2^{1-h/2}\prod_{j=0}^{h-1}\sqrt{1+2\delta_j}
 \leq\frac{r_0}{2}
       \exp\left(\sum_{j=0}^{h-1}\delta_j\right)
 <r_0.
\]
This contradicts the zero band of $A$.
Therefore \eqref{eq:sphere-small-space-assumption} is false, and
\[
 s>\frac\eta4\exp\left(\frac{(d-1)r_0^2}{1000}\right),
\]
which proves \eqref{eq:spherical-space-main}.

This choice of thresholds is what preserves the dependence on $c$.
The squared deviations $\delta_j^2$ grow like $2^j$, exactly the
growth of $-\ln(\ell_j/(4s))$. Meanwhile $\sum_j\delta_j$ stays
bounded, which is enough to reach the zero band in $h$ steps.

\subsection{Conversion to the streaming parameters}
\begin{proof}[Proof of Theorem~\ref{thm:main-lower}]
For $d\geq4$, one has $d-1\geq d/2$.  The net upper bound gives
$d\geq\ln U/\ln(802g)$, hence
\[
 S+\lceil\log_2(U+1)\rceil
 >\frac\eta4 U^{\beta_g},
 \qquad
 \beta_g=\frac{2401}{5000000g^2\ln(802g)}.
\]
Since $g\leq c+2\leq3c$,
\[
 \beta_g=\Omega\!\left(\frac1{c^2\log(c+1)}\right).
\]
For fixed $c$, the logarithmic clock term is negligible compared with
$U^{\beta_g}$.  Setting $N=U+1$ gives the same power lower bound in
maximum active cardinality, with stream length at most $2N-1$ and
dimension $\Theta_c(\log N)$.

For every sufficiently large prescribed capacity $N$, rather than
only the cardinalities of this hard family, choose
$d=\lfloor\ln(N-1)/\ln(802g)\rfloor$. The same construction then has
$U\leq N-1$, length at most $2N-1$, and
$d=\Theta_c(\log N)$. Its lower bound exponential in $d$ therefore
gives $N^{\Omega(1/(c^2\log(c+1)))}$ for every such capacity.
The displayed numerical exponent $\beta_g$ above is the direct
hard-family conversion; this padding statement need not preserve
that exact numerical constant.

On these instances the dimension is $O_c(\log N)$, the coordinates
have $O_c(\log\log N)$ bits, and the stream has $O(N)$ updates.
Thus a standard word has $O_c(\log N)$ bits. Dividing the
information lower bound by this word length, and absorbing
polylogarithmic factors into the exponent for fixed $c$, proves
Theorem~\ref{thm:main-lower}.
\end{proof}
An estimate in $[D,cD]$ gives the convention used here upon division
by $c$. For the symmetric convention $[D/c,cD]$, dividing by $c$
gives a one-sided factor $c^2$. The conclusion that every fixed factor
needs polynomial space also holds under that convention.

\section*{AI Use Statement}
This work resulted from extensive interaction between the authors and OpenAI
Astra, with the authors providing suggestions and critically checking Astra's responses through
iterative exchanges. OpenAI Astra was also used in writing and revising the manuscript through
detailed prompts from the authors, with Gemini Argon assisting in some parts. The authors directed the process and evaluated the proposed ideas and arguments.
\ifx\LeanRepository\empty
Lean formalization and a GitHub link are to be added.
\else
Lean files formalizing the main theorem are available on
\href{\LeanRepository}{GitHub}.
\fi

\bibliographystyle{alpha}
\bibliography{ref}
\appendix
\section{Appendix}
\subsection{Choosing the constants for factor two}
\label{sec:combined-quantitative-two}

We now specify all approximation and exponent parameters.
The proof of Theorem~\ref{thm:main-upper} requires only strict inequalities.
Here we verify one rational choice. The resulting bounds are
\[
 \boxed{\begin{aligned}
 \text{Space:}\quad &N^{\,1-10^{-10}}\operatorname{poly}(\mathcal L),\\
 \text{Worst-case update:}\quad &\operatorname{poly}(\mathcal L).
\end{aligned}}
\]
The algorithm maintains a cached score $L$. Every nonzero score has
a witness pair of current input points; an unsuccessful copy may report zero.  With probability at least $1-\delta$, simultaneously after every
update,
\[
                L\le\operatorname{diam}(P)<2L
\]
whenever the diameter is positive; zero diameter is handled exactly.
The resource bounds and the lower inequality hold on every random
outcome.  All constants below are fixed, independently of
$N,d,\Delta,T$.  Table~\ref{tab:constants} lists the inequalities that determine the approximation and the space saving.

\begin{table}[htbp]
\centering\small
\renewcommand{\arraystretch}{1.15}
\begin{tabular}{@{}ll@{}}
\toprule
Required inequality & Verified choice\\
\midrule
Hull factor below two & $2a^2-1=3.999122<4$\\
Mixed pair no worse than hull & $(1+\eta_{\rm in})^2+a^2\kappa^2=3.9898284241<3.999122$\\
Candidate factor no worse than hull & $(1+\eta_{\rm in})^2+1=2.0201<3.999122$\\
Oracle accuracy & $\kappa(1-\nu-2e_0)=1.084441>1.0801$\\
Bucket exponent & $\theta(1+G)\ln C_g<\rho$\\
Oracle work per update & $s(B_\kappa+1)-1=-2/2003<0$\\
Recovery saving & $\epsilon_0>1/6{,}345{,}504{,}000>10^{-10}$\\
\bottomrule
\end{tabular}
\caption{The factor-two parameter check. All parameters are fixed
independently of $N,d,\Delta,T$. The exponent is deliberately
conservative.}
\label{tab:constants}
\end{table}

\subsubsection{Geometric and recovery constants}

We use the following hull parameter, radius-scale ratio, and target ratios
\[
 a=\frac{1581}{1000},\qquad q_{\rm scale}=\frac{5001}{5000},
 \qquad \lambda_0=\frac{7903}{5000},\qquad
 \lambda=\frac{79}{50}.
\]
They satisfy $a/q_{\rm scale}>\lambda_0>\lambda>\sqrt2$ and
\[
                      2a^2-1=3.999122<4.
\]
The masking slack is $1/20$, the radii are $r_1=1,r_2=4/5$, and the potential is
\[
 \phi(t,q)=q\exp\left(\frac{(\lambda-t)^2}{q}-1\right),
 \qquad t^2+q\le1.
\]
The following two-cap calculation shows that the first mask
$\phi\le21/20$ lies in the radius-$4/5$ ball centered at
$(3/5)$ times the target direction, and that the second mask
$\phi\le(21/20)(4/5)^2$ is empty.  Indeed,
$F(3/5)>79/75>21/20$, while the global potential lower bound is
$\phi>16133/24000>84/125=(21/20)(4/5)^2$.

For each of the two independent Gaussian-grid stages choose
\[
 \xi=10^{-6},\qquad \delta_g=10^{-4},\qquad
 \beta=\frac{1023}{1000},\qquad C_g=42300,\qquad
 q_0=\frac{1999}{2000},\qquad \tau=-\ln q_0.
\]
The signed target-to-bulk projection gap is at least $\lambda-t$:
\[
 \sqrt{1-\xi}(\lambda_0-t)-\delta_g r_i
 \ge(1-10^{-6})(\lambda_0+1)-1-\delta_g-t
 >\lambda-t.
\]
The core-cell bound is $C_g^{k_i}$ because
$4.132<\sqrt{2\pi e}<4.133$ and
$4.133\cdot10231=42284.723<C_g$.  A point outside the stage mask
has same-cell probability at most $q_0^{k_i}/C_g^{k_i}$, since
\[
 \frac{C_g\delta_g}{\sqrt{2\pi e}\sqrt{21/20}}
 <\frac{2115/2066}{\sqrt{21/20}}
 <\frac{10575000}{10584118}<q_0.
\]
For the conditional projected-core tail,
$x=\beta^2/(1+\xi)-1$ lies in $(0.0465,0.047)$, and its Chernoff
rate satisfies
\[
 J\ge\frac{x^2}{4(1+x)}>
       \frac{0.0465^2}{4(1.047)}>
       \frac1{1950}>\frac1{1999}>\tau.
\]
Thus both terms in the first-safe-stage argument have rational decay
$q_0^{k_i}$.

\subsubsection{Verification of the two masks}
\label{sec:two-level-explicit}
As in the hierarchy proof, write a bulk point in its original unit ball
as $t u+v$, with $q=\|v\|^2$ and $t^2+q\leq1$, and set
\[
 \phi(t,q)=q\exp\left(\frac{(\lambda-t)^2}{q}-1\right).
\]
Set $\eta_{\rm mask}=1/20$. The first mask is $\phi\leq21/20$.  Its possible longitudinal
coordinates are determined by
\[
 F(t)=(1-t^2)
       \exp\left(\frac{(\lambda-t)^2}{1-t^2}-1\right).
\]
Differentiation gives
\[
 (\log F)'(t)=\frac{2g(t)}{(1-t^2)^2},\qquad
 g(t)=t\lambda^2-\lambda t^2+t^3-\lambda,
\]
\[
 g'(t)=(\lambda-t)^2+2t^2>0,
 \qquad g(3/5)=-0.43496<0.
\]
Hence $F$ decreases throughout $(-1,3/5]$.  At its right endpoint,
\[
 F(3/5)=\frac{16}{25}\exp(0.500625)
 >\frac{16}{25}
       \left(1+\frac12+\frac18+\frac1{48}\right)
 =\frac{79}{75}>\frac{21}{20}.
\]
Therefore every first-mask point has $t>3/5$.  The entire spherical
cap $t\geq3/5$, $t^2+q\leq1$, lies in
$B((3/5)u,4/5)$.  This is the analysis-only ball for the second level.

The second mask is empty.  Indeed,
\[
 (\lambda-t)^2-q\geq
 m:=\lambda^2/2-1=\frac{1241}{5000},
\]
and thus
\[
 \phi(t,q)\geq q e^{m/q}\geq em.
\]
The last inequality follows by minimizing $q e^{m/q}$ over $q>0$,
whose minimum is at $q=m$.  Since
$e>1+1+1/2+1/6+1/24=65/24$,
\[
 em>\frac{16133}{24000}>\frac{84}{125}
       =\left(1+\eta_{\rm mask}\right)r_2^2.
\]
This proves the two-stage terminal condition with explicit slack.

\subsubsection{Coupling the earlier density with the actual later core radius}

For a normalized original-ball point $z=tu+v$, let $u$ be the target
direction, $q=\|v\|^2$, $t^2+q\le1$, and write $\phi=\phi(t,q)$.
Set
\[
 D_0=C_g\delta_g/\sqrt{2\pi e}<1.024,\qquad
 \sigma(\phi)=\max\{0,\ln(D_0/\sqrt\phi)\}.
\]
The conditional same-cell probability at stage $i$ is at most
$C_g^{-k_i}(D_0r_i/\sqrt\phi)^{k_i}$.
The global potential lower bound $\phi>84/125$, together with
$D_0<128/125$ and
$(128/125)^2<(5/4)^2(84/125)$, gives
$\sigma(\phi)<\ln(5/4)<9/40$.
We set $A_0=200$ and let $B_i$ be the least power of two at least
$A_0C_g^{k_i}$. Conditional on $A_1u$ in its prescribed norm range,
the first-stage pass probability is at most
\[
             \frac{C}{B_1}e^{\sigma(\phi)k_1}.
\]
Indeed, distinct cells pass with probability $1/B_1$, while a shared
cell contributes at most $C_g^{-k_1}e^{\sigma(\phi)k_1}$. This gives
$C=401$ uniformly. The exponent will be paired with the tail probability
at the point's actual distance from the second core center.

\paragraph{Two sharper lower bounds for the potential.}

For every real $w$, the inequality $e^x\geq1+x$ implies
\[
 q e^{(\lambda-t)^2/q-1}
 \geq e^w\big((\lambda-t)^2-wq\big).
\]
Taking $w=7/8$, using $q\leq1-t^2$, and minimizing the resulting
quadratic in $t$ gives
\[
 \phi\geq e^{7/8}\frac78
       \left(\frac{\lambda^2}{15/8}-1\right)
 >\frac{119}{50}\frac{21749}{75000}
 =\frac{2588131}{3750000}>\frac{69}{100}.
\]
For completeness, $e^{7/8}>119/50$ follows by retaining the first
five nonnegative terms of its Taylor series.

Also, $q\mapsto\phi(t,q)$ decreases on $0<q\leq1-t^2$,
because $(\lambda-t)^2-q\geq\lambda^2/2-1>0$.
The identity
\[
 (\lambda-t)^2-(\lambda^2-1)(1-t^2)=(\lambda t-1)^2
\]
therefore gives
\begin{equation}\label{eq:combined-coupled-potential-latitude}
 \phi\geq e^{\lambda^2-2}(1-t^2)
             >\frac{41}{25}(1-t^2).
\end{equation}
The last strict inequality follows, for example, from
$\lambda^2-2=0.4964>0.495$ and the first five Taylor terms of
$e^{0.495}$, whose sum exceeds $1.6402>41/25$.

\paragraph{The actual distance from the second-core center.}

Consider a point in the second departure class $G_2$.  The first-mask
geometry already proves $t>0.6$.  Its actual squared distance from
the second-core center is
\[
 v_*^2=(t-0.6)^2+q\leq1.36-1.2t<0.64.
\]
Put
\[
 f_*=(41/25)(16/25)=1.0496,
 \qquad u_*=(f_*-\phi)_+.
\]
When $\phi\leq f_*$, \eqref{eq:combined-coupled-potential-latitude} and
$t>0$ imply
\[
 v_*^2\leq h(\phi):=
 1.36-1.2\sqrt{1-\phi/1.64}.
\]
Here $h(f_*)=0.64$, and throughout $0.69<\phi\leq f_*$,
\[
 h'(\phi)=\frac{0.6}{1.64\sqrt{1-\phi/1.64}}
 >\frac{0.6}{1.64(0.77)}>0.47.
\]
The bound $\sqrt{1-0.69/1.64}<0.77$ follows by squaring.
For $\phi>f_*$ use the independent inequality $v_*^2<0.64$.
Both cases give the uniform pointwise estimate
\begin{equation}\label{eq:combined-coupled-actual-core-radius}
                 v_*^2\leq0.64-0.47u_*.
\end{equation}

Because $D_0^2<1.048576<f_*$, the prior exponent vanishes when
$\phi\geq f_*$.  Otherwise $\ln(1+x)\leq x$ and $\phi>0.69$
give
\begin{equation}\label{eq:combined-coupled-prior-excess}
 \sigma(\phi)\leq\tfrac12\ln(f_*/\phi)
       \leq\frac{f_*-\phi}{2\phi}<0.75u_*.
\end{equation}

\paragraph{A norm-tail rate using the actual radius.}

We condition on the second-stage target-axis norm event.  Write
$g=A_2u$; then $\|g\|^2\leq(1+\xi)k_2$ and
\[
 A_2(z-0.6u)=(t-0.6)g+\sqrt q\,Z,
                 \qquad Z\sim N(0,I_{k_2}).
\]
For $v_*>0$, apply the calculation of Section~\ref{sec:grid-tail}
to $(z-0.6u)/v_*$. The squared projected norm has moment-generating
function bounded by that of $(1+\xi)v_*^2\chi^2_{k_2}$.
Thus, for
\[
 a_0=\frac{\beta^2r_2^2}{1+\xi},\qquad
                    0.6697<a_0<0.67,
\]
the probability of lying outside the projected second core is at most
$\exp(-J_* k_2)$, where, for $v_*^2>0$,
\[
 J_* =\frac12\left[\frac{a_0}{v_*^2}-1
                       -\ln\frac{a_0}{v_*^2}\right]
 \geq\frac{(a_0-v_*^2)^2}{4a_0v_*^2}
 \geq\widehat J(u_*):=
                    \frac{(0.0297+0.47u_*)^2}{1.7152}.
\]
When $v_*^2=0$, the projected point is at the core center and the
tail probability is exactly zero, so the same upper bound with
$\widehat J$ remains valid.  The latter finite function is used in
all rounding estimates below; no unbounded rate is needed.

In particular,
\[
 \widehat J(u_*)\geq\frac{0.0297^2}{1.7152}>\frac1{1950}.
\]
The elementary square inequality gives
\[
 (0.0297+0.47u_*)^2\geq4(0.0297)(0.47)u_*.
\]
Since
\[
 0.75(1.7152)=1.2864
       <24\cdot4(0.0297)(0.47)=1.340064,
\]
\eqref{eq:combined-coupled-prior-excess} implies the key coupled bound
\begin{equation}\label{eq:combined-coupled-density-tail}
                        \sigma(\phi)<24\widehat J(u_*).
\end{equation}
It holds also when $u_*=0$, since then $\sigma=0$ and
$\widehat J>0$.

\paragraph{Combining the two independent stages.}

The second-stage same-cell term has a separate stronger uniform rate:
\[
 \frac{D_0r_2}{\sqrt\phi}
  <\frac{1.024(0.8)}{\sqrt{0.69}}<0.99.
\]
The last comparison is equivalent to
$0.67108864<0.676269$.  Put $\nu_{\rm cell}=-\ln(0.99)>1/100$.
On second-core isolation, a $G_2$ point passes stage two only by
sharing the target cell or by being a projected tail point whose
different cell hashes to the target bucket.  Therefore its probability
of reaching the target leaf while both isolation events hold is at
most
\[
 \frac{C}{B_1B_2}\left[
 e^{\sigma(\phi)k_1-\nu_{\rm cell} k_2}
 +e^{\sigma(\phi)k_1-\widehat J(u_*) k_2}\right].
\]
The first-stage factor bounds passage without imposing first-stage
isolation.  Independence of the two stage seeds permits
the multiplication after conditioning only on their separate axis
norm events.  No conditioning on the isolation event is used.

We take $G=32$ and
\[
 k_1=\lceil\theta\ln(N+2)\rceil,
 \qquad k_2=\lceil32\theta\ln(N+2)\rceil.
\]
Uniformly over the second class,
\[
 32\nu_{\rm cell}-\sigma>0.32-0.225=0.095>\tau/2,
\]
\[
 32\widehat J-\sigma>8\widehat J>8/1950>\tau/2.
\]
The ceiling in $k_1$ costs at most a factor $e^{\sigma}<5/4$;
the ceiling in $k_2$ only improves the bound.  Hence both terms are
$O(N^{-\tau\theta/2})/(B_1B_2)$ with a uniform constant, including
the case of zero actual core radius.  The first departure class
retains its earlier $O(N^{-\tau\theta})/(B_1B_2)$ bound.

The same indicator-weighted Markov argument, with norm and isolation
failures charged separately, consequently yields
\[
 B_1B_2K\leq C\left[
              B_1B_2(b+1)+N^{1-\epsilon_0}\right],
 \qquad\epsilon_0=\tau\theta/2.
\]
For an entirely rational capacity parameter, take
\begin{equation}\label{eq:factor-two-capacity}
 K=\left\lceil C'\left[
 b+1+\frac{Nq_0^{\lfloor k_1/2\rfloor}}{B_1B_2}
                         \right]\right\rceil
\end{equation}
with the explicit constant $C'=10^9$.
To verify this choice, the first-stage pass coefficient is at
most $401$, and the two second-stage coefficients are at most
$400$ and $1$. Each exponential product is at most
$2q_0^{\lfloor k_1/2\rfloor}$, including the ceilings and an
upward logarithm approximation within one. The sum of both
departure-class bounds is therefore at most
$3\cdot401^2<10^6$ times
$Nq_0^{\lfloor k_1/2\rfloor}/(B_1B_2)$.
Markov's inequality with capacity coefficient $10^9$ gives
failure probability at most $10^{-3}$.  Indeed,
$q_0^{\lfloor k_1/2\rfloor}$ differs from
$\exp(-\tau\theta\ln(N+2)/2)$ by at most a fixed factor, so this
capacity dominates the required load on the same success event and
has the displayed total-space exponent.  All its inputs are finite
integers or rational constants.
More explicitly, for $t_0=\theta\ln(N+2)$ and $k_1=\lceil t_0\rceil$,
the difference $|\lfloor k_1/2\rfloor-t_0/2|$ is at most $1/2$,
giving a factor between $q_0^{1/2}$ and $q_0^{-1/2}$.
Bounded-error rational logarithm approximations add only another
fixed factor.
This pointwise argument replaces the product of two unrelated worst
cases; it does not change any maintained sketches or their updates.

The power-of-two bucket counts used in the coefficients above admit
the affine binary hashes of Appendix~\ref{sec:finite-bit-gaussian-implementation}.

\subsubsection{An explicit preprocessing exponent of 2000}

Take
\[
 \kappa=\frac{109}{100},\qquad b_0=\frac{27}{25},\qquad
 \nu=\frac1{10000},\qquad e_0=\frac1{400}.
\]
Then
\[
 \kappa(1-\nu-2e_0)=1.084441>1.0801=b_0+\nu,
\]
so the approximation condition holds with fixed slack. Moreover,
\[
 I(b_0)\ge\frac{0.1664^2}{4(1.1664)}>\frac1{170},
\]
because $0.02768896\cdot170=4.7071232>4.6656$. Choose
\[
                 k=120\left\lceil\log_2(240m)\right\rceil.
\]
Since $\ln2<0.7$, we have
$kI(b_0)>(120/119)\ln(240m)$; the simultaneous ideal upper-tail
failure for the $m$ query vectors is at most $1/240$.

A near query has $\|q-s_0\|<(100/9)D$. Its rounded projected site
lies at distance less than
\[
             1.0801(100/9)D+D/800<13D
\]
from $\widetilde A s_0$. We use a rational table radius between $13D$ and $14D$.
The grid-step bounds then make every table key satisfy
$\|j\|\le11200\sqrt k$. At enumeration depth $i$, disjoint unit
cubes around feasible integer prefixes lie in an $i$-dimensional ball
of radius $11200\sqrt k+\sqrt i/2$. The volume bound
$v_i\le(2\pi e/i)^{i/2}$ and $\sqrt{2\pi e}<4.14$ give at most
\[
       (50000\sqrt{k/i})^i\le50000^k e^{k/(2e)}
\]
prefixes, where $i\ln(k/i)\le k/e$ was used. Since $e>8/3$ and
$\ln(5/4)\ge1/5>3/16$,
\[
                 50000 e^{1/(2e)}<62500<65536=2^{16}.
\]
All prefixes can therefore be enumerated in
$(k+1)2^{16k}$ steps up to polynomial factors, and
\[
                 2^{16k}\le2^{1920}(240m)^{1920}=C m^{1920}.
\]
One sample scan per grid site adds one factor of $m$. Exact sample
diameter computation and all coordinate and key arithmetic fit the
same bound. Thus the deliberately rounded preprocessing exponent
\begin{equation}\label{eq:one-sided-B109-improved}
                             B_\kappa=2000
\end{equation}
is valid. Remaining factors are polynomial in
$d,\log m,\log\Delta,b_q$ and the logarithmic repetition parameter.
The query time remains polynomial in these parameters, and every
returned point is an actual member of $S$.

\subsubsection{Oracle and cached score}

We take the candidate-certificate accuracy
$\eta_{\rm in}=1/100$.  At each valid active node center $c$ with
variance $V$, query the frozen sample's static $1.09$ farthest-point
oracle.  If it returns $s_*\in S$, score the actual pair
\[
                       \max_{z\text{ an anchor}}\|z-s_*\|.
\]
This squared score is at least $V+\|c-s_*\|^2$.  The center-query
variance argument therefore gives squared mixed-case factor
\[
 (1+\eta_{\rm in})^2+a^2(1.09)^2
       =3.9898284241<3.999122.
\]
Both-inside-hull pairs have squared factor $2a^2-1=3.999122$.
Both-outside pairs have squared factor
$(1+\eta_{\rm in})^2+1=2.0201$.  Thus the hull factor remains the
largest, and it is strictly below two.

Thus the choices proved in \eqref{eq:one-sided-B109-improved} and
\eqref{eq:one-sided-oracle-accuracy} apply to every center comparison
in Algorithm~\ref{alg:update}.

\subsubsection{Sample exponent, deadlines, and final space exponent}

We choose the rational exponents and dimension ratio
\[
 s=\frac1{2003},\qquad \rho=\frac s2,\qquad
 G=32,\qquad
 \theta=\frac{\rho}{12(G+1)}.
\]
We maintain a live sample of rate at most $N^{s-1}$, freeze it at global
intervals $H_{\rm epoch}=\max\{1,\lfloor N^{1-s}/100\rfloor\}$, and build its oracle over
one interval before serving it for the next.  The sample cap and
ball-net theorem give
$b=N^{1-s}\operatorname{poly}$ exceptions; the additional at most
$2H_{\rm epoch}$ intervening updates have the same order and are included in $b$.
The bounded-$N$ convention applies when $N^{1-s}<100$.
In a finite implementation the sampling probability may be a dyadic
number between $N^{s-1}/4$ and $N^{s-1}$, obtained from the integer
binary length of $N$ and the rational constant $s$.  This multiplies
the exception bound by at most four, preserves the sample-size upper
bound and survival probability, and changes no exponent.  The epoch
length can be computed by integer-power comparisons because $s$ is
rational.

Set
\[
       k_1=\lceil\theta\ln(N+2)\rceil,\qquad
       k_2=\lceil G\theta\ln(N+2)\rceil.
\]
Rational upper approximations to the logarithm within additive one
may be used before taking these ceilings.  This changes dimensions
only by fixed constants.  The bucket product is $O(N^\rho)$ because
$\theta(G+1)\ln C_g<\rho$, using $\ln C_g<11<12$.
The pointwise coupling just proved supplies exponent saving
\[
 \epsilon_0=\frac{\tau\theta}{2}
 >\frac1{96000\cdot2003\cdot33}
 =\frac1{6345504000}>10^{-10}.
\]
The remaining space exponents are
\[
           sB_\kappa=\frac{2000}{2003},
           \qquad
           1-s+\rho=1-\frac1{4006}.
\]
All three are strictly smaller than $1-10^{-10}$.
Oracle construction fits a global epoch with polynomial work per
original update because
\[
                 s(B_\kappa+1)-1=-\frac2{2003}<0.
\]

Substituting these exponents into \eqref{eq:cached-composition-space}
and the per-update bound \eqref{eq:frozen-oracle-work-budget} proves
the explicit factor-two claim. All repetitions, local reconstructions
and finite-precision costs enter through the polynomial factors.

\subsection{Bounded-precision implementation}
\label{sec:finite-bit-gaussian-implementation}

The Gaussian arguments can be implemented with a prescribed finite
number of random bits and a worst-case polynomial amount of arithmetic.
For the recovery grids we couple the implementation to the ideal
continuous construction so that every streamed point has exactly the
same cell identifier.  The static farthest-point oracle instead needs
only a uniform bound on the perturbation of projected distances.  Its
lookup cells need not agree with an ideal lookup table.

Throughout this section the stream is fixed independently of the
algorithm's random seeds.  This is the same oblivious-stream assumption
as in the upper theorem.  All arithmetic on the resulting rational
coordinates, grid indices, and original squared distances is exact.

\subsubsection{Budgets and the number of quantities to approximate}

Recall that $|p_j|\le\Delta$ for every input coordinate. Now
let $\ell_{\min}>0$ be a known lower bound on every recovery-grid side
length.  The grids in the construction have rational side lengths
\[
                  \ell_i=\delta_g\widetilde r_i w,
\]
where $w$ belongs to the prescribed finite rational scale list.
The encodings of the scale parameters and the magnitude of
$\log(1/\ell_{\min})$ are polynomially bounded in the input parameters.

Let $K$ be an upper bound on the total number of scalar grid rows,
counting a row again if it is used at a different scale or with a
different shift.  There are at most $T$ point identities appearing in
a stream of $T$ updates, even when an identity is deleted and reinserted.
Consequently
\[
                    E=\max\{1,TK\}
\]
is a sufficient bound on the number of scalar point/grid membership
tests that need to be preserved.  One may union-bound over every input
identity and every maintained grid, whether that particular pair is
ever evaluated or not.

Let $R\geq1$ bound the number of independent scalar Gaussian entries
generated throughout the computation, including the static-oracle
maps in every frozen-sample generation and every repetition.  Such a
bound is known in advance.  Recovery matrices are shared by all leaves
at their stage; they are not generated separately for every leaf.
There are polynomially many recovery matrices in the logarithmic
parameters.  Every static sample generation uses polynomially many
maps, each with $d\,O(\log N)$ entries, and there are at most $T+1$
generations per copy.  Enumerating its large lookup table draws no
additional Gaussian entries.  Thus a sufficiently generous bound is
\[
 R\leq (N+T+d+2)^{C}
       \bigl(1+\log\Delta+\log(1/\delta)\bigr)^{C}
\]
for a fixed constant $C$ depending only on the fixed approximation
parameters.  The relevant fact is that $\log R$ is polynomial in the
allowed dimension and logarithmic parameters.  Using a larger bound
only increases precision logarithmically.

Gaussian maps can be indexed by their copy, scale, stage, and sample
generation, reserving the largest allowed row count and ignoring
unused rows.  This supplies a fixed family of at most $R$ entries to
couple.  The structural update schedule never depends on a static
oracle's returned answer.  Alternatively, fresh entries can be drawn
sequentially with the same deterministic upper bound on their number;
the conditional union bound below is unchanged.

\subsubsection{Generating a rounded Gaussian with bounded work}

Fix an entry error $h=2^{-b}$, $b\geq1$, and a desired failure budget
$\varepsilon_{\mathrm{cpl}}\in(0,1)$.  Put
\[
 L=\left\lceil\log_2(8R/\varepsilon_{\mathrm{cpl}})\right\rceil,
 \qquad M=\left\lceil\sqrt{2L}\right\rceil.
\]
Only integer comparisons are needed to obtain these upper bounds.
For a standard Gaussian $G$,
\[
 \Pr[|G|>M]\leq2e^{-M^2/2}
       \leq2\,2^{-L}\leq\varepsilon_{\mathrm{cpl}}/(4R).
\]
Thus all $R$ ideal Gaussian entries lie in $[-M,M]$ except with
probability at most $\varepsilon_{\mathrm{cpl}}/4$.

Here is a finite construction coupled to one ideal entry.  Let
$U$ be uniform on $(0,1)$ and set $G=\Phi^{-1}(U)$, where $\Phi$ is
the standard Gaussian distribution function.  The implementation
reads only the first
\[
                       r=b+M^2+8
\]
bits of $U$, obtaining a uniform integer $J\in\{0,\ldots,2^r-1\}$,
and uses $u_J=(J+1/2)2^{-r}$.  This is just a coupling description:
the actual algorithm draws the $r$ independent bits directly.

The clipped inverse
\[
 g_M(u)=\max\{-M,\min\{M,\Phi^{-1}(u)\}\}
\]
is globally Lipschitz with constant at most
\[
 \sqrt{2\pi}e^{M^2/2}\leq3\,2^{M^2}.
\]
It follows that
\[
 |g_M(U)-g_M(u_J)|\leq3\,2^{M^2-r-1}<h/8.
\]
We compute $g_M(u_J)$ to additive error at most $h/4$ by the bounded
bisection described below, and round the result to a dyadic number
with mesh at most $h/4$.  The resulting rational $\widetilde G$ obeys
\[
 |\widetilde G-g_M(U)|<h,
 \qquad |\widetilde G|\leq M+h,
\]
on every outcome.  In particular
$|\widetilde G-G|<h$ whenever $|G|\leq M$.
Each entry uses exactly $r$ random bits and a prescribed polynomial
upper bound on its arithmetic work.  There is no rejection loop.

\paragraph{A certified inverse-CDF calculation.}
On $[-M,M]$ the Gaussian density is at least
\[
                    m_0=\tfrac13\,2^{-M^2}.
\]
At a bisection midpoint $x$, evaluate $\Phi(x)$ with absolute error
at most $h m_0/100$.  If its resulting interval is entirely below
or above $u_J$, keep the appropriate half interval.  If the interval
contains $u_J$, the midpoint is within $h/50$ of the clipped inverse,
by the density lower bound; return it.  If no such early termination
occurs, at most $\lceil\log_2(32(M+1)/h)\rceil$ iterations make the
remaining bracket shorter than $h/8$.  This argument also applies
when the unclipped inverse lies outside $[-M,M]$: bisection converges
to the appropriate endpoint.  The number of iterations is bounded
independently of all random outcomes.

For completeness, the needed evaluations of $\Phi$ also take
polynomial time.  For rational $x\in[-M,M]$ use
\[
 \Phi(x)=\frac12+\frac1{\sqrt{2\pi}}
       \int_0^x e^{-t^2/2}\,dt
\]
and integrate the Taylor polynomial
\[
 \sum_{j=0}^{J_0}\frac{(-1)^j t^{2j}}{2^j j!}
\]
term by term.  The pointwise absolute remainder is at most
$(M^2/2)^{J_0+1}/(J_0+1)!$.  Taking
$J_0+1\geq3M^2$ and
$J_0+1\geq\log_2(8(M+1)/\epsilon)$ makes the integral error below
$\epsilon/8$, using $j!\geq(j/e)^j$ and $e<3$.
The required number of terms is therefore
$O(M^2+\log(1/\epsilon)+\log(M+1))$.
Exact rational intermediate values have polynomial bit length,
even though cancellation can occur.  The constant $1/\sqrt{2\pi}$
can be enclosed to the additional required precision in polynomial
time: for example, use the geometrically convergent arctangent
series in
$\pi=16\arctan(1/5)-4\arctan(1/239)$, followed by rational
bisection for the square root.  These formulas give explicit
certified error bounds and fixed polynomial operation counts.

Normalized maps $A=G/\sqrt{k}$ are handled the same way.  Approximate
each $G$ sufficiently finely, approximate $1/\sqrt{k}$ by rational
bisection, and multiply, reserving half of the desired entry error
for each operation.  This adds only
$O(\log k+\log(M+1))$ precision bits.  In either normalization the
implementation can guarantee a prescribed entrywise error $h$ on
the common truncation event.

\subsubsection{Preserving all recovery-grid identifiers}

Write a scalar ideal cell identifier as
\[
             \left\lfloor\frac{a\cdot p}{\ell}+U_0\right\rfloor,
\]
where $a$ is a standard Gaussian row and $U_0$ is an independent
uniform number in $[0,1)$.  It is equivalent to the usual uniformly
shifted grid of side $\ell$.  Couple the implementation's shift
$\widetilde U_0$ to the first $b_s$ bits of $U_0$, rounded down.
Then
$|\widetilde U_0-U_0|\leq2^{-b_s}$ deterministically, and the
implemented physical shift $\ell\widetilde U_0$ is rational.

Set
\[
 \eta=\frac{\varepsilon_{\mathrm{cpl}}}{8E},
 \qquad
 h\leq\frac{\eta\ell_{\min}}{2d\Delta},
 \qquad 2^{-b_s}\leq\eta/2.
\]
The entry mesh may be rounded downward to the next dyadic value.
On the common Gaussian truncation event, for every input point and
every grid row,
\[
 \left|\frac{\widetilde a\cdot p}{\ell}
          +\widetilde U_0
       -\frac{a\cdot p}{\ell}-U_0\right|
 \leq\frac{d\Delta h}{\ell_{\min}}+2^{-b_s}\leq\eta.
\]
For a fixed input point, conditional on the ideal Gaussian matrix,
the fractional part of $a\cdot p/\ell+U_0$ is uniform.  Its distance
to an integer is at most $\eta$ with probability at most $2\eta$.
Union-bounding over the at most $E$ scalar tests gives probability
at most $2E\eta=\varepsilon_{\mathrm{cpl}}/4$ that any relevant ideal
coordinate is too close to a boundary.  Independence between these
boundary events is unnecessary.  On their complement every floor,
and hence every full grid key and outer bucket identifier, is
exactly unchanged.

Combining the boundary and Gaussian-truncation bounds, with
probability at least $1-\varepsilon_{\mathrm{cpl}}$ all streamed grid
memberships agree with the ideal continuous construction.  Repeated insertions and deletions use the same stored finite
matrix and shift, so the key of a point is a fixed function of its
identity on every outcome, including a failed coupling outcome.
Strict-turnstile legality inside every bucket is consequently
preserved without any probability qualification.

This use of uniform shifts is why no anti-concentration assumption
on the point coordinates or on a particular Gaussian dot product
is needed.  The fixed oblivious input sequence is essential to the
displayed boundary union bound; it does not assert the same guarantee
for points chosen after inspecting the grid seeds.

\subsubsection{Rational-center queries in the static oracle}

The static oracle is built and queried using its implemented rational
matrix $\widetilde A$.  Its distance proof does not require the lookup
cell of a query to agree with that under the ideal matrix.  It is
enough to transfer the ideal norm upper and lower events to
$\widetilde A$ with the reserved allowance $\nu$.

Let $k_{\max}$ be the largest number of rows in a normalized oracle map.
We use entrywise error $h\le\nu/\sqrt{k_{\max}d}$ on the common truncation
event. For a $k\times d$ map and every $v\in\mathbb R^d$,
\[
 \|(\widetilde A-A)v\|_2
 \le \|\widetilde A-A\|_{\rm F}\,\|v\|_2
 \le \sqrt{kd}\,h\,\|v\|_2\le\nu\|v\|_2 .
\]
This also holds for $v=0$. Hence any ideal upper bound $b_0\|v\|$
transfers to $(b_0+\nu)\|v\|$, and ideal noncontraction transfers to
$(1-\nu)\|v\|$. The perturbation estimate is uniform over all vectors.
The probabilistic norm bounds for the ideal map apply to queries
independent of its seeds, as in Lemma~\ref{lem:static-oracle}.
A certificate center is a rational convex combination of input points
in $[-\Delta,\Delta]^d$. Its polynomial bit bound $b_q$ controls the
cost of the projected query, grid rounding, and exact distance comparisons.

The table grid step and its enclosing radius can be chosen rationally
within the intervals specified in the static-oracle lemma by integer
squared comparisons or rational square-root bisection.  The strict
constant slack in those intervals makes this a polynomial-bit task.
All table entries are original sample identities, and final scores
are exact original squared distances, so projection rounding cannot
create a false witness or an overestimate of the current diameter.

\subsubsection{Resource and probability accounting}

For a common precision, take the largest dyadic $h\le1/2$ satisfying
both entry-error restrictions (a smaller dyadic value within a factor
of two also suffices). The required number of bits is at most
\[
 \begin{aligned}
 O\!\big(&\log(E+2)+\log(R+2)
             +\log(1/\varepsilon_{\mathrm{cpl}})\\
         &+\log(d\Delta+2)+\log(k_{\max}+2)
          +\log(1+1/\ell_{\min})+\log(1/\nu)\big).
 \end{aligned}
\]
The shift precision, Gaussian truncation length, random-bit count
per entry, and arithmetic precision are polynomial in the same
parameters.  In the diameter construction these are all polynomial
in $d$, the input coordinate bit bound, $\log N$, and
$\log(T/\delta)$, with constants depending only on the chosen
approximation factor.  Exact grid indices have polynomial bit
length as well, since
\[
 \abs{\left\lfloor\widetilde a\cdot p/\ell
                   +\widetilde U_0\right\rfloor}
       \leq 2+d\Delta(M+1)/\ell_{\min}.
\]
They can be encoded injectively as bounded binary strings for the
outer pairwise-independent affine hashes.

More explicitly, choose an integer $Q_0$ above the displayed bound
and encode each grid key in $[-Q_0,Q_0]^k$ using a fixed-length signed
integer encoding.  An affine binary-linear map from that finite bit
domain to the bucket bits is pairwise independent: for two distinct
keys, their difference is nonzero, so each independently sampled
matrix row and its independent affine offset give two independent
uniform output bits.  For the ideal comparison on the full lattice, extend this random
hash by giving each out-of-domain cell an independent uniform bucket
label. The extended family is pairwise independent, so the ideal
isolation and collision estimates apply on every grid. The actual
algorithm generates only the finite affine map. On the coupling event
all streamed cell keys belong to the implemented domain and agree in
the two processes. On every outcome the clipped implementation has
bounded keys.

Generating each finite matrix takes a fixed polynomial amount of
work per entry.  Initial recovery matrices are generated in the
allowed preprocessing; frozen static-oracle maps are generated
within the already budgeted background construction job.  This does
not introduce a positive power of $N$ into an original update and
does not change any space exponent.  There is no assumption of
constant-time real arithmetic or an unlimited random tape.

Allocate, for example, $\varepsilon_{\mathrm{cpl}}=\delta/10$ to the
entire finite-precision coupling and distribute the remaining error
budget among the ideal probabilistic lemmas. Apply the independent-copy
amplification in the ideal experiment before charging the coupling failure.  On the coupling event,
all original-point recovery memberships are those analyzed in the
continuous proof, and every static norm event transfers with its
reserved slack.  A union bound supplies the desired simultaneous
guarantee over all $T$ updates.  Resource bounds, fixed bucket
membership, exact sparse-recovery certification, and validity of
every reported nonzero current-pair distance hold on every random
outcome, independently of whether this coupling succeeds.

\subsection{The Gaussian-embedding baseline}
\label{sec:two-thirds-baseline}

For comparison, consider an update bound of the form
$N^\alpha\operatorname{poly}(d,\log N)$ for a fixed $\alpha<1$.
There is a substantially simpler baseline at exponent $2/3$ when both
space and update time may have that exponent.  This baseline is based
on Indyk's Gaussian embedding, in the dynamic-streaming formulation
of Padaki, Singal, and Waingarten~\cite{psw}.
The implementation below makes its update and reporting costs
explicit using the certified sparse-recovery primitive in this paper.

\begin{theorem}[Two-thirds-power baseline]
In the same bounded-integer, oblivious, strict-turnstile model, there
is a one-pass factor-two diameter estimator with
\[
 \begin{aligned}
 \text{space}&=N^{2/3}\operatorname{poly}(d,\log(N+2),
                                  \log(\Delta+2),\log((T+2)/\delta)),\\
 \text{worst-case update time}&=N^{2/3}\operatorname{poly}(d,\log(N+2),
                                  \log(\Delta+2),\log((T+2)/\delta)).
 \end{aligned}
\]
It caches a scalar answer after every update.  With probability at
least $1-\delta$, all answers satisfy $D_t/2\le L_t\le D_t$.
Both approximation inequalities are conditioned on this success event.
\end{theorem}

\subsubsection{The Gaussian maximum embedding}

Put
\[
 q=\lceil\log_2(N+2)\rceil,\qquad
 \gamma=\frac1{100q},\qquad
 a=\frac2{1+8\gamma},\qquad
 k=\left\lceil N^{2/(a^2-1)}q^{16}\right\rceil.
\]
The expansion $2/(a^2-1)=2/3+O(1/q)$ shows that
$k=N^{2/3}\operatorname{poly}(\log(N+2))$.
Let $G$ have $k$ independent standard-Gaussian rows in $\mathbb R^d$,
and put
\[
                  \rho=\frac1a\sqrt{2\ln(40kN^2)}.
\]
For any fixed point set $P$ of size at most $N$, with probability
at least $0.9$,
\begin{equation}\label{eq:baseline-embedding}
 \rho\|x-y\|\le\|Gx-Gy\|_\infty
                    \le a\rho\|x-y\|\qquad(x,y\in P).
\end{equation}
For completeness, for a fixed nonzero difference the normalized
row inner products are independent standard Gaussians.  A union bound and
$\Pr(|Z|\ge t)\le2e^{-t^2/2}$ give upper failure at most
$1/(20N^2)$.  The lower Gaussian tail is at least
$C\rho^{-1}e^{-\rho^2/2}$ when $\rho\ge2$.  Thus the expected number
of coordinates exceeding $\rho$ is at least
\[
 C\rho^{-1} k^{1-1/a^2}N^{-2/a^2}
       =\Omega\bigl(q^{16(1-1/a^2)-1/2}\bigr).
\]
The exponent is uniformly greater than one, since $a>1.9$.
For all sufficiently large $N$ this exceeds $\ln(20N^2)$, so the
lower failure for that difference is at most $1/(20N^2)$.
Union-bound over pairs.  Bounded values of $N$ are handled by storing
the active set, with a fixed constant absorbed in the bounds.

We use the finite-bit Gaussian construction in Appendix~\ref{sec:finite-bit-gaussian-implementation}
to replace $G$ by a dyadic matrix $A$ of bounded entries, with
\[
                 |A_{ij}-G_{ij}|\le h,
                 \qquad 2d\Delta h\le\gamma\rho.
\]
Its coupling failure can be made below $1/100$ for one whole matrix.
Every distinct pair of integer points has distance at least one.
Consequently, on the embedding and rounding events,
\begin{equation}\label{eq:baseline-rounded-embedding}
 (1-\gamma)\rho\|x-y\|
 \le\|Ax-Ay\|_\infty
 \le(a+\gamma)\rho\|x-y\|.
\end{equation}
The absolute entry precision and truncation bounds have polynomial
bit length in the displayed parameters.  No union bound over the
stream is used to increase $k$; amplification will be outside this
embedding.

\subsubsection{Maintaining one projected coordinate range}

Fix a row of $A$ and write $z(x)=\langle A_i,x\rangle$.
These are exact dyadic rationals with a common denominator $2^b$.
Maintain, for every dyadic length $\ell$ in a fixed range, the
frequency vector of interval identifiers
\[
                         j(x)=\lfloor z(x)/\ell\rfloor.
\]
One original update changes exactly one interval count for that row
and scale.  All these counts are nonnegative.  Set
$K=\lceil4/\gamma+3\rceil$ and maintain the certified $K$-sparse recovery sketch of Section~\ref{sec:exact-recovery} on interval identifiers.  Use
$O(\log(kK(J+1)))$ independent hash tables, where $J+1$ is the
number of lengths.  At a fixed time, every vector with at most $K$
occupied intervals is recovered with sufficiently high probability
that all such rows and scales succeed together with probability
at least $0.99$.  This follows conditionally on $A$, since the sparse
sketch seeds are independent of $A$.

For a certified, nonempty interval dictionary, let $j_{\min}$ and
$j_{\max}$ be its extreme identifiers and put
\[
              U_{i,\ell}=(j_{\max}-j_{\min}+1)\ell.
\]
This is always an upper bound on the actual coordinate range $w_i$.
If the decoder does not certify completeness, ignore that scale.
Also maintain the exact total mass and first two moments of $z$;
the zero-variance identity certifies $w_i=0$, including any number
of original points with the same projected coordinate.

Let the smallest length be the largest dyadic number at most
$\gamma2^{-b}/4$, and let the largest length be the least dyadic
number at least twice the absolute projected-coordinate bound.  The number
$J+1$ is polynomial in the logarithmic and dimension parameters.
For every $w_i>0$ there is a length
\[
                    \gamma w_i/4\le\ell\le\gamma w_i/2.
\]
Its occupied-interval count is at most $w_i/\ell+2\le K$, and
its recovered bound satisfies $U_{i,\ell}\le w_i+2\ell
\le(1+\gamma)w_i$.
Let $\mathcal C_i$ be the set of lengths with a certified interval
dictionary for row $i$. If $w_i>0$ and $\mathcal C_i\ne\varnothing$,
define
\[
 U_i:=\min_{\ell\in\mathcal C_i}U_{i,\ell}.
\]
We use $U_i=0$ when the support is empty or the exact moment test gives
$w_i=0$. If $w_i>0$ and $\mathcal C_i=\varnothing$, use twice the
known absolute projected-coordinate bound. On the simultaneous
recovery event the scale above belongs to $\mathcal C_i$ whenever
$w_i>0$, and hence $w_i\le U_i\le(1+\gamma)w_i$ in all cases.

The sparse decoder costs $K\operatorname{poly}$, and here $K$ is
itself polynomial in $\log N$.  We may therefore decode every scale of every row immediately after
an update and maintain $U=\max_i U_i$.
There is no postponed large decoding job and no query-time scan.

\subsubsection{Normalization, amplification, and the actual update bound}

We choose a rational $\widetilde\rho$ with
$\rho\le\widetilde\rho\le(1+\gamma)\rho$, and cache
\[
                    L=\frac{U}{(1+\gamma)(a+\gamma)
                                             \widetilde\rho}.
\]
On the preceding good events, if $D=\operatorname{diam}(P)>0$,
\[
 \frac{1-\gamma}{(1+\gamma)^2(a+\gamma)}D
                         \le L\le D.
\]
The lower coefficient is greater than $1/2$.  Indeed, multiplying
the desired inequality by $1+8\gamma$ reduces it to positivity of
\[
                       9\gamma-28\gamma^2-17\gamma^3-8\gamma^4,
\]
which holds for $0<\gamma\le1/100$.

At every fixed time, one whole copy succeeds with probability greater
than $0.8$. We run $O(\log((T+2)/\delta))$ independent whole copies and cache their median.  A Chernoff bound and a union bound over the
prescribed horizon give the stated simultaneous guarantee.  The median tolerates scalar errors in either direction. Taking a
maximum has a different justification in Algorithm~\ref{alg:update},
where every positive score has current witnesses.

An original point update computes $k$ dot products, updates
polynomially many interval sketches per dot product, decodes only
polynomial-size sketches, and updates the cached maximum and median.
Every operation therefore costs $k\operatorname{poly}$, in the
worst case.  The storage is also $k\operatorname{poly}$.  Counts have
$O(\log(T+2))$ bits, and interval keys, Gaussian approximations,
moment counters, and normalization values have polynomial bit length.
No step uses expiration times, an original-point replay, or an
uncharged copy of the input.

All parameter choices are effectively computable with polynomial
bit overhead.  In particular, $2/(a^2-1)$ is rational with numerator
and denominator polynomial in $q$; its power and ceiling can be
found by integer-power comparisons.  Certified rational bounds
on the logarithm and square root suffice for $\widetilde\rho$ and
the entry-precision choices.

This baseline has a much better memory exponent than our polynomial-update construction, but it pays the same $N^{2/3}$
power in update time.  The two guarantees answer different versions
of the time--space tradeoff.

\end{document}